\documentclass[useAMS,usenatbib,referee]{mn2e}
\usepackage{graphicx}        % For eps figures, newer & more powerfull
\usepackage{amssymb}        % useful mathematical symbols
\usepackage{color}           % For color text: \color command
\usepackage{url}             % For breaking URLs easily trough lines
\usepackage{epstopdf}
\newcommand{\rin}{r_{\rm in}}
\newcommand{\rout}{r_{\rm out}}
\newcommand{\rcz}{r_{\rm cz}}
\newcommand{\pran}{{\rm Pr}}

\begin{document}

\title[Dynamics of the solar tachocline]{Dynamics of the solar tachocline III: Numerical solutions of the Gough \& McIntyre model }
\author[L. A. Acevedo-Arreguin, P. Garaud \& T. S. Wood]{L. A. Acevedo-Arreguin$^1$, P. Garaud$^1$\thanks{E-mail: pgaraud@ucsc.edu (PG)} \& T. S. Wood$^1$ \\
$^1$Department of Applied Mathematics and Statistics, Baskin School of Engineering, \\
University of California Santa Cruz, 1156 High Street, Santa Cruz, CA 95064, USA}
\date{}

\maketitle

\begin{abstract}
We present the first numerical simulations of the solar interior to exhibit a tachocline consistent with the Gough \& McIntyre (1998) model. We find nonlinear, axisymmetric, steady-state numerical solutions in which: (1) a large-scale primordial field is confined within the radiation zone by downwelling meridional flows that are gyroscopically pumped in the convection zone (2) the radiation zone is in almost-uniform rotation, with a rotation rate consistent with observations (3) the bulk of the tachocline is magnetic free, in thermal-wind balance and in thermal equilibrium and (4) the interaction between the field and the flows takes place within a very thin magnetic boundary layer, the tachopause, located at the bottom of the tachocline. We show that the thickness of the tachocline scales with the amplitude of the meridional flows exactly as predicted by Gough \& McIntyre. We also determine the parameter conditions under which such solutions can be obtained, and provide a simple explanation for the failure of previous numerical attempts at reproducing the Gough \& McIntyre model. Finally, we discuss the implications of our findings for future numerical models of the solar interior, and for future observations of the Sun and other stars. 
%Checked, ok.
 \end{abstract}


\section{Introduction}
     \label{intro} 

The rotation profile of the solar interior, which is nearly uniform within the radiation zone yet strongly differential within the convection zone \citep{JCDSchou88,Brownal89},
presents a serious challenge to theoreticians \citep[see reviews by][]{Zahn07,Garaud07}.
In particular, the thinness of the tachocline between these two zones implies that the transport of angular momentum below the radiative--convective interface must be predominantly \emph{latitudinal}, yet the absence of rapid core rotation also requires significant \emph{radial} transport throughout the radiation zone. At a cursory glance, this appears to be at odds with helioseismic sound-speed inversions, which suggest that the bulk of the radiation zone undergoes very little compositional mixing \citep{ElliottGough99}. This apparent contradiction can be resolved if the radiation zone harbors a global-scale magnetic field, which transports angular momentum without engendering significant mixing. Such a field can also explain the uniform rotation of the radiation zone \citep{Ferraro37,MestelWeiss87,RudigerKitchatinov97}, as long as it remains confined strictly below the tachocline \citep{GoughMcIntyre98,MacGregorCharbonneau99}.
The key problem to developing a self-consistent model of the solar interior within this framework
is to explain how the confinement of this field is maintained.
%Checked ok.

The first model to address the ``magnetic confinement problem''
was proposed by \citet[][GM98 hereafter]{GoughMcIntyre98}. 
They argued that the field is confined by a large-scale meridional 
circulation, gyroscopically pumped\footnote{Gyroscopic pumping is a mechanism whereby any azimuthal forcing, by conservation of angular momentum, also drives meridional flows. Gyroscopic pumping occurs in the solar interior by way of the turbulent stresses exerted by the convection zone onto the radiation zone.} by the convection zone, which converges at high latitudes and 
downwells into the radiation zone. In this model, the downwelling flows confine the magnetic field across a very thin layer located, by construction, 
at the bottom of the tachocline, which they called the ``tachopause''.
The thickness of the tachopause is determined by a balance between the downward burrowing of the flows
and the upward diffusion of the field. At the same time, the remaining shear in the tachopause winds up the confined field lines,
providing a Lorentz torque that pumps the meridional flow equatorward. %Thus, the strength of the meridional flow in the tachopause is linked to the strength of the confined magnetic field.
The flows returns to the convection zone in a mid-latitude upwelling region,
whose dynamics were not explicitly addressed in the GM98 model.
%Checked ok.

In this picture, the bulk of the tachocline (excluding the tachopause) is ``magnetic free". Its dynamics are regulated by a balance of forces between thermal buoyancy, pressure, and the Coriolis force, leading to the so-called thermal-wind relation \citep{Pedlosky-GFD}.
The differential rotation of the tachocline thus implies the presence of latitudinal temperature and
entropy gradients, with a ``hot spot'' at the pole \citep[GM98;][]{Rempel05,Miesch-etal06}.
The diffusion of the polar temperature anomaly is balanced by the advection of the background %subadiabatic 
stratification by the downwelling flow.
In this way, the stratification and thickness of the tachocline constrain the mass flux allowed through it. Since the same mass flux must 
flow through the tachopause, and is regulated by the Lorentz torque therein (see above discussion), the thickness of the tachocline is tied to the strength of the internal magnetic field. 
%Checked ok.

Since 1998, a number of attempts have been made to obtain the GM98 tachocline in a self-consistent numerical model. Among them are the time-dependent, either axisymmetric or fully three-dimensional (3D) global numerical simulations of \citet{BrunZahn06}, \citet{Rogers11} and \citet{Strugarekal11}.
These simulations are initialized with a global-scale dipolar magnetic field confined within a uniformly rotating radiation zone, and follow the resulting interaction of the field
with differential rotation and meridional flows driven in the overlying layers.
\citet{BrunZahn06} used a 3D model, but their computational domain only included the radiation zone, the top of which was modeled as an impenetrable boundary with an imposed latitudinal differential rotation. The more recent model of \citet{Rogers11} includes both the radiation and convection zones, but in a 2D meridional slice. Lastly, \citet{Strugarekal11} have extended the 3D work of \citet{BrunZahn06} by including the convection zone in their computational domain.
%Checked ok.

In all cases, the initially confined magnetic field gradually connects to the convection zone by magnetic diffusion. 
As a result, none of these studies recovers the picture of the solar interior envisaged by GM98.
In the models of \citet{BrunZahn06} and \citet{Strugarekal11}, the differential rotation propagates rapidly into the radiation zone once the field lines connect to the convection zone, as expected from Ferraro's isorotation law \citep{Ferraro37}. The resulting angular-velocity profile then bears little resemblance with observations. \citet{Rogers11} also finds that the field ultimately spreads throughout the whole domain, but that the radiative region remains mostly in solid-body rotation. The absence of Ferraro's isorotation probably results from the fact that the magnetic field is relatively weak in that model. Indeed, a relevant measure of the field strength is the Elsasser number, defined as $\Lambda = B^2/4\pi \rho \eta \Omega$ (where $B$ is the field amplitude, $\rho$ is the density, $\eta$ is the magnetic diffusivity and $\Omega$ is the rotation rate). In the simulations of \citet{Rogers11}, $\Lambda < 1$ throughout the radiation zone, which may explain why the magnetic field appears to play no dynamical role.
%Checked ok.

The failure of global numerical simulations to obtain solar-like dynamics have led some to conclude that a primordial magnetic field cannot explain the uniform rotation of the solar interior \citep{Strugarekal11}. However, we believe that this conclusion is premature. The numerical model parameters used in all simulations to date are necessarily very far from their corresponding solar values, because of computational limitations. In particular, the values used for the magnetic diffusivity and viscosity are typically many orders of magnitude larger than their microscopic solar counterparts. In each of the simulations reported above, the transport of magnetic field and angular momentum is dominated by these (artificially large) diffusivities.
In the original model of GM98, on the other hand,
viscosity is assumed to be entirely negligible, and magnetic diffusion is important
only within the very thin tachopause. One should therefore consider very carefully the physical parameter conditions under which the GM98 model
might be expected to apply. As we now show, this depends not only on the absolute magnitude of the diffusivities, but also on their ratios.
%Checked ok.

As demonstrated by \citet{WoodBrummell12}  \citep[see also][]{GaraudBrummell08,GaraudAA09,GaraudGaraud08},
the importance of viscosity in the radiation zone can be described in terms of the dimensionless parameter $\sigma$,
which is defined as
\begin{equation}
\sigma = \sqrt{\frac{\nu}{\kappa}} \frac{\bar N}{\Omega_\odot} \mbox{   ,}
%Checked ok.
\end{equation}
where $\nu$ is the viscosity, $\kappa$ is the thermal diffusivity, $\bar N$ is the Brunt--V\"ais\"al\"a frequency, and $\Omega_\odot$ is the Sun's mean rotation rate. This parameter can be interpreted as the ratio of the timescales for angular-momentum transport by Eddington--Sweet meridional flows and by viscous stresses respectively, across the same region. An analogous parameter appears in geophysical studies of stratified rotating flows
\citep[e.g.][]{Lineykin55,BarcilonPedlosky67}.
%Checked ok.

The condition for viscous effects to be negligible in a magnetic-free tachocline of thickness $\Delta$ is $\Delta \ll r_{\rm cz}/\sigma$, where $r_{\rm cz}$ is the radius of the base of the convection zone. It is satisfied in the solar interior, where $\sigma$ varies from 0 to about 0.5 within the tachocline \citep{GaraudAA09}, but is demonstrably not satisfied in any of the global numerical simulations described above. Indeed, the latter use realistic values for $\Omega_\odot$ and $\bar N$, but have unrealistic diffusivities (and more importantly, an
unrealistically large Prandtl number $\pran =\nu/\kappa$), leading to values of $\sigma$ that are significantly larger than one. We demonstrate in this paper that, in the regime $\sigma \gg 1$, meridional flows downwelling from the convection zone are strongly suppressed and therefore unable to confine an interior magnetic field.
%Checked ok.

There are essentially two different routes toward achieving the $\sigma < 1$ regime in simulations: by decreasing Pr  while keeping $\bar N/\Omega_\odot$ constant, or by decreasing $\bar N/\Omega_\odot$ while keeping Pr constant.
The studies described above all took the first route, using solar profiles for $\bar N/\Omega_\odot$
and making Pr as small as computationally possible.  Unfortunately, to achieve $\sigma < 1$ in the tachocline would then require
$\pran \lesssim 10^{-5}$, which is far beyond the capabilities of any 2.5D or 3D numerical code.
Here, we favor the second route: we make Pr as small as possible in our numerical model and artificially decrease $\bar N/\Omega_\odot $ to achieve $\sigma <1$. Although in this approach neither $\bar N/\Omega_\odot$ nor Pr take their ``true'' solar values, we argue that this shortcoming is superseded by the need to achieve a non-viscous dynamical regime. 
% Checked ok .

A ``proof-of-concept'' of this approach was recently presented by \citet{WoodBrummell12}, using 3D numerical simulations in a local Cartesian domain straddling the radiative--convective interface. They show that an angular-velocity shear forced in the convection zone propagates into the radiation zone either by viscous diffusion or by advection by meridional flows,
depending on whether $\sigma > 1$ or $\sigma < 1$. In cases with
$\sigma > 1$, they find that 
meridional flows decay exponentially with depth beneath the radiative--convective interface on a lengthscale $\sim r_{\rm cz}/\sigma$, as predicted by \citet{GaraudBrummell08} and \citet{GaraudAA09},
and that viscous stresses dominate angular-momentum transport.
This explains why the global numerical simulations described above, which have $\sigma \gg 1$ in the tachocline, are all dominated by viscous effects, and most likely also explains why the meridional flows are unable to confine the magnetic field. A similar conclusion had earlier been reached by \citet[][GG08 hereafter]{GaraudGaraud08} albeit through more idealized simulations.
By contrast, in cases with $\sigma < 1$, \citet{WoodBrummell12} find that viscous stresses are essentially negligible. Angular momentum is transported by large-scale meridional flows that burrow into the radiation zone on a local Eddington--Sweet timescale, as first discussed by \citet{SpiegelZahn92}.
These flows retain a significant amplitude in the radiation zone, as expected from the work of \citet{GaraudBrummell08} and \citet{GaraudAA09}. 
% Checked ok .

In short, \citet{WoodBrummell12} showed that it is possible to obtain solar-like dynamics without using true solar parameters, by identifying the appropriate parameter regime -- in the case of the tachocline this requires having $\sigma < 1$. However, the numerical challenge involved in modeling simultaneously the convection zone and the radiation zone, in a full 3D MHD simulation and in a parameter regime where $\sigma < 1$, remains quite formidable. To prepare for a time in the not-so-distant future where such simulations are possible, we have already begun to study the problem using reduced models, which enable us to refine our understanding of the dynamics of the system and provide insight into the appropriate selection of other parameters. In \citet{Woodal11}, we studied a Cartesian and laminar toy model of the solar interior. Its analytical simplicity enabled us to understand in greater detail the structure and global force balance of the tachocline and the tachopause, leading us to propose a set of diffusion coefficients (viscosity, magnetic, and thermal diffusivity) that can realistically be used in simulations to yield a GM98-like tachocline. The next step is to use these proposed parameters in two concurrent and complementary models: in an axisymmetric, steady-state, nonlinear model of the full solar interior (this paper), and in a 3D, fully nonlinear local Cartesian model of the tachocline (Wood \& Brummell, in prep). 
% Checked ok .

In what follows, we consider a global quasi-steady axisymmetric numerical model of the solar interior, presented in detail in Section \ref{ourmodel}. Section \ref{issues} discusses an important yet subtle issue discovered in the process of searching for tachocline-like solutions, namely the difference between magnetic confinement of the type considered by GM98, which occurs at the bottom of the tachocline, and confinement by flows in the convection zone, which occurs at the radiative--convective interface. This second type of magnetic confinement is similar to that proposed by \citet{KitchatinovRudiger06},
and depends sensitively on the details of the model at the bottom of the convection zone.
In Section \ref{results}, we present and discuss the first numerical simulation of the solar interior to exhibit the layered tachocline/tachopause structure anticipated by GM98. Section \ref{varyparams} then examines how various properties of the tachocline and tachopause vary with magnetic and thermal diffusivities, and ultimately validates the predictions of the GM98 model. We also run a set of simulations in a parameter regime similar to that used by \citet{Strugarekal11}, and recover their results -- namely, that the magnetic field is not confined by the tachocline flows -- hence showing that using the correct value of $\sigma$ is indeed necessary if one wishes to obtain a GM98-like solution. Finally, we discuss the implications of our results in view of future 3D simulations of the solar interior, and within the greater context of stellar astrophysics in general, in Section \ref{conclusion}.
% Checked ok .

\section{The Model} 
\label{ourmodel}

Our goal is to study how the upper layers of the radiation zone respond to forcing by the differential rotation in the convection zone above and to the presence of a large-scale primordial field below. To do so, we use a model and numerical algorithm that is
an extension of the one developed by GG08.

We focus primarily on getting the dynamics of the radiative interior (radiation zone and tachocline) right, at the cost of having to simplify the dynamics of the convection zone quite dramatically. This sacrifice is necessary, as no numerical algorithm today can model at the same time the very fast timescales and short lengthscales associated with convection, and the very long timescales and global lengthscales appropriate within the deep interior. Our convection zone model is described in detail in Section \ref{forcing1}. 
% Checked ok .

The timescale of principal interest in the magnetic confinement problem is the timescale for magnetic diffusion across the tachocline, which in the Sun is approximately 10 Myr. This is much shorter than the global-scale Eddington--Sweet and magnetic diffusion times, but much longer than the oscillation periods of inertia, gravity, and Alfv\'en waves.  On this 10 Myr timescale, we may assume the tachocline to be in a quasi-steady force and thermal balance, as in GM98. The effect of the fast dynamics (including the mean effects of waves and turbulence) can be parameterized, if desired, through the inclusion of additional terms (e.g. Reynolds stresses, turbulent thermal diffusivity, turbulent magnetic diffusivity, etc..). In this paper we use such parameterizations only within the convection zone (see below). This is for simplicity, and because the effect of fast dynamics on the tachocline is neither particularly well understood, not well constrained. We discuss their potential impact on our results in Section \ref{conclusion}.  %Checked ok. 

The thermodynamic structure of the solar interior is accurately known from helioseismology, and is not greatly affected by the flows in the radiation zone. We therefore linearize all the governing equations around a hydrostatic background state derived from the 1-D solar model of \citet{JCD96}. We then solve numerically the resulting set of MHD equations, assuming axisymmetric dynamics.
%Checked ok. 

GG08 designed a numerical algorithm to search for steady-state solutions of this system for various input parameters, such as the assumed viscosity, thermal diffusivity, magnetic diffusivity, etc. In this work, we use the same algorithm, but with a few notable modifications. We now describe the original method and the modifications in detail for completeness.
%Checked ok. 

\subsection{The background state}

As in GG08, our background state assumes a spherically symmetric  
Sun in hydrostatic equilibrium. All quantities are expressed in spherical coordinates $(r,\theta,\phi)$, where $\theta$ is colatitude. 
Whereas GG08 modeled the radiation zone only, with the radial coordinate $r$ ranging from $\rin = 0.35 R_{\odot}$ to $\rout = 0.71 R_{\odot}$, we now include a significant portion of the convection zone as well in our numerical domain, and probe deeper towards the center. In what follows, $r$ spans the interval $[\rin = 0.05 R_\odot, \rout = 0.9 R_\odot]$. We interpolate Model S \citep{JCD96} in that range to obtain the density $\bar{\rho}$, temperature $\bar{T}$, pressure $\bar{p}$, gravity $\bar{g}$, and heat capacity at constant pressure $\bar{c}_p$ of the background fluid as a function of $r$. 
%Checked ok. 

\subsection{The model equations}
\label{basicmodeleqs}

We work in a frame that rotates with angular velocity  
\begin{equation}
\Omega_{\odot} = \Omega_{\rm eq} \left[ 1 - \frac{a_2}{5}  - \frac{3a_4}{35}  \right] %\mbox{ , }
%Checked ok.
\end{equation}
about the vertical axis,
where $\Omega_{\rm eq} = 2\pi \times 463$nHz is the equatorial velocity of the Sun near the base of the convection zone, $a_2 = 0.17$, and $a_4 = 0.08$ \citep{Schou-etal98,Gough07}.
In this frame, the total specific angular momentum of the convection zone is zero \citep{Gilmanal89}. 
%Checked ok. 

We first expand each of the
thermodynamical variables $q$ as $q(r,\theta) = \bar{q}(r) + \tilde{q}(r,\theta)$, where the bar indicates the spherically symmetric background state and the tilde refers to the axisymmetric perturbation. Under the steady-state assumption and once linearized in the thermodynamical fields around the background state, the system of governing equations can be written as:
\begin{eqnarray}
\label{numeqs0}
2\bar{\rho}(r) {\bf \Omega}_\odot \times {\bf u} + \bar{\rho}(r){\bf u} \cdot \nabla {\bf u}&=& - \nabla \tilde{p} - \tilde{\rho}\bar{g}(r) {\bf e}_r +  {\bf j} \times {\bf B}  +  \nabla \cdot {\bf \Pi} - \frac{\bar{\rho}(r) }{\tau(r)}({\bf u} - {\bf u}_{\rm cz}) \label{eq:mom1}\\
\bar{\rho}(r)\bar{T}(r) {\bf u} \cdot \nabla \bar{s}(r) &=&  \nabla \cdot (k(r) \nabla \tilde{T}) \label{eq:Teq}\\
\nabla \times ({\bf u} \times {\bf B}) &=& \nabla \times [ \eta(r) (\nabla \times {\bf B} - 4 \pi \bf{j}_0) ] \label{eq:indeq}\\
\frac{\tilde{p}}{\bar{p}(r)} &=& \frac{\tilde{\rho}}{\bar{\rho}(r)}+\frac{\tilde{T}}{\bar{T}(r)} \\
\nabla \cdot (\bar{\rho}(r){\bf u}) &=& 0 \\
\nabla \cdot {\bf B}  &=&  0  \mbox{ , }
%Checked ok.
\end{eqnarray}
where ${\bf u}$ is the velocity of the fluid, ${\bf B}$ the magnetic field, ${\bf j}$ the electric current density, $\eta$ the magnetic diffusivity, and $k$ the thermal conductivity. The viscous stress tensor $\bf{ \Pi}$ incorporates the contribution of the viscosity $\nu$:
\begin{equation}
{\bf \Pi} = \bar{\rho}(r) \nu(r) \left[ \nabla \bf{u} + (\nabla {\bf u})^T - \frac{2}{3} (\nabla \cdot {\bf u}) {\bf I}\right ] \mbox{ , } 
%Checked ok.
\end{equation}
where ${\bf I}$ is the identity matrix. Note that the advection of the entropy perturbations has been neglected. Finally, and as in GG08, we
retain only those terms in ${\bf u} \cdot \nabla {\bf u}$ that include the azimuthal velocity $u_\phi$, since the other (meridional) velocity components are generally much weaker. We do this to accelerate numerical convergence, but have checked that this has very little effect on the final solution in the parameter regime of interest.  
%Checked text, need to check equation. 

This system of equations is very similar to the one used by GG08, with two notable differences.
First, we parametrize the driving of the differential rotation in the convection zone by the body-force term $-\frac{\bar \rho}{\tau}({\bf u} - {\bf u}_{\rm cz})$ where $1/\tau(r)$ vanishes in the radiation zone (see Section \ref{forcing1} for more detail).
Second, we have introduced a source term $ 4\pi {\bf j}_0 = \nabla \times {\bf B}_0 $ in the magnetic induction equation to maintain an assumed primordial magnetic field ${\bf B}_0$ against diffusion (see Section~\ref{forcing2} for detail). 
 %Checked ok. 

\subsection{The diffusivity profiles}
\label{diffsect}

Although our steady-state model is able to achieve much lower values for the various diffusivities (viscosity $\nu$, magnetic diffusivity $\eta$ and thermal diffusivity $\kappa = k / \bar \rho \bar c_p$) than direct numerical simulations, they must still be artificially increased by several orders of magnitude to yield a numerically tractable system. As a result, there is little benefit in using realistic profiles for these quantities, so we choose them instead to be as simple as possible.
 %Checked ok. 

\begin{figure}    
%    \centerline{\includegraphics[width=8cm]{fig1.eps}}
    \centerline{\includegraphics[width=8cm]{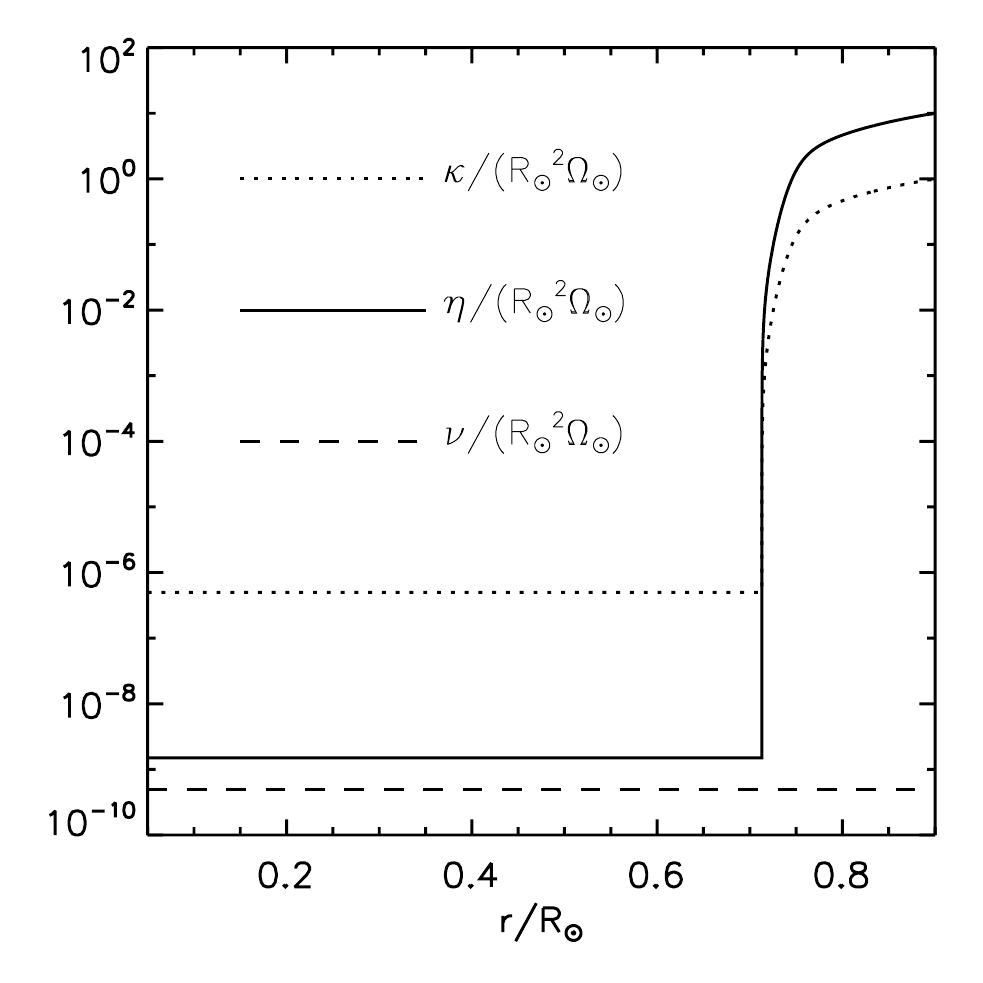}}
    \caption{Non-dimensional diffusivity profiles $\nu/R_\odot^2 \Omega_\odot$ (dashed line), $\kappa/R_\odot^2 \Omega_\odot$ (dotted line) and $\eta/R_\odot^2 \Omega_\odot$ (solid line) as a function of $r$ for our reference model (see Section \ref{results}). The constant non-dimensional values in the radiation zone are $E_{\nu} = 5.0 \times 10^{-10}$, $E_{\eta}= 1.5 \times 10^{-9}$, and $E_{\kappa}= 5.0 \times 10^{-7}$. }
    \label{diff_params}
  \end{figure}

In what follows, we use $\rcz= 0.7127R_\odot$ to denote the radius of the radiative--convective interface. For simplicity, we take the diffusivities to be constant within the radiation zone ($r < \rcz$), with values $\nu_{\rm rz}$, $\eta_{\rm rz}$, and $\kappa_{\rm rz}$ respectively. In the convection zone, we model the effects of the turbulence on the magnetic field and heat transport as an increase in the effective diffusivities.\footnote{We do not increase the viscosity in the convection zone because we have already introduced a body-force to parameterize the turbulent transport of momentum.} For reasons that are explained in Section~\ref{issues}, we construct the profiles to ensure that this increase occurs slightly above the radiative--convective interface, at a radius $r_1 > \rcz$. The choice of $r_1$ is discussed in Section ~\ref{issues} as well.
Thus, 
\begin{eqnarray}
\nu(r) &=& \nu_{\rm rz} \\
\eta(r) &=& \eta_{\rm rz} + H(r-r_1) \eta_{\rm t}(r) \mbox{ , } \\
\kappa(r) &=& \kappa_{\rm rz} + H(r-r_1) \kappa_{\rm t}(r) \mbox{ , }
%Checked ok.
\end{eqnarray}
where $H$ is a Heaviside function, and 
\begin{eqnarray}
\label{Eetaprofile}
\eta_{\rm t}(r) &=& \frac{1}{2}(\eta_{\rm cz}-\eta_{\rm rz})\frac{r - r_1}{\rout-r_1}\left[1 +\tanh\left(\frac{r- r_2}{\Delta_2}\right) \right]  \mbox{ , }\\
\label{Ekappaprofile}
\kappa_{\rm t}(r) &=& \frac{1}{2}(\kappa_{\rm cz}-\kappa_{\rm rz})\frac{r - r_1}{\rout-r_1}\left[1 + \tanh\left(\frac{r- r_2}{\Delta_2}\right) \right] \mbox{ . }
%Checked ok.
\end{eqnarray}
The tanh terms in (\ref{Eetaprofile}) and (\ref{Ekappaprofile}) smooth the transition between the laminar and turbulent regions over a layer of thickness $\Delta_2$ centered at radius $r_2$.
In all that follows, we take $r_2 = \rcz + 0.03R_\odot$ and $\Delta_2 = 0.02 R_{\odot}$. The resulting profiles, as used in our reference model, are shown in Figure \ref{diff_params}. The selected values of $r_1$, $\kappa_{\rm cz}$ and $\eta_{\rm cz}$ for this model are reported in Table 1. While $r_1$ has significant influence on the solution (see Section \ref{issues} for detail), $\kappa_{\rm cz}$ and $\eta_{\rm cz}$ do not as long as they are large enough. 
 %Checked ok. 

%%%%%%%%%%%%%%%%%% FIGURE XX
 %Checked ok. 
%%%%%%%%%%%%%%%%%% FIGURE XX

\subsection{The parameter $\sigma$ and the selection of $N^2$}
\label{sigma}

In Section \ref{intro}, we introduced and discussed the importance of the parameter $\sigma = \sqrt{\rm{Pr}}\bar N/\Omega_{\odot}$ in setting the vertical scale of penetration of the meridional flows into the radiative interior. 
We therefore have to ensure that $\sigma$ is of the same order as in the Sun. Since our Prandtl number is {\it not} solar, we choose to modify the background stratification profile $\bar N$ in the simulation in such a way as to have $\sigma(r)$ be the actual solar profile $\sigma_{\odot}(r)$, as computed from Model S \citep{JCD96}. To do so, we proceed as follows.

\begin{figure}    
                                % includes only one panel
%    \centerline{\includegraphics[width=8cm]{fig2.eps}}
    \centerline{\includegraphics[width=8cm]{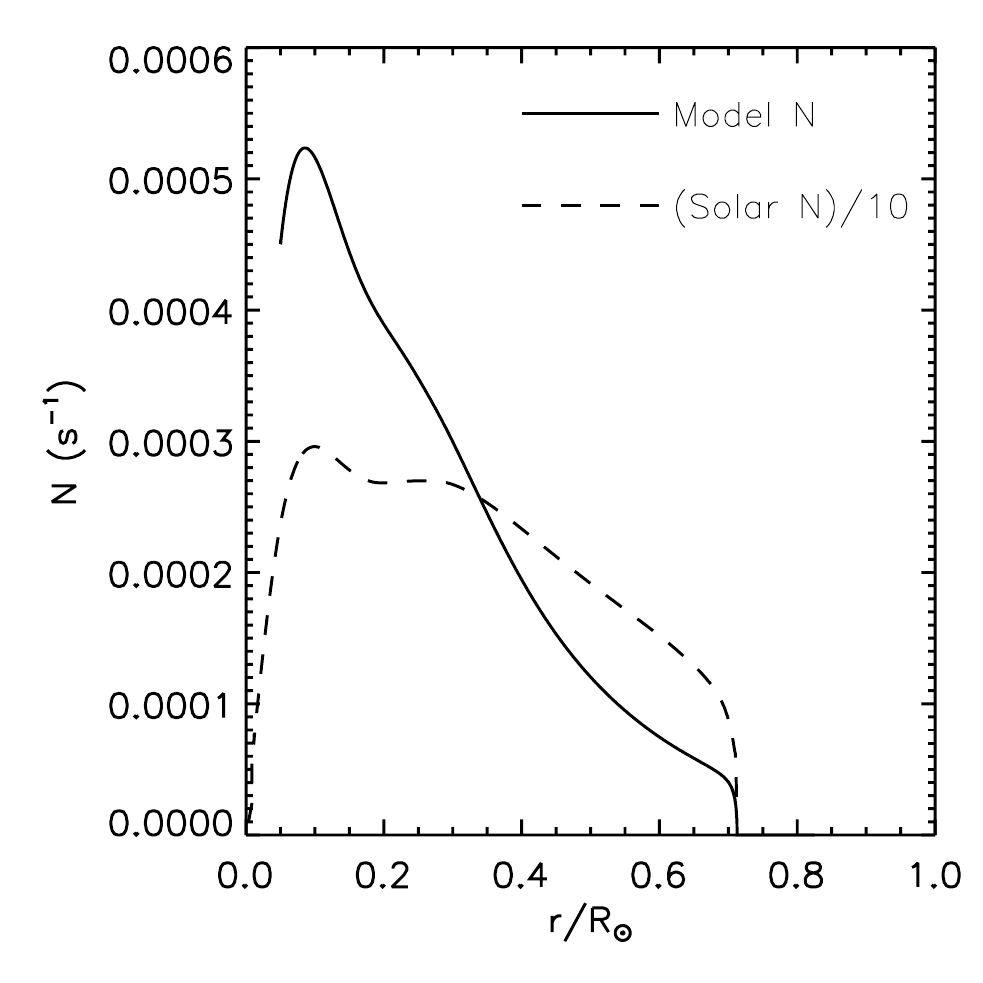}}
   \caption{Background buoyancy frequency profile $\bar N(r)$ in our reference model (see Section \ref{results}), compared with the solar profile. A lower value of $\bar N$ is needed to ensure that $\sigma(r)$ is exactly solar in our numerical model.  }
    \label{buoy_diff}
   \end{figure}

We first compute the ``true'' background diffusivities $\nu_{\odot}(r)$ and $\kappa_{\odot}(r)$ in the radiation zone from the formulae given by \citet{Gough07}. We then compute the solar 
stratification parameter
\begin{equation}
\sigma_{\odot}(r) = \sqrt{\frac{\nu_{\odot}(r)}{\kappa_{\odot}(r)}} \left(\frac{N_{\odot}(r)}{\Omega_{\odot}}\right) \mbox{ , }
%Checked ok.
\end{equation}
 where $N_{\odot}(r)$ is interpolated from Model S.
Once the numerical diffusivity profiles $\nu(r)$ and $\kappa(r)$ have been selected (see Section \ref{diffsect}), we construct an artificial background buoyancy frequency profile $\bar{N}$ satisfying:
\begin{equation}
\bar N(r) = \left \{ \begin{array}{lc} 
\sigma_{\odot}(r) \Omega_{\odot}\displaystyle \sqrt{\frac{\kappa_{\rm rz}}{\nu_{\rm rz}}}  & \mbox{for } r < r_{\rm cz} \nonumber \\   
0 &  \mbox{otherwise. }     \end{array} \right .
%Checked ok.
\end{equation}
Figure \ref{buoy_diff} shows $\bar N(r)$ as used in our reference model, and compares it with the solar profile $N_{\odot}(r)$. In order to have a solar $\sigma$, we have to take $\bar N(r)$ to be a factor of about 20 smaller than in the solar tachocline.
 %Checked ok. 

As we demonstrate in Sections \ref{results} and \ref{varyparams}, using a realistic value of $\sigma$ is crucial to obtain confined-field solutions (see also Wood \& Brummell 2013, in prep). But what are the consequences of choosing a lower-than-solar value of $\bar N$ on other aspects of the problem? In our steady-state, laminar model of the solar interior, choosing a lower value of $\bar N$ can only influence the large-scale meridional flows, and does so in such a way as to improve the dynamical realism of the results. In 3D time-dependent simulations, however, a lower value of $\bar N$ would also increase the depth of penetration of overshooting convective plumes, and change the frequency of oscillation of gravity waves, which may have detrimental effects on the solutions. However, it is clear that no simulations can ever yield an exactly solar tachocline until such a time when they can be run at exactly solar parameters. 
In the meantime, one can still meaningfully study the dynamics of the tachocline even with a thicker overshoot layer, by selecting other input parameters to ensure that the tachocline is correspondingly thicker as well. %One of the central results of the present study is to identify the factors that determine the tachocline thickness.
 %Checked ok. 

%%%%%%%%%%%%%%%%%% FIGURE XX
   %Checked ok. 
%%%%%%%%%%%%%%%%%% FIGURE XX

\subsection{The forcing terms}

The main differences between the model used here and that of GG08 lies in the forcing terms. In  GG08, forcing was applied through boundary conditions. At the outer boundary of their domain, differential rotation and radial inflows/outflows were imposed to model the driving of meridional flows by differential rotation in the overlying convection zone. A primordial magnetic field was maintained by a point-dipole at $r=0$. This method, however, has two disadvantages. First, it cannot take into account the back-reaction of the radiation zone on the convection zone dynamics, and introduces artificial Ekman layers 
at the outer boundary. Secondly, under the point-dipole assumption, the primordial field ${\bf B}_0$ reaches unrealistically large amplitudes as $r\rightarrow 0$, leading to numerical convergence difficulties and driving artificial MHD instabilities near the lower boundary. Both problems ultimately prevented GG08 from obtaining solutions at low-enough diffusivities to exhibit a satisfactory GM98-like tachocline.
 %Checked ok. 

In order to model the dynamics of the complete system, including both the convection zone and the radiation zone, we have extended the computational domain to $r\in[0.05,0.9]R_\odot$, as discussed earlier. In order to improve the stability of the numerical scheme, we have also modified GG08's algorithm to drive the system through body-forcing terms rather than boundary conditions. These terms are now described in more detail. 
 %Checked ok. 

\subsubsection{Convection zone forcing}
\label{forcing1}

In the solar convection zone, gyroscopic pumping is caused by the Reynolds stresses associated with strongly anisotropic eddies, which drive the system away from uniform rotation \citep{Rudiger89,KitchatinovRudiger93,Rempel05}. To model this in detail either requires the use of a 3D time-dependent code as in \citet{Strugarekal11} or \citet{Rogers11}, where angular-momentum transport arises naturally from the convective dynamics, or the use of a parametric model for the convective Reynolds stresses \citep{KitchatinovRudiger93,Rempel05,Garaudal10}. Both approaches have well-known pros and cons. The former is computationally expensive, and of course cannot be implemented within the steady-state axisymmetric approach we have elected to follow here. The latter, by contrast, can, and has been used with some degree of success by \citet{Rempel05} to model the differential rotation profile and meridional flows within the convection zone, but its reliability is inherently tied to the reliability of the turbulence model used, a factor which is difficult to evaluate objectively. 
 %Checked ok. 

Here, we use a much simpler model for the Reynolds stresses, which takes the form of a Darcy friction law \citep{BrethertonSpiegel68,GaraudAA09,Woodal11}:
\begin{equation}
 -\frac{\bar \rho(r)}{\tau(r)}({\bf u} - {\bf u}_{\rm cz}) \mbox{ , }
\end{equation} 
where ${\bf u}_{\rm cz}$ is the observed azimuthal velocity of the solar convection zone (expressed in the rotating frame),
\begin{equation}
{\bf u}_{\rm cz} = u_{\rm cz}(r,\theta) {\bf e}_{\phi} = r \sin\theta\, \left[ \Omega_{\rm eq} (1-a_2\cos^2\theta - a_4 \cos^4\theta) - \Omega_\odot \right]  {\bf e}_{\phi} \mbox{ , }
%Checked ok.
\label{eq:diffrot}
\end{equation}
and where the relaxation timescale $\tau(r)$ is defined such that 
\begin{equation}
\frac{1}{\tau(r)} = \frac{1}{\tau_{\rm cz}}\left[\frac{r-r_{\rm cz}}{\rout-r_{\rm cz}} \right] H(r-r_{\rm cz}) \mbox{ . }
%Checked ok.
\end{equation}
The quantity $\tau_{\rm cz}$, which we take to be $0.1\Omega_\odot^{-1}$, can be interpreted as a turnover time for the convective flows. We have chosen the $\tau$ profile so that $1/\tau(r)$ starts increasing linearly exactly from the base of the convection zone upward. By increasing the forcing continuously, we avoid the formation of artificial boundary layers at the radiative--convective interface \citep{GaraudAA09}. In other respects, the particular choice of the function $\tau(r)$ has relatively little impact on the resulting long-term dynamics in the radiative interior \citep{Woodal11}, see below for detail. Finally, note that the forcing that maintains the differential rotation also produces a persistent gyroscopic pumping of meridional flows in the convection zone \citep{GaraudAA09,Woodal11,WoodBrummell12}, with downwelling in polar regions and upwelling near the equator.
 %Checked ok. 

The reason for our choice of forcing is mere simplicity. However, as shown by \citet{Woodal11}, the dynamics of the tachocline itself are fairly independent from those of the convective zone, as long as, to leading order, (1) the exchange of angular momentum between the convection and radiation zones is less efficient than the redistribution of angular momentum within the convection zone (2) the convective heat transport is efficient enough that the convection zone is very nearly adiabatic and (3) the meridional flows within the tachocline remain significantly weaker than the meridional flows in the convection zone. These three conditions are naturally satisfied in the real convection zone, and were implicitly assumed in the models of \citet{SpiegelZahn92} and GM98.
 %Checked ok. 

The first two conditions can be satisfied in our Darcy friction model provided $\tau_{\rm cz}\Omega_{\odot} < 1$ and $\kappa_{\rm cz} > (R_\odot-\rcz) \Omega_{\odot}^2$. This guides our choice of $\tau$, and $\kappa_{\rm cz}$, although their exact values have little-to-no influence on the resulting system dynamics as long as the inequalities above are satisfied. The third condition is automatically satisfied in the Sun, which has a relatively massive convection zone and strong differential rotation (both aspects are represented fairly accurately in our model), but may not be in higher-mass stars. 
 %Checked text, but are conditions actually satisfied in the code? 

\subsubsection{Forcing of the primordial magnetic field }
\label{forcing2}

\begin{figure}    
%            \centerline{\includegraphics[width=8cm]{fig3.eps}}
            \centerline{\includegraphics[width=8cm]{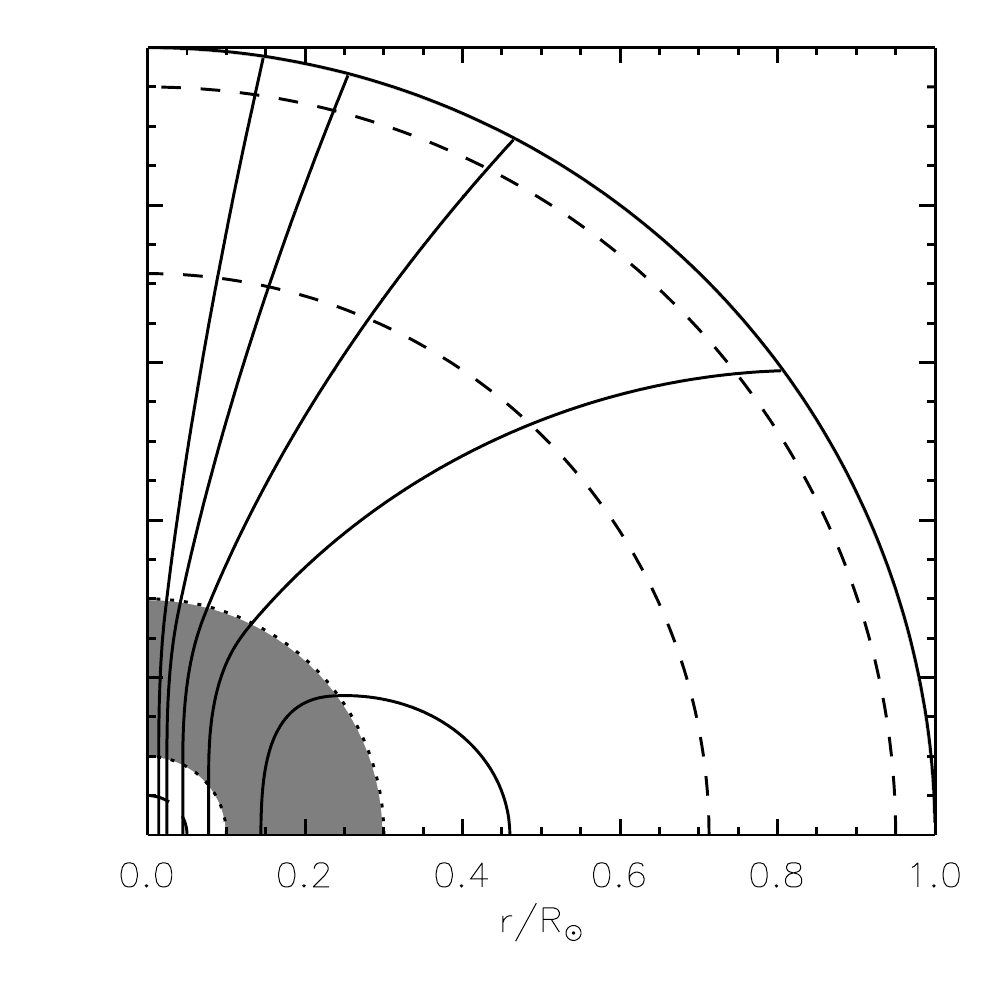}}
\caption{Geometry of the computational domain and of our assumed primordial magnetic field ${\bf B}_0$. The latter is maintained by the electric current density ${\bf j}_0$ (see Equation (\ref{eq:J0def}) and Section \ref{forcing2} for detail). The solid curves represent a few selected field lines. The outermost circle (solid line) marks the solar surface. The dashed circles, from the outermost to the innermost one, mark $\rout$, $r_{\rm cz}$ and $\rin$. The shaded area is the interval between $r_a$ and $r_b$, and marks the region where the azimuthal current ${\bf j}_0$ is imposed. }
   \label{magneticforcing}
   \end{figure}

The primordial field ${\bf B}_0$ is maintained against diffusion by
a source term in the induction Equation (\ref{eq:indeq})
that imposes a permanent azimuthal electric current ${\bf j}_0 = \nabla \times {\bf B}_0/4\pi$.
To avoid artificially altering the tachocline dynamics, the imposed current is localized within a region $r_a \leq r \leq r_b$ deep in the radiation zone:
\begin{equation}
{\bf j}_0(r,\theta) = \left \{ \begin{array}{lc}
J_0\frac{(r-r_a)(r-r_b)}{R_\odot^2}\sin \theta {\bf e}_{\phi} & \mbox{for } r_a < r <r_b \nonumber \\
0 & \mbox{otherwise } \end{array} \right. 
 %Checked ok. 
\label{eq:J0def}
\end{equation}
where $r_a = 0.1R_\odot$ and $r_b = 0.3R_\odot$.
In the absence of any induction by fluid motions, this current would generate an open dipolar magnetic field ${\bf B}_0$, whose exact expression is derived in Appendix~\ref{appA}.
We do not impose any current close to the core, in order to avoid the numerical difficulties described by GG08,
and so ${\bf B}_0$ is uniform within the sphere $r < r_a$.
The amplitude of the current density, $J_0$, is chosen so that the magnitude of this uniform field is equal to $B_0$.
Figure \ref{magneticforcing} shows selected magnetic field lines of ${\bf B}_0$ to illustrate its structure.  Note that the actual structure of the 
poloidal field calculated from the simulation is only similar to ${\bf B}_0$ when the magnetic Reynolds number is much smaller than one. 
%Checked ok.

%%%%%%%%%%%%%%%%%% FIGURE XX
%Checked ok.
%%%%%%%%%%%%%%%%%% FIGURE X

\subsection{Boundary conditions}

The inner core ($r < \rin$, not modeled explicitly) is assumed to be a thermally and electrically conducting solid with the same thermal and magnetic diffusivity as the fluid just above $r=\rin$. As such, the bottom boundary at $\rin$ is impermeable. The solid core has a uniform angular velocity, $\Omega_{\rm in}$, chosen in such a way as to guarantee that it exerts a zero total torque on the fluid above. Altogether, these conditions imply:
\begin{eqnarray}
u_r (\rin,\theta) &=& 0 \\
\frac{\partial u_{\theta}(\rin,\theta)}{\partial r} &=& 0 \\
\int_{-\pi/2}^{\pi/2} \left( \bar{\rho}\nu \rin^2 \frac{\partial\tilde{\Omega}}{\partial r}\sin^2 \theta + \frac{B_r B_{\phi}}{4 \pi} \rin \sin \theta \right) \sin \theta\, \rm{d}\theta & = & 0 \mbox{ . }
\end{eqnarray}
The temperature perturbations in the core satisfy Laplace's equation:
\begin{equation}
\label{Laplace1}
\nabla^2 \tilde{T}  = 0 \mbox{ . }
\end{equation}
The solutions of Equation (\ref{Laplace1}) are matched onto the solutions obtained in the computational domain at the boundary $r=\rin$, requiring continuity of $\tilde{T}$ and $\partial \tilde{T}/\partial r$. Similarly, the magnetic field in the core satisfies:  
\begin{equation}
\nabla^2 {\bf B} = 0 \mbox{ , }
\end{equation}
and the solutions in the computational domain are matched onto the core solution by requiring continuity of $B_r$ and $B_{\theta}$ at $r=\rin$.

At the outer boundary, $r=\rout$, we impose the following boundary conditions:
\begin{eqnarray}
u_r(\rout,\theta)  &=& 0 \\
u_{\theta}(\rout,\theta)  &=& 0 \\
%%% \frac{\partial u_{\theta}(r_{out},\theta)}{\partial r} &=& 0 \\
u_{\phi}(\rout,\theta)  &=& u_{\rm cz}(\rout,\theta) 
\end{eqnarray}
which represent an impermeable, no-slip outer boundary, where $u_{\rm cz}$ is given in Equation (\ref{eq:diffrot}). As with the inner boundary conditions, we assume that the temperature perturbations and the magnetic field satisfy Laplace's equation outside the domain, and require the continuity of $\tilde{T}$, $\partial \tilde{T}/\partial r$, $B_r$ and $B_\theta$ across the boundary. 

\subsection{Non-dimensional parameters}

For comparison with other works, we now present a non-dimensional version of the governing parameters. By normalizing distances with respect to the solar radius $R_{\odot}$, time to the inverse of the mean solar rotation rate $\Omega_{\odot}$, velocities to $R_{\odot} \Omega_{\odot}$,  density to $\rho_0 = 1$g/cm$^{3}$, and the amplitude of the magnetic field ${\bf B}$ to $B_0$ (the constant value of the imposed primordial field in the core, see Section \ref{forcing2}), the following non-dimensional parameters naturally arise:
\begin{eqnarray}
E_{\kappa} &=& \frac{\kappa_{\rm rz}}{R_{\odot}^2 \Omega_{\odot}} \nonumber \\
E_{\eta} &=& \frac{\eta_{\rm rz}}{R_{\odot}^2 \Omega_{\odot}} \nonumber \\
E_{\nu} &=& \frac{\nu_{\rm rz}}{R_{\odot}^2 \Omega_{\odot}} \nonumber \\
\Lambda &=& \frac{B_0^2}{4\pi \rho_0 \eta_{\rm rz} \Omega_{\odot}} \mbox{  ,} 
\label{eq:paramsdef}
 %Checked ok. 
\end{eqnarray}
where we take $\rho_0 = 1$g/cm$^{3}$. The first three numbers are the ``Ekman'' numbers for the radiation zone, which are dimensionless measures of the three diffusivities.
The last is an Elsasser number, which measures the relative strength of the Lorentz force and the Coriolis force associated with azimuthal flows, in the core. The effective Elsasser number in the tachocline is typically significantly smaller than $\Lambda$, since the magnitude of ${\bf B}_0$ drops rapidly with $r$ (see Appendix A for detail). 
%Checked ok.

\subsection{The numerical method}

The system of partial differential equations and boundary conditions discussed in this Section is solved numerically using the nonlinear relaxation Newton--Raphson--Kantorovich algorithm developed by GG08. The details on how these equations are solved and implemented in our model are described in GG08 (see their Appendices).  
%XXX Paragraph is a bit short -- get rid of it ? XXX
%Here, we give a brief description of the method for completeness.
%The system of governing equations and boundary conditions is expanded in terms of Chebyshev polynomials and projected onto the radial coordinate to generate a system of ordinary differential equations. Numerical solutions of these coupled nonlinear ODEs are obtained by the parallel implementation of the Newton-Raphson-Kantorovich method developed by \citeauthor{GG08} (\citeyear{GG08}). T

\section{Model sensitivity to turbulent magnetic diffusivity profile in the convection zone}
\label{issues}

\begin{figure*}    
                                % includes the two top panels 
   \centerline{\hspace*{0.015\textwidth}
                  \includegraphics[width=8cm]{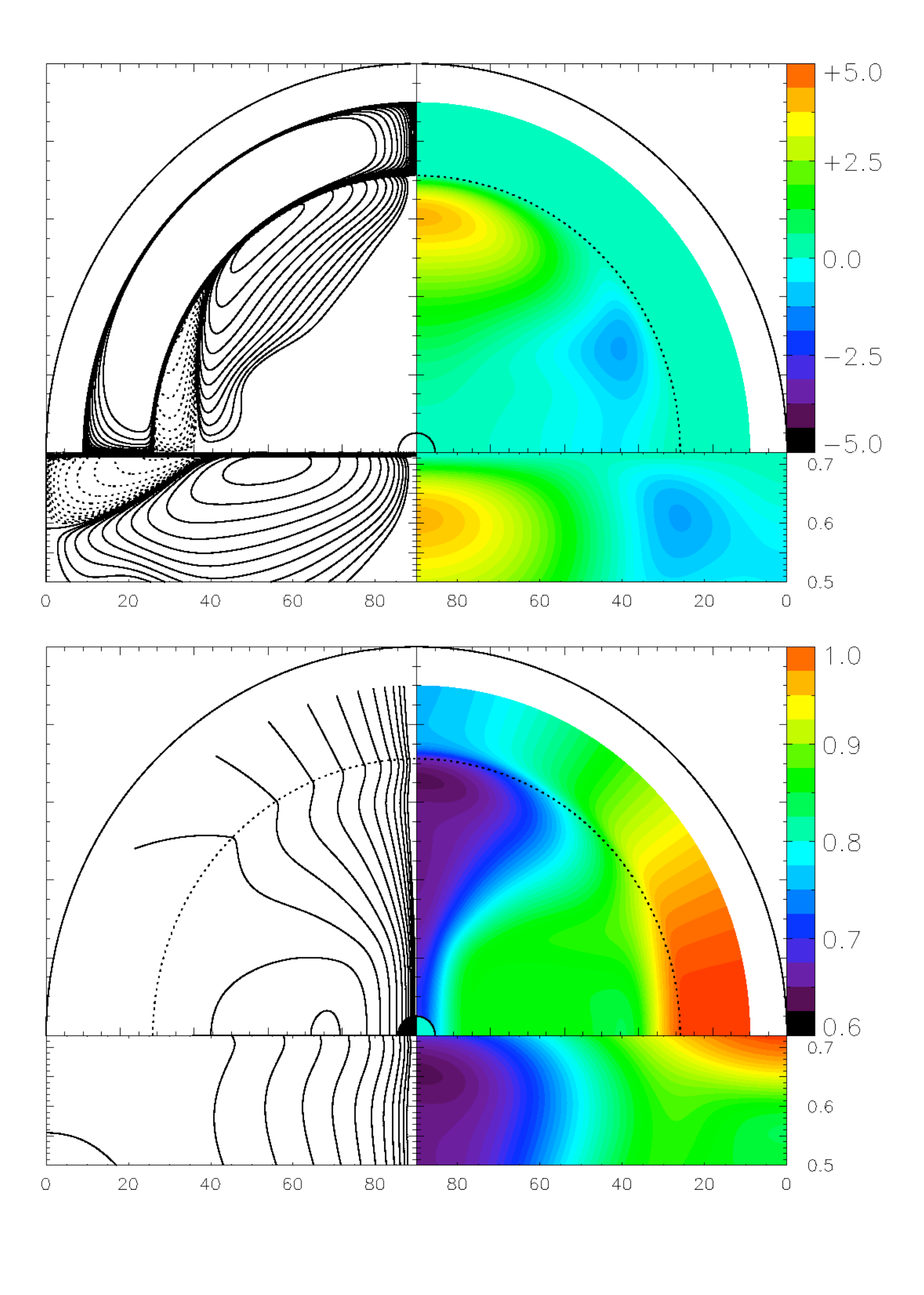}
               \hspace*{-0.03\textwidth}
               \includegraphics[width=8cm]{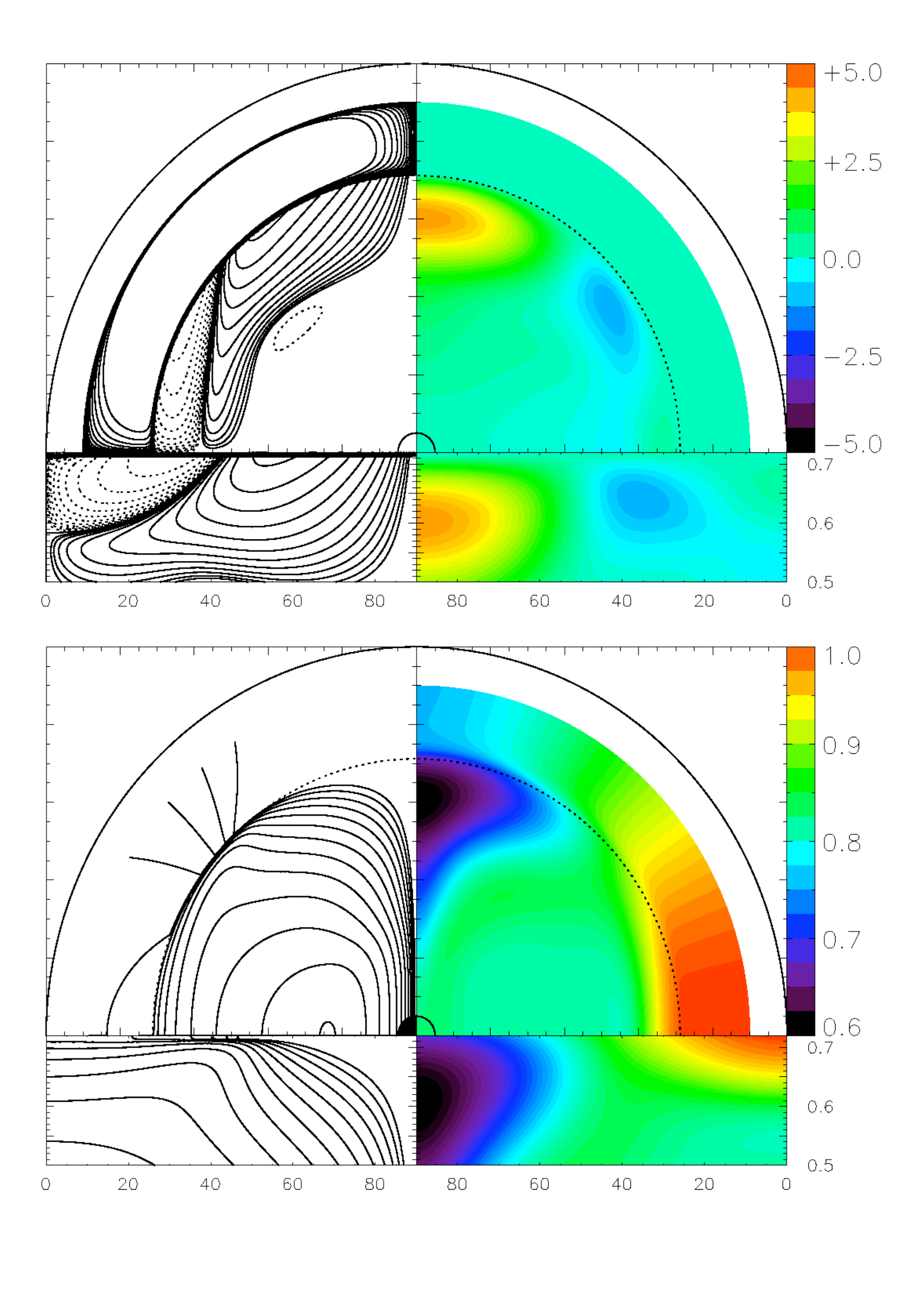}
              }
     \vspace{-0.35\textwidth}   % Shift close to the panel top 
     \centerline{\Large \bf     % Includes the labels (here needs the color 
                                %   package, see beginning of this file)
      \hspace{0.0 \textwidth}  % \color{blue}{(a)}
      \hspace{0.415\textwidth} % \color{blue}{(b)}
         \hfill}
     \vspace{0.31\textwidth}    % Shift back to the panel bottom 

\caption{Comparison of two simulations with and without pre-confinement (see Section \ref{issues} for detail), produced using $E_{\nu} =  5 \times 10^{-9}$, $E_{\eta} = 1.5 \times 10^{-8}$, and $E_{\kappa} = 5 \times 10^{-6}$ (i.e. diffusivities that are all 10 times higher than the reference model studied in Section \ref{results}), a solar $\sigma$ profile, and $\Lambda = 0.66 \times 10^5$. On the left $r_1 = r_{\rm cz}$, and on the right $r_1-r_{\rm cz} = 0.002 R_{\odot}$. In both figures, starting from the top left panel and going clockwise, we show selected flow streamlines (with solid lines for clockwise flow, and dotted lines for anticlockwise flow), temperature perturbations (in units of Kelvin), angular velocity (in units of $\Omega_{\rm eq}$), and magnetic field lines. The strip below each panel zooms into the region between $r=0.5R_\odot$ and $r_{\rm cz}$.}
   \label{open_conf}
   \end{figure*}

In Sections \ref{results} and \ref{varyparams}, we study the model results in detail and discuss their implications for the dynamics of the solar tachocline and radiation zone. First, however, we must discuss an important point regarding the model sensitivity to the magnetic diffusivity profile near the radiative--convective interface. Figure \ref{open_conf} illustrates this issue by showing the results of two simulations that differ only in the location $r_1$ at which the magnetic diffusivity\footnote{Changing $r_1$ also changes the position where the thermal diffusivity increases, see Equation (\ref{Ekappaprofile}); however, this has a negligible effect on this discussion, since the thickness of the thermal diffusion layer is much larger than the thickness of the magnetic diffusion layer.} increases from its laminar radiation zone value to its turbulent convection zone value (see Equation (\ref{Eetaprofile}) in Section \ref{diffsect}). On the left, $r_1 = r_{\rm cz}$, while on the right, $r_1 = r_{\rm cz} + 0.002 R_{\odot}$.
%Checked ok. 

When $r_1= r_{\rm cz}$, we see that the field is unconfined at high latitudes, but when $r_1$ is moved up even a tiny distance into the convection zone then the magnetic field is confined by large-scale meridional flows at the radiative--convective interface. For the sake of brevity, in all that follows we call this effect ``pre-confinement'', to contrast it with confinement by meridional flows deeper in the radiation zone. Pre-confinement is quite different from the dynamics described by GM98 for the reasons pointed out below. 
%Checked ok. 

%%%%%%%%%%%%%%%%%% FIGURE XX
%Checked ok.
%%%%%%%%%%%%%%%%%% FIGURE XX

\begin{figure*}    
                                % includes only one panel 
%              \centerline{\includegraphics[width=\textwidth]{fig5.eps}}
    \centerline{\includegraphics[width=0.9\textwidth]{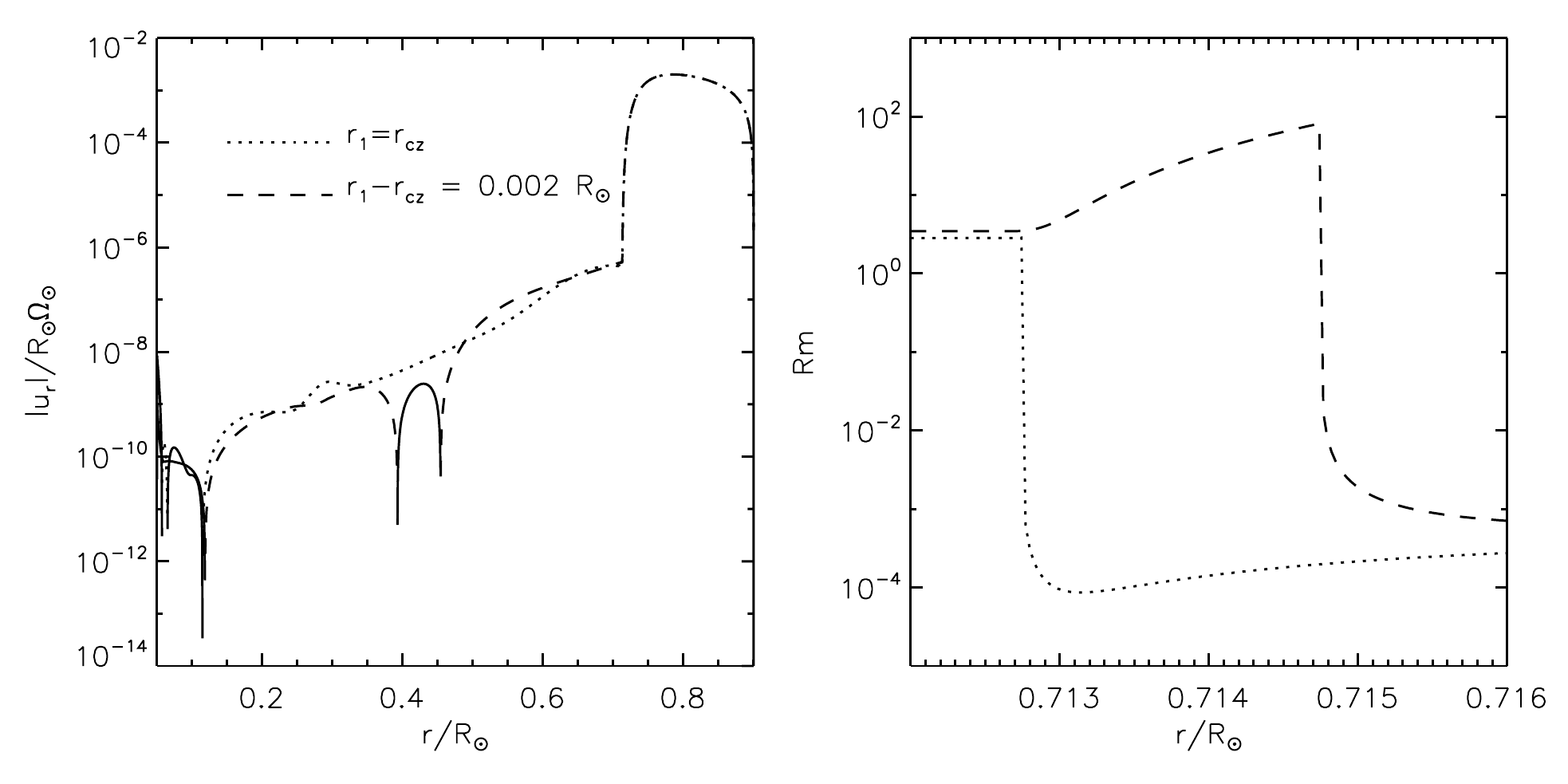}}
    \caption{Left: Radial flow velocity $|u_r|$ at $80^\circ$ latitude, for the two simulations shown in Figure \ref{open_conf}. The solid line shows upwelling flows, and the patterned line shows downwelling flows (dotted for the $r_1=r_{\rm cz}$ case, and dashed for the $r_1-r_{\rm cz} = 0.002R_\odot$ case). The vertical velocities in the convection zone are indistinguishable. Right: 
Magnetic Reynolds number, ${\rm Rm} = |u_r| L/\eta$, computed at $80^\circ$ latitude, where $L = 0.1 R_{\odot}$. The same linestyle as the left-side figure is used. The two kinks in the curves correspond to $\rcz = 0.7127R_\odot$ and $r_1 = 0.7147R_\odot$. Just above the base of the convection zone, ${\rm Rm} \gg 1$ for the case where $r_1 > r_{\rm cz}$ while ${\rm Rm} \ll 1$ for case where $r_1 = r_{\rm cz}$. }
   \label{Rm_openconf}
   \end{figure*}

But first, in order to understand why such a small change in $r_1$ results in such a dramatic effect on the magnetic field configuration, we compute the magnetic Reynolds number ${\rm Rm} = |u_r| L/\eta$, where $L$ is (somewhat arbitrarily) selected to be $L= 0.1R_{\odot}$. Since the value of $u_r$ in the lower convection zone is essentially identical in both cases (see Figure \ref{Rm_openconf}a), ${\rm Rm}$ is mostly controlled by the magnetic diffusivity profile. When $r_1= r_{\rm cz}$, the flow velocity drops more-or-less at the same rate as $\eta_{\rm t}$ (see Equation (\ref{Eetaprofile})) while approaching the base of the convection zone, leading to ${\rm Rm} < 1$ for all $r > r_{\rm cz}$, as shown in Figure \ref{Rm_openconf}b. On the other hand, when $r_1-r_{\rm cz}= 0.002 R_{\odot}$, a shallow region $r_{\rm cz} < r < r_1$ appears where the same flows, this time combined with a sufficiently low magnetic diffusivity, yield ${\rm Rm} \gg 1$. Such flows can then substantially affect the magnetic field, advecting it horizontally just above the base of the convection zone, which results in the aforementioned magnetic ``pre-confinement".
%Checked ok. 

%%%%%%%%%%%%%%%%%% FIGURE XX
%Checked ok. 
%%%%%%%%%%%%%%%%%% FIGURE XX

The possibility of magnetic confinement by latitudinal flows at the bottom of the convection zone has previously been discussed by \citet{KitchatinovRudiger06}. They argued that viscosity (either microscopic or turbulent) prevents meridional flows from penetrating more than a short distance into the radiation zone, leading to a strong radial gradient in the latitudinal flow velocity at the radiative--convective interface. This shear, combined with an assumed weak magnetic diffusivity, leads to field confinement much as in Figure \ref{open_conf}b. We note, however, that the model of \citet{KitchatinovRudiger06} does not explicitly consider angular-momentum balance, instead treating magnetic confinement as a purely kinematic problem. It is therefore not a fully self-consistent model for the tachocline.  Moreover, as discussed in Section \ref{intro}, and illustrated in Figure \ref{open_conf}, viscosity does not prevent meridional flows from penetrating into the solar radiation zone. % In the GM98 model and in this work, these flows are instrumental in determining the structure of the tachocline.
%Checked ok. 

What is interesting, however, is that pre-confinement clearly facilitates confinement deeper in the radiation zone, at least within the scope of our numerical model.  With pre-confinement, we are able to find deeply confined solutions (see Figure \ref{open_conf}b and Section \ref{results}), while similar solutions are much more elusive in models which are not pre-confined\footnote{This does not mean they don't exist. There are strong indications that they do at lower diffusivities than the ones for which we are able to obtain fully resolved, well-converged solutions.}.
%Checked ok. 

 Is magnetic field pre-confinement by the convection zone flows a possible scenario in the Sun? We believe it is, although not in the manner described above: in the solar convection zone, transport of magnetic field is probably dominated by the action of small-scale turbulent convective plumes, rather than advection by large-scale mean meridional flows. However, it is often argued that such turbulent transport also promotes magnetic confinement \citep{Zeldovich57,Radler68,Weiss66,DrobyshevskiYuferev74}. In the case of the solar tachocline, this has been demonstrated in a series of numerical simulations by Tobias et al. (2001) \citep[see also][]{GaraudRogers07} who showed that overshooting convective plumes produce a net pumping of magnetic field out of the convection zone. In other words, whether by large-scale laminar flows, or by small-scale turbulent flows, pre-confinement is likely to take place. 
%Checked ok. 

Pre-confinement alone, however, cannot explain all tachocline-related observations. Indeed, if  magnetic pumping were the only confinement mechanism in operation, then the tachocline would be only as deep as the convective overshoot layer, which is thought to be significantly smaller than the observed tachocline thickness \citep{JCDal95,Brummellal02,RogersGlatzmaier06}; such a model is not consistent with helioseismology. Moreover, turbulent magnetic pumping is only known to be effective at confining a horizontal magnetic field. Recent simulations by Wood and Brummell (2013, in prep.) strongly suggest that large-scale  flows similar to those proposed by GM98 are needed to confine the field near the poles, as previously proposed by \citet{WoodMcIntyre11}. %To what extent magnetic pumping contributes to the confinement of the field at other latitudes remains to be determined, but it appears that the GM98 model is the only explanation for magnetic confinement near the poles
%Checked ok. 

In summary, since turbulent pre-confinement via magnetic pumping is likely to be present in the Sun, and since pre-confinement clearly facilitates our task of finding deeply confined solutions reminiscent of the GM98 model, in all that follows we let it take place (in the laminar sense) by choosing $r_1 - r_{\rm cz} = 0.0003 R_{\odot}$ (an even smaller separation than that illustrated in Figures \ref{open_conf} and \ref{Rm_openconf}) unless otherwise indicated. 
%Checked ok. 

%%%%%%%%%%%%%%%%%% FIGURE XX  REMOVED, COMBINED WITH FIGURE 4.
% \begin{figure}    
%                                % includes the two top panels 
%   \centerline{\hspace*{0.015\textwidth}
%               \includegraphics[width=8cm]{ploturrho3single_01_xocz_0.7127_60m.eps}
%               \hspace*{-0.03\textwidth}
%               \includegraphics[width=8cm]{ploturrho3single_01_xocz_0.7147_80m.eps}
%              }
%     \vspace{-0.35\textwidth}   % Shift close to the panel top 
%     \centerline{\Large \bf     % Includes the labels (here needs the color 
%                                %   package, see beginning of this file)
%      \hspace{0.0 \textwidth}  % \color{blue}{(a)}
%      \hspace{0.415\textwidth} %  \color{blue}{(b)}
%         \hfill}
%     \vspace{0.31\textwidth}    % Shift back to the panel bottom 
%           
%\caption{These plots show the mass flux at $80^\circ$ latitude for the two cases depicted in Figure \ref{open_conf}, respectively. First, we see in the left panel how the downwelling mass flux remains high in amplitude even at very deep zones in the radiation zone. This happens when the magnetic field remains open at the radiative--convective interface. On the other hand, we see in the right panel how a pre-confined magnetic field at the radiative--convective interface affects the downwelling mass flux to the extent that the fluid in the radiation zone stops its burrowing at depths not so far from the interface.}
%   \label{massflux_openconf}
%  \end{figure}
%%%%%%%%%%%%%%%%%% FIGURE XX 

\section{Reference model results} %%%%%%%%%%%%%%%%%%%%%%%%%%%%%%%%%%%%%%%%
      \label{results}

\begin{table*}
\centerline{
\begin{tabular}{lll} 
\hline
Parameter & Radiation zone $(r<\rcz)$ & Convection Zone $(r>\rcz)$ \\
\hline
$\nu/R_\odot^2 \Omega_\odot$ & $5 \times 10^{-10}$ & $5 \times 10^{-10}$ \\
$\eta/R_\odot^2 \Omega_\odot$ & $1.5 \times 10^{-9}$ & $1.5 \times 10^{-9}+ 5.0 \frac{r - r_1}{\rout-r_1}\left[1+\tanh\left(\frac{r- r_2}{\Delta_2}\right)  \right] H(r-r_1) $ \\
$\kappa/R_\odot^2 \Omega_\odot$ & $5 \times 10^{-7}$ & $5 \times 10^{-7}+ 0.5 \frac{r - r_1}{\rout-r_1}\left[1+\tanh\left(\frac{r- r_2}{\Delta_2}\right)  \right] H(r-r_1)$ \\
$\sigma$ & $\sigma_{\odot}(r)$ & $0$ \\
$\frac{1}{\tau}$ & $0$ & $\frac{1}{\tau_{\rm cz}}\left[\frac{r-r_{\rm cz}}{\rout-r_{\rm cz}} \right]$ \\
$\Lambda$ & $0.66 \times 10^{6}$ & $0.66 \times 10^{6}$ \\
\hline
\end{tabular}
}
\caption{Non-dimensional parameters for the reference model. In addition, $r_1-r_{\rm cz} = 0.0003 R_{\odot}$, and $r_2$ and $\Delta_2$ were defined in Section \ref{diffsect}. Finally, $\tau_{\rm cz} \Omega_\odot = 0.1$. }
\label{table1} 
\end{table*}
%Checked ok.

We now present and analyse in detail the results of our ``reference'' model. The governing parameters for this simulation were selected after a series of experiments starting from a parameter regime similar to the one proposed by \citet{Woodal11} and gradually reducing all diffusivities simultaneously. We found it necessary to reduce all diffusivities by four orders of magnitude, relative to those suggested by \citet{Woodal11}, in order to obtain a solution with a well-formed tachocline and tachopause . 
The reference model parameters are summarized in Table \ref{table1}.  We first describe its qualitative properties and then analyze it more quantitatively by studying the balance of forces in the tachocline. 
% Checked ok. 

\subsection{A first glimpse of the solar tachocline}

\begin{figure}    
                                % includes only one panel 
%              \centerline{\includegraphics[width=8cm]{fig6.eps}}
              \centerline{\includegraphics[width=8cm]{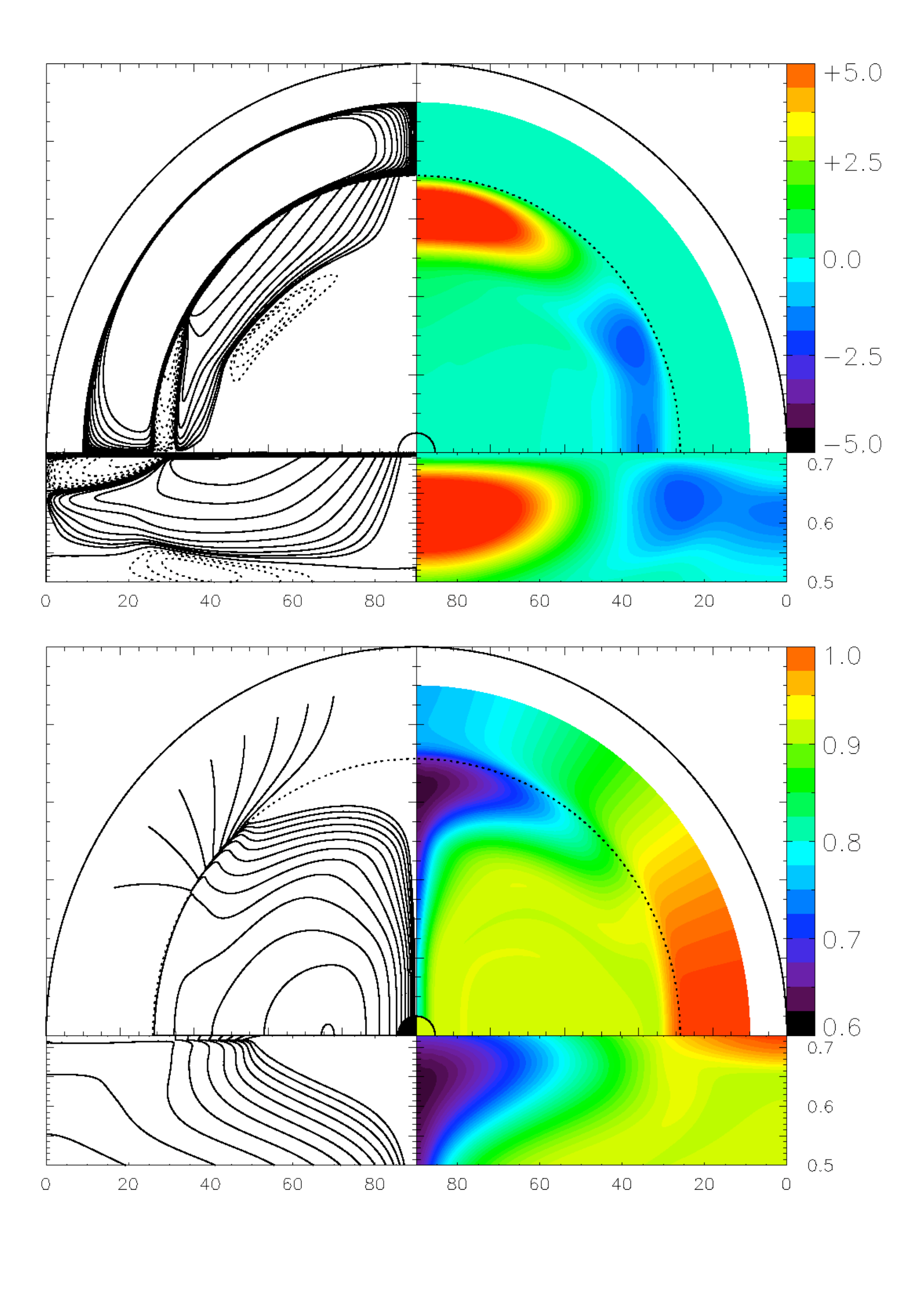}}
\caption{Steady-state solution for our reference model. The four panels show the same variables as in Figure \ref{open_conf}. Large-scale meridional flows generated in the convection zone penetrate into the radiation zone at high latitudes. These downwelling flows are deflected by the internal magnetic field in the tachopause and, in turn, confine the field. The confined field enforces solid-body rotation of the radiative interior below the tachopause.}
   \label{sigma_bestsim_ref}
   \end{figure}

Figure \ref{sigma_bestsim_ref} shows the flow field, temperature perturbation, magnetic field, and angular velocity in the reference model, demonstrating that magnetic confinement strictly below the radiative--convective interface, i.e. distinct from ``pre-confinement'', is possible.

%%%%%%%%%%%%%%%%%% FIGURE XX
% Checked ok. 
%%%%%%%%%%%%%%%%%% FIGURE XX

The top left figure shows the steady-state meridional flows in the reference model. From the surface to the center, we identify three main regions. In the convection zone, the flows are driven by the gyroscopic pumping effect of the imposed differential rotation. Fluid is pumped toward the poles near the surface, and returns equatorward near the base of the convection zone. The tachocline is located below the radiative--convective interface. Since the radiative region in our reference model is weakly stratified in terms of the parameter $\sigma$, a fraction of the meridional mass flux driven in the convection zone enters the tachocline. There, we observe two cells. The main tachocline cell downwells at the poles, is deflected equatorward at about $0.5 R_{\odot}$, then returns to the convection zone in a thin upwelling region around 30$^\circ$ in latitude. Close to the equator, a small cell with meridional flows circulating in the opposite direction is also visible. The downwelling flows in the large cell, although weak in amplitude, are sufficiently strong to distort the magnetic field (see below). Below the tachocline, we observe a thin meridional counter-cell, which is part of the tachopause. Deeper within the radiation zone, meridional flows become negligibly weak.
%Checked ok.

Note that the tachocline in this simulation is much thicker than in the Sun. This is  not surprising, since the model thermal and magnetic diffusivities are much larger than in the Sun, and we expect  the thickness of the tachocline and the tachopause to depend sensitively on these parameters. In Section \ref{varyparams}, we test the dependence of this thickness on the diffusivities, and demonstrate that our results become consistent with observations once extrapolated to solar parameter values. 
%Checked ok.

%%%%%%%%%%%%%%%%%% FIGURE XX
%  \begin{figure}    
%                                % includes only one panel 
%   \centerline{\includegraphics[width=8cm]{var_N91_RZ.eps}}
%\caption{This plot shows the magnetic Reynolds number, ${\rm Rm} = u_r L/\eta$, with $L = 0.10 R_{\odot}$. Thick contour lines correspond to ${\rm Rm} = \{1.0, 10.0, 100.0\}$, and emphasize the region below the radiative--convective interface where the advection of the field is expected to dominate over diffusion. XXXX Is this figure necessary? XXX }
%   \label{Rmplot}
%  \end{figure}
%%%%%%%%%%%%%%%%%% FIGURE XX

The bottom left figure shows the effect of these flows on the poloidal magnetic field. In the upper part of the convection zone, the magnetic diffusivity is high and the flows do not affect the field. By contrast, we see that the field lines are strongly distorted by the flows in the lower part of the convection zone, where the magnetic diffusivity drops and the magnetic Reynolds number increases above one in a manner very similar to that shown in Figure \ref{Rm_openconf}b (dashed line). The equatorward flows advect field lines from high to low latitudes, so their geometry in the vicinity of the base of the convection zone is mostly horizontal. This effect is the pre-confinement process discussed in Section \ref{issues}. 
%Checked ok.

Interestingly, we find that the magnetic Reynolds number remains relatively high in the tachocline, even though the meridional flow velocity drops significantly. Polar field lines are bent horizontally, pushed downward and advected toward mid-latitudes (roughly 30$^\circ$) as predicted by GM98 and studied by \citet{WoodMcIntyre11}. Below the tachopause at $r=0.5 R_{\odot}$, by contrast, we see that the field is not affected by the flows and is more or less equal to the imposed primordial field (see Figure \ref{magneticforcing}).
%Checked ok.

The bottom right panel of Figure \ref{sigma_bestsim_ref} shows the angular velocity profile of the reference model. The convection zone, as expected, rotates with the imposed differential rotation given by Equation (\ref{eq:diffrot}). Most of the radiation zone (for $r<0.5R_\odot$), by contrast, exhibits a uniform angular velocity with $\Omega_{\rm rz} \simeq 0.92\Omega_{\rm eq}$, which extends from the base of the tachocline to the core. The fact that $\Omega_{\rm rz}$ is so close to the observed rotation rate in the solar radiation zone is interesting, but could be a coincidence. 
%Checked ok.

The tachocline lies between these two regions. We note that its rotation rate (in the reference model) is not a monotonic function of radius at all latitudes. Indeed, the polar region rotates even more slowly than the convection zone above, a feature that is not seen in the solar tachocline. 
This sub-rotating region is caused by the extraction of angular momentum by the equatorward meridional flow. Because of  our unrealistically high thermal diffusivity, the strength of the meridional flow in our model is much larger than expected in the Sun, which in turn causes this sub-rotating feature to be unrealistically wide and strong. We show in Section \ref{varyparams} that extrapolating our results towards more realistic parameters leads to weaker meridional flows, and that the amplitude and extent of the sub-rotation decreases.
In other words, the presence of this feature in our reference simulation should not be viewed as an inherent problem with the GM98 model, but instead, as a natural limitation of simulations that have to be run, for computational feasibility, at non-solar parameters. 
%Checked ok.

The remaining panel in Figure \ref{sigma_bestsim_ref} (top right) shows the temperature perturbation profile in the reference model. The latter is negligible in the magnetically dominated part of the radiative region, below the tachopause. By contrast, the tachocline exhibits significant radial and latitudinal gradients consistent with thermal equilibrium and thermal-wind balance (see Section \ref{thermal}).

We now analyse the results of the reference model more quantitatively to determine the dominant force balance in each region, and compare our findings with the models of GM98 and \citet{Woodal11}. 
%Checked ok.

%%%%%%%%%%%%%%%%%% FIGURE XX
  \begin{figure}    
                                % includes only one panel 
%    \centerline{\includegraphics[width=8cm]{fig7.eps}}
    \centerline{\includegraphics[width=8cm]{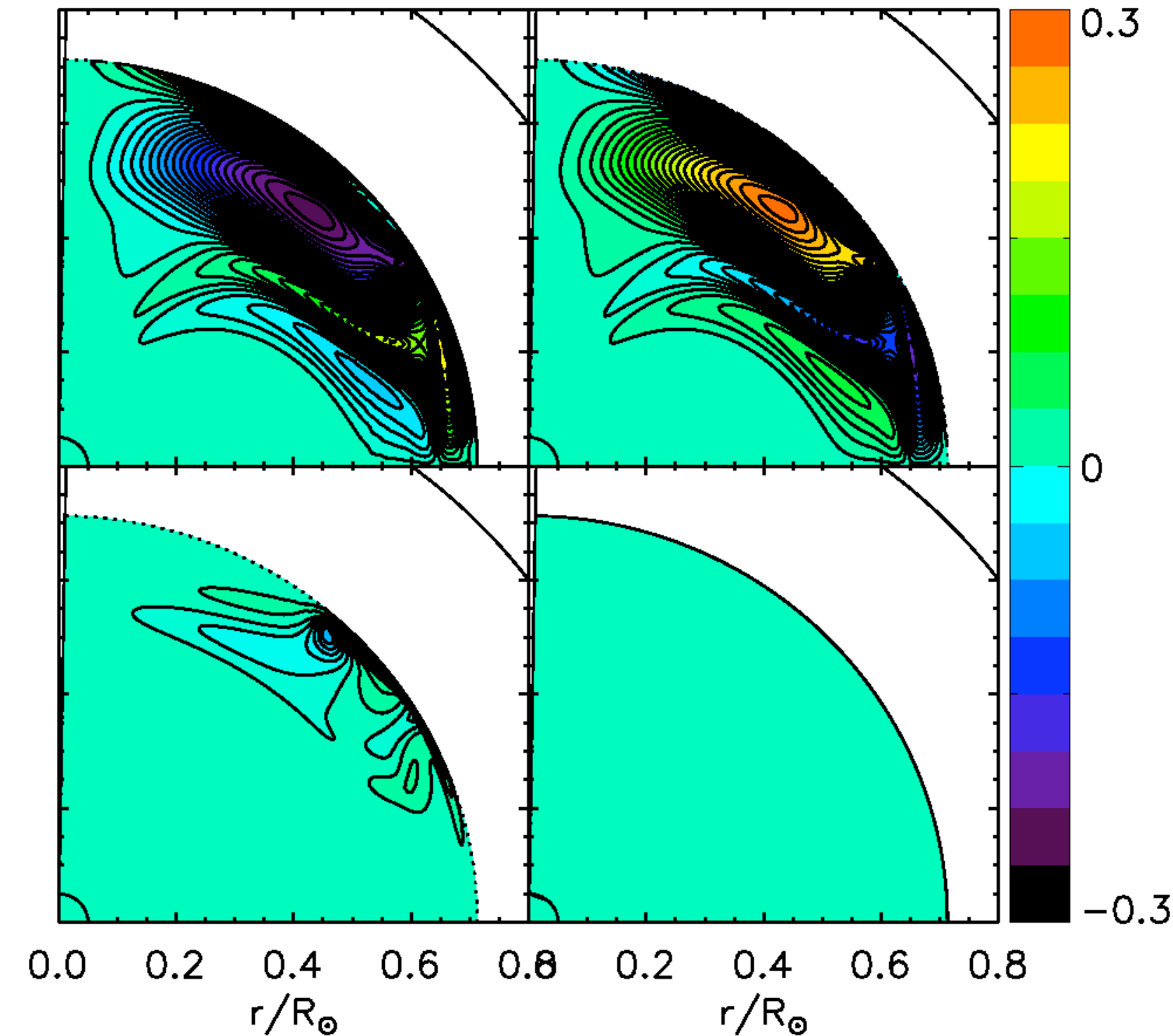}}
\caption{Comparison of the various terms contributing to the azimuthal vorticity equation in our reference model, in units of $\Omega_{\odot}^{2}$. Only the radiation zone is shown. The rotational shear (top left panel) balances the baroclinicity terms (top right panel). Both viscous (bottom left) and magnetic (bottom right) vorticity production terms are negligible. }
   \label{vorticityterms}
   \end{figure}
%Checked ok.
%%%%%%%%%%%%%%%%%% FIGURE XX

\subsection{Thermal wind balance in the radiative interior}
\label{thermal}

GM98 begin their scaling analysis by assuming that  the tachocline and tachopause are in thermal-wind balance.  We can easily verify this assumption in our model. Thermal-wind balance occurs when the system satisfies both hydrostatic and geostrophic equilibrium. Taking the curl of the steady-state momentum equation divided by the background density $\bar{\rho}$, we obtain:
\begin{eqnarray}
\label{thermalwind} % Checked
& & 2(\Omega_\odot +\tilde{\Omega}) r \sin \theta \left[ \cos \theta \frac{\partial \tilde{\Omega}}{\partial r} - \frac{\sin \theta}{r}\frac{\partial \tilde{\Omega}}{\partial \theta} \right]  + \frac{1}{r \bar{\rho}^2}\frac{\partial \bar{\rho}}{\partial r} \frac{\partial \tilde{p}}{\partial \theta} + \frac{\bar{g}}{r\bar{\rho}}\frac{\partial \tilde{\rho}}{\partial \theta}\nonumber \\
& &  + \left[\nabla \times \left( \frac{1}{\bar{\rho}} {\bf j} \times {\bf B}\right)\right]_{\phi} + \left[\nabla \times \left( \frac{1}{\bar{\rho}} \nabla \cdot {\bf \Pi}\right)\right]_{\phi} - \left[\nabla \times \left( \frac{{\bf u}-{\bf u}_{cz}}{\tau} \right) \right]_{\phi} = 0 \mbox{ , }
\end{eqnarray}
where the first term measures the rotational shear along the rotation axis\footnote{The two terms in the bracket on the left-hand side can be expressed more concisely as $\mathbf{e}_z\cdot\nabla\Omega$.}, and the next two terms account for the total baroclinicity. The balance between rotational shear and total baroclinicity is the thermal-wind equation. The remaining terms in this equation are the magnetic and viscous vorticity-production terms as well as the curl of the gyroscopic pumping term. The latter is included for completeness but vanishes in the radiative interior (see Section \ref{forcing1}).
% Checked ok.

The two top panels in Figure \ref{vorticityterms} show that the rotational shear (left) and the total baroclinicity term (right) clearly balance each other in the tachocline and tachopause region. The bottom panels show the viscous (left) and magnetic (right) terms, and confirm that their respective contributions to the azimuthal vorticity equation is negligible. In our solutions, thermal-wind balance holds throughout most of the radiative interior, as assumed by GM98. 
In the model of \citet{Woodal11}, on the other hand, thermal-wind balance is broken by Lorentz forces in the tachopause at the bottom of the tachocline.  Our results show that this is not necessary in order for a tachopause to form.  We note, however, that the strength of the primordial magnetic field in our reference model is much weaker than the typical field strengths considered by \citet{Woodal11}. With a stronger magnetic field, thermal-wind balance might well be broken in the tachopause.  
% Checked ok.

%%%%%%%%%%%%%%%%%% FIGURE XX
  \begin{figure}    
                                % includes only one panel 
%              \centerline{\includegraphics[width=12cm]{fig8.eps}}
              \centerline{\includegraphics[width=8cm]{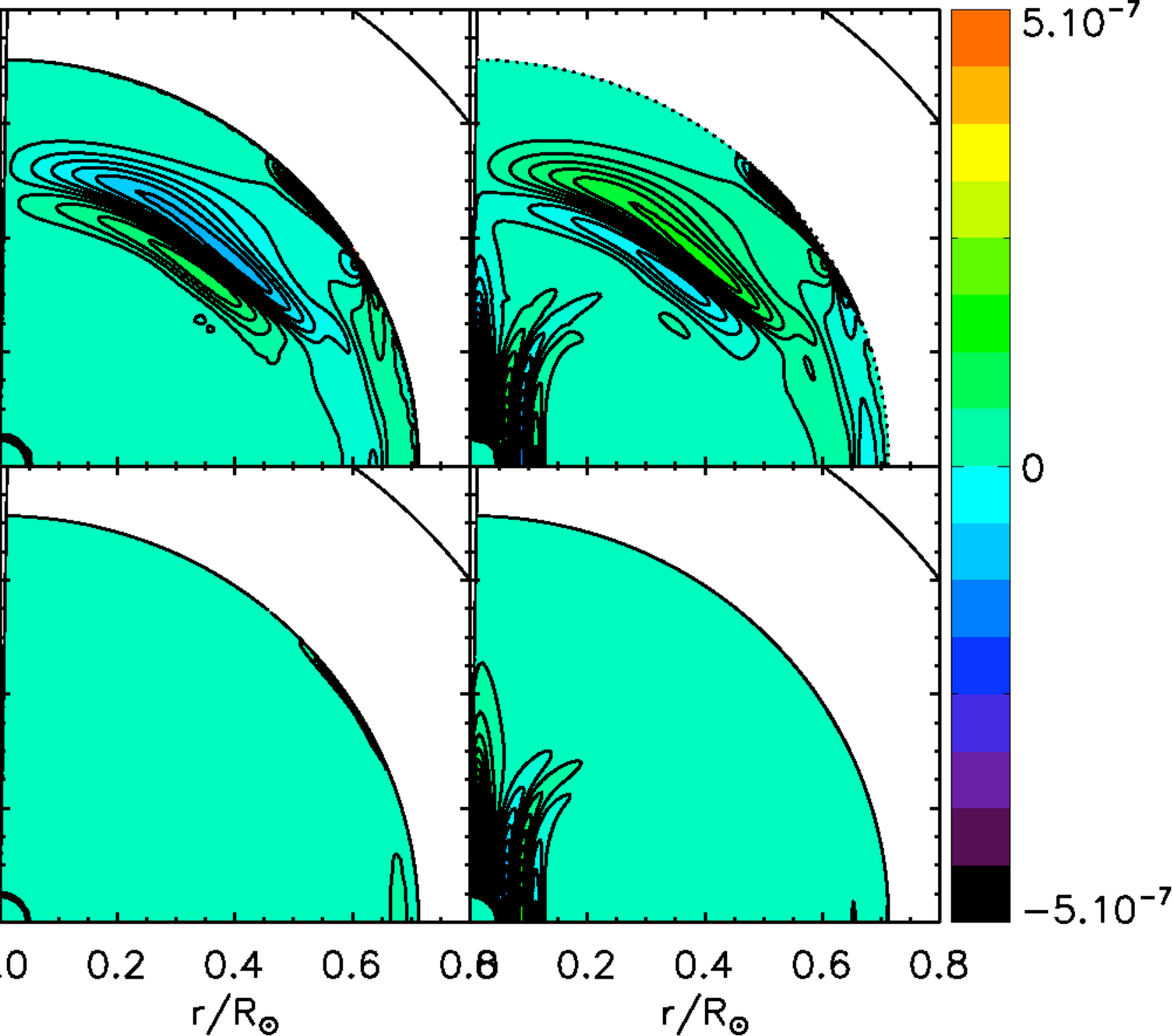}}
\caption{Comparison of the various terms that contribute to the azimuthal force balance in the radiation zone in our reference model, in units of $\rho_0 R_\odot \Omega^2_\odot$. In the tachocline and tachopause (for $r>0.5R_\odot$), the Coriolis force (top left) and Lorentz force (top right) are in balance, while the non-linear advection term (bottom left) and the viscous torque (bottom right) are negligible. Note that the tachocline is mostly force-free while the tachopause stands out as the region where strong Coriolis and Lorentz forces balance. }
   \label{momentumforces}
   \end{figure}
% Checked ok.
%%%%%%%%%%%%%%%%%% FIGURE XX

\subsection{The azimuthal force balance in the tachocline}
\label{azimuthalforces}

We now examine the second ingredient of the GM98 model, namely the angular-momentum balance, by inspecting the azimuthal component of the momentum equation. The latter can be expanded as:
\begin{eqnarray} % Checked
-2 \bar \rho \Omega_{\odot}(u_{\theta} \cos \theta + u_r \sin \theta) & & \mbox{ (Coriolis) }\nonumber \\
-\bar \rho \left[ u_r \frac{\partial u_{\phi}}{\partial r} + \frac{u_{\theta}}{r} \frac{\partial u_{\phi}}{\partial \theta} + \frac{1}{r}u_{\phi}u_r + \frac{\cos \theta}{r \sin \theta}u_{\phi}u_{\theta}\right] & & \mbox{ (Inertial) }\nonumber \\
+ j_rB_{\theta} - j_{\theta}B_r & & \mbox{ (Lorentz) }\nonumber \\
-\frac{\bar \rho}{\tau} [ u_{\phi}-u_{\rm cz}] & & \mbox{ (Gyroscopic pumping) }\nonumber \\
+ (\nabla \cdot {\bf \Pi})_{\phi}  = 0 & & \mbox{ (Viscous stresses) }
\label{phimom}
\end{eqnarray}
where ${\bf \Pi}  = \bar \rho \nu [\nabla {\bf u} +(\nabla {\bf u})^T - \frac{2}{3}(\nabla \cdot {\bf u}) {\bf I}]$.
Because neither gravity nor pressure enters into the angular-momentum balance, forces that are insignificant in the meridional directions can be of leading order here.
Figure \ref{momentumforces} shows each of these terms individually (except the pumping term, which does not act in the radiation zone).
%Checked text, not eqs.

First and foremost, note that the viscous stresses are negligible in the tachocline, confirming the theoretical expectation that viscosity should play no role in its dynamics \footnote{Viscous stresses are significant in the proximity of the polar axis, where they balance the Lorentz forces in a thin Ekman--Hartmann-type boundary layer that is produced by the artificial lower boundary at $r=\rin$.  Such a boundary layer is not present in the Sun.}. This is in contrast with the results of the direct numerical simulations of \citet{Strugarekal11} and \citet{Rogers11},
in which $\sigma \gg 1$ and angular-momentum transport is viscously dominated.  

The top panels of Figure \ref{momentumforces} show, from left to right, the azimuthal component of the Coriolis and Lorentz forces. Both are in balance in most of the radiation zone (except close to the polar axis near the inner boundary), strongest in the tachopause, and significantly weaker in the tachocline. This is consistent with the GM98 model, in which the tachocline is essentially force-free. Lorentz forces become important only in the tachopause, where they provide the angular momentum necessary for the flows to be deflected equatorward. %
%Checked ok. 

\subsection{The thickness of the tachopause and the tachocline}
\label{tachocline_tachopause}

%%%%%%%%%%%%%%%%%% FIGURE XX
  \begin{figure}    
                                % includes only one panel 
%    \centerline{\includegraphics[width=8cm]{fig9.eps}}
    \centerline{\includegraphics[width=8cm]{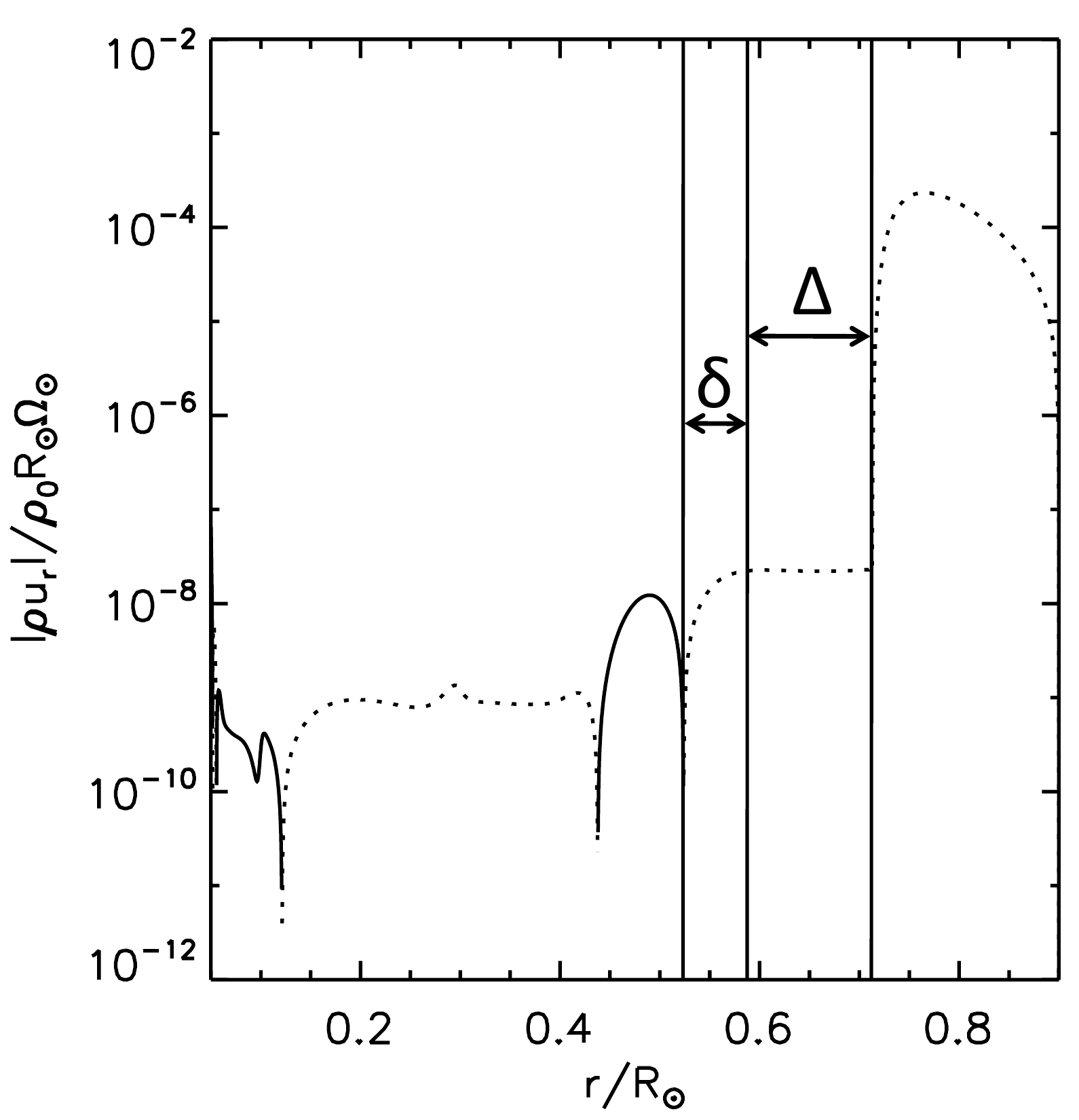}}
\caption{Radial mass flux $|\bar{\rho} u_r|$ at 80$^\circ$ latitude. A dotted line is used when $u_r < 0$, and a solid line is used when $u_r > 0$. We observe depth-independent downwelling  from the radiative--convective interface down to about $r=0.58 R_{\odot}$, a region we identify as the tachocline. Below $0.58 R_{\odot}$, $|\bar{\rho} u_r|$ decreases to zero then oscillates about it. We identify the region between the base of the tachocline and the first zero of $|\bar{\rho} u_r|$ as the tachopause (here at $r = 0.525 R_{\odot}$).}
   \label{massflux}
   \end{figure}
%Checked ok. 
%%%%%%%%%%%%%%%%%% FIGURE XX

The force-free nature of the tachocline constrains the flow of material downwelling from the convection zone \citep[see][for a detailed discussion]{Woodal11}. Indeed, if viscous torques, magnetic torques and inertial terms are neglected in Equation (\ref{phimom}), then  
\begin{equation}
({\bf \Omega}_\odot \times {\bf u})_\phi = 0
%Checked ok. 
\end{equation}
which implies that the meridional flow velocity must be parallel to the rotation axis. Combining this result with mass conservation (in a cylindrical coordinate system) we then have
\begin{equation}
\frac{\partial}{\partial z}(\bar{\rho} u_z) = 0 \mbox{ , }
%Checked ok. 
\end{equation}
where $z= r \cos \theta$. At high latitudes, $u_r \simeq u_z$, so the radial mass flux must be roughly constant along the polar axis. Within the tachopause, by contrast, the prograde torque from the Lorentz force is significant and gyroscopically pumps the fluid equatorward, allowing the meridional flow to turn over and return to the convection zone. As a result, $\bar \rho u_r$ is no longer constant.

This qualitative picture is verified in Figure~\ref{massflux}, which shows the mass flux $|\bar \rho u_r|$ at 80$^\circ$ latitude. We see that $|\bar \rho u_r|$ is constant between the radiative--convective interface and $r \approx 0.58 R_{\odot}$ and then decreases rapidly to zero, where the magnetic torque is strongest.
%Checked text, not eqs.

This observation provides an objective measure of the location of the tachopause, and hence of the thickness of the tachocline. In all that follows, we define the base of the tachocline $r_T$ (which is also the top of the tachopause) to be located at the depth where $|\bar{\rho} u_r|$ drops by $5 \%$ from its value
at $r=\rcz$, as measured at $80^\circ$ latitude.
In Figure~\ref{massflux}, we find that $r_T = 0.575 R_{\odot}$ in the reference model.
We then define the tachopause as the region located between the base of the tachocline and the depth at which the vertical mass flux first equals zero. From Figure \ref{massflux}, we find that this happens at $r_t = 0.525 R_{\odot}$. The thickness of our reference model tachocline is then roughly three times that of the tachopause.
%Checked ok.

Although the particular way in which we measure the tachocline and tachopause thicknesses (e.g. the latitude at which the measurement is made, the percentage drop in $|\bar{\rho} u_r|$) are somewhat arbitrary, the concepts behind the definitions themselves are robust. This provides a means to test how the thicknesses of the tachocline and tachopause vary with the governing parameters (in particular, $E_{\kappa}$ and $E_{\eta}$) and is used in Section \ref{varyparams} to test the predictions of the GM98 model. 
  \begin{figure}%[ht]   
                                % includes only one panel 
%    \centerline{\includegraphics[width=8cm]{fig10.eps}}
    \centerline{\includegraphics[width=8cm]{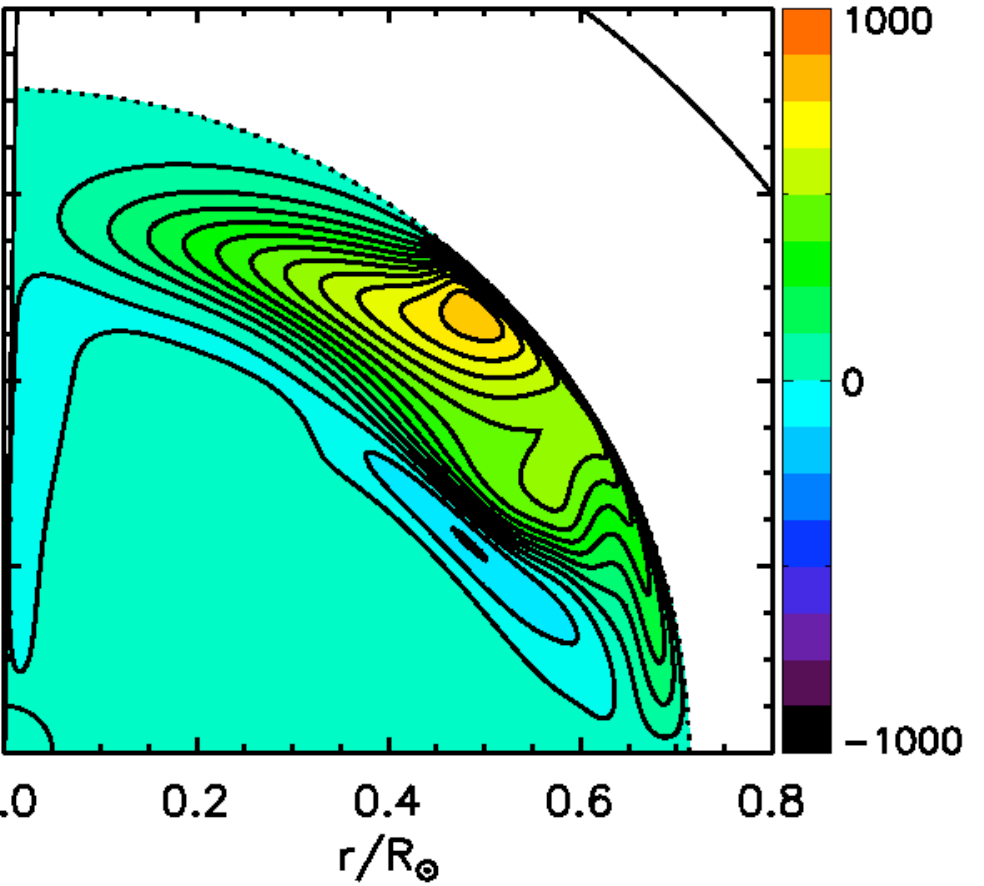}}
\caption{The toroidal field in the radiation zone, in units of $B_0$. Note how the strongest field concentrations are found within the tachopause, and within a narrow region near the base of the convection zone around 45$^\circ$ latitude. }
   \label{toroidalplot}
   \end{figure}
%Checked ok.
%%%%%%%%%%%%%%%%%% FIGURE XX

\subsection{Toroidal field}

The final aspect of the GM98 theory that we investigate is the distribution of toroidal field in the tachocline and in the tachopause.  Strong toroidal fields are generated when the poloidal component of the magnetic field is twisted by the azimuthal flows. GM98 argue that this effect should be most important in the tachopause, which lies at the interface between a region of strong angular-velocity shear (the tachocline) and a region dominated by a strong poloidal field (the deep interior). They also mention the possibility that toroidal fields could be produced in the mid-latitude upwelling region (where the flow drags poloidal field lines back into the tachocline), and might thus play a role in the solar dynamo.

Figure \ref{toroidalplot} shows the distribution of toroidal field in the radiative region for our reference model. As expected, we find that strong fields are indeed generated in the tachopause.  We also find them in a localized region of the tachocline, around $45^\circ$ latitude, which is somewhat closer to the poles than the center of the upwelling region (which lies around $30^\circ$ latitude). %However, this result should not be surprising: by inspection of the angular velocity profile and meridional flow structure in Figure \ref{sigma_bestsim_ref} we see that the upwelling region is actually free of shear. The maximum of the toroidal flux is thus located somewhat north of the upwelling instead, where stronger toroidal shear can be found.
%Checked ok.

\section{Numerical experiments varying the parameters}
\label{varyparams}

Having established that our reference model satisfies the various force balances predicted by the GM98 model, we now test the scaling laws implied by these balances. We do so by varying the parameters governing the system around
the reference model values. 
%Checked ok.

\subsection{The effect of the thermal diffusivity}
\label{kappa_section}

We first vary the thermal diffusivity, keeping all other parameters fixed -- in particular, without changing $\nu$ or $\bar{N}$. We do this by multiplying the reference model function $\kappa(r)$ (see Table 1 and Section \ref{diffsect}) by a constant factor $f_\kappa$. This implies that $\kappa(r)$ changes in the convection zone as well, but it can be shown by inspection of the meridional flow velocity profile for $r>r_{\rm cz}$ that this alone has a negligible effect on the solution. Note that varying $\kappa$ without varying $\nu$ or $\bar{N}$ implies that $\sigma$ changes as well, but in all cases presented we made sure that $\sigma$ remains smaller than 1 to guarantee that the effects of viscosity are indeed negligible in the tachocline and tachopause. 
%Checked ok.

%%%%%%%%%%%%%%%%%% FIGURE XX
  \begin{figure*}    
                                % includes the two top panels 
   \centerline{\hspace*{0.015\textwidth}
               \includegraphics[width=8cm]{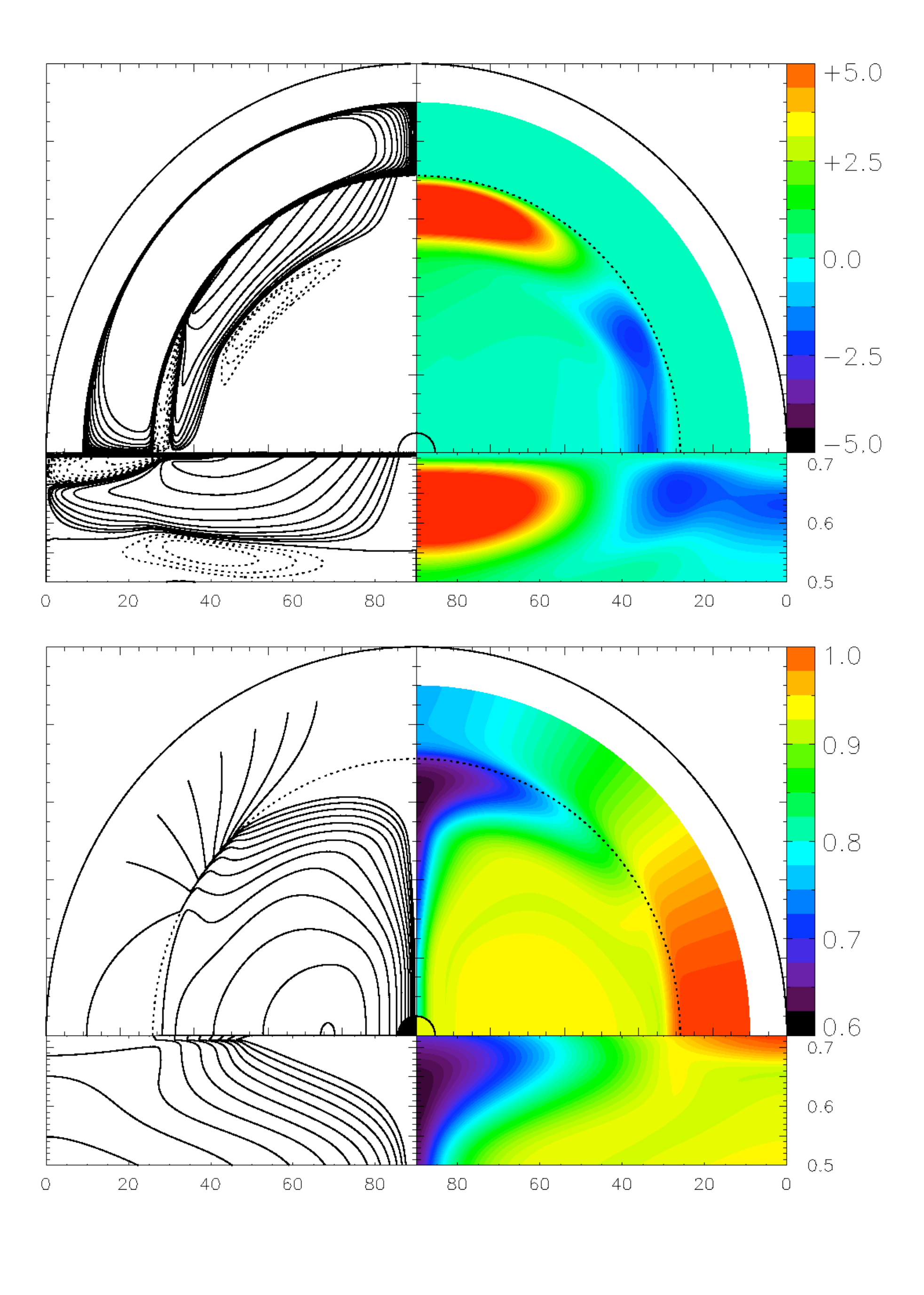}
               \hspace*{-0.03\textwidth}
               \includegraphics[width=8cm]{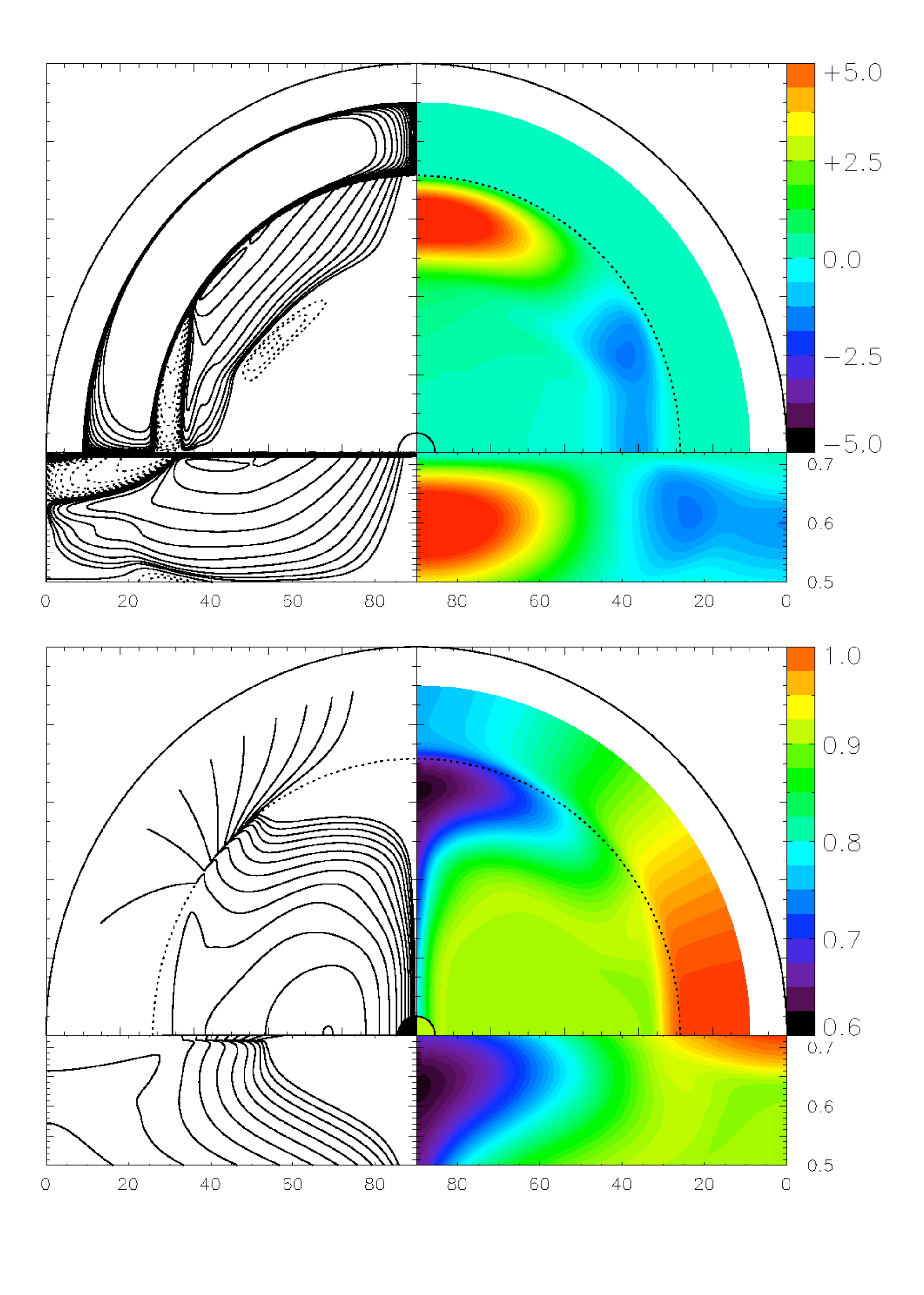}
              }
     \vspace{-0.35\textwidth}   % Shift close to the panel top 
     \centerline{\Large \bf     % Includes the labels (here needs the color 
                                %   package, see beginning of this file)
      \hspace{0.0 \textwidth}  % \color{blue}{(a)}
      \hspace{0.415\textwidth} % \color{blue}{(b)}
         \hfill}
     \vspace{0.31\textwidth}    % Shift back to the panel bottom 

\caption{Simulations with a thermal diffusivity $\kappa$ that is half that of the reference model (left, $f_\kappa = 0.5$), and 2.25 times that of the reference model ($f_\kappa = 2.25$, right). All other parameter values are otherwise unchanged from those of Table 1. The higher diffusivity simulation has a thicker tachocline, and field lines are more strongly distorted by the downwelling flows. }
   \label{sigma_bestsim1}
   \end{figure*}
%Checked ok.
%%%%%%%%%%%%%%%%%% FIGURE XX
 
We are interested in finding out how varying $\kappa$ in the radiation zone and the tachocline affects the dynamics of these regions. The results are shown in Figure \ref{sigma_bestsim1}. The left panel shows the effect of reducing $\kappa$ from the reference model by a factor of $2$. Because of thermal equilibrium, this directly affects the amplitude of the downwelling meridional flows in the tachocline and therefore the depth at which these flows confine the field. We see in the streamline plot that the tachocline is now thinner than in the reference case (see Figure \ref{sigma_bestsim_ref} for comparison). The field lines are also less distorted in the tachocline than in the reference model. The right panel shows the effects of an increase in $\kappa$ from the reference model value by a factor of $2.25$. This time, the meridional flows are stronger, which leads to a thicker tachocline, and greater distortion of the magnetic field lines.
%Checked ok.

Another consequence of the respectively weaker and stronger meridional flows is a clear change in the angular-velocity profile of the polar tachocline. As discussed in Section \ref{results}, we find that the subrotating region becomes both weaker and smaller as the meridional flow velocity in the tachocline decreases. 
%Checked ok.

We now study these trends more quantitatively. As discussed by \citet{SpiegelZahn92} and GM98, thermal-wind balance and thermal equilibrium imply specific scalings for the tachocline. Indeed, thermal-wind balance means that 
\begin{equation}
2 r \sin \theta \Omega_\odot \frac{\partial \tilde{\Omega}}{\partial z} \approx \frac{\bar{g}}{r\bar{T}}\frac{\partial \tilde{T}}{\partial \theta} \mbox{ , }
%Checked ok.
\end{equation}
(see Section \ref{thermal}) where $z= r \cos \theta$. If we further assume that the tachocline thickness $\Delta \ll R_\odot$ then:
\begin{equation}
\frac{\tilde{T}}{\bar{T}} \approx \frac{ \alpha \Omega_\odot^2 r_{\rm cz}^2}{\bar{g}l \Delta} \mbox{ , }
%Checked ok.
\end{equation}
where we have approximated the rotational shear as $\partial \tilde{\Omega}/\partial z \approx \alpha \Omega_\odot/\Delta$, where $\alpha$ characterizes the level of differential rotation. The latitudinal temperature gradient is approximated as $\partial \tilde{T} / \partial \theta \approx l \tilde{T}$, where $l$ is a latitudinal wavenumber of order unity.  If the tachocline is also in thermal equilibrium, then
\begin{equation}
\label{thermal3}
\frac{\bar{N}^2 \bar{T}}{\bar{g}} u_r   \approx  \kappa \frac{\partial^2\tilde{T} }{\partial r^2}  \mbox{ , }
\end{equation}
so that 
\begin{equation}
u_r \approx  \frac{\beta^2 \kappa_{\rm rz} \bar{g}}{\bar{N}^2 \Delta^2} \frac{\tilde{T}}{\bar{T}} \mbox{ , }
%Checked ok.
\end{equation}  
using $\partial/\partial r \approx  \beta/\Delta$, where $\beta$ is a geometrical constant of order unity.  From these two equilibria, GM98 deduce that
\begin{equation}
\label{tacho_ur3}
u_r \approx  \frac{\alpha \beta^2}{l} \frac{\Omega_\odot^2}{\bar N^2} \frac{ \kappa_{\rm rz} }{r_{\rm cz} } \frac{   r_{\rm cz}^3}{ \Delta^3}  = K \frac{\Omega_\odot^2}{\bar N^2} \frac{ \kappa_{\rm rz} }{r_{\rm cz} }  \frac{   r_{\rm cz}^3}{ \Delta^3}    \mbox{ ,}
%Checked ok.
\end{equation}
where $K$ is a geometrical constant. Using $\alpha = 0.07$, $\beta = \pi$ and $l = 4.5$, they estimate that $K \simeq 0.15$. 
%CHecked ok

Equation (\ref{tacho_ur3}) relates the amplitude of the downwelling meridional flows in the tachocline to its thickness under robust physical assumptions; it is a generic property of many models \citep{SpiegelZahn92,GoughMcIntyre98,Woodal11}. We can now check its validity using our simulations in conjunction with the measurement technique for the thickness of the tachocline and tachopause described earlier. However, the comparison is not entirely trivial, since the tachocline and tachopause in our simulations are not necessarily as thin as required by the assumptions made by GM98 (recall that our diffusion parameters are still significantly larger than those of the Sun). This has two consequences. First, the background density and buoyancy frequency vary by a significant amount within the tachocline. This implies that although $\bar \rho u_r$ is constant with depth, $u_r$ is not. One must then choose at which radial position Equation (\ref{tacho_ur3}) should be used to estimate $u_r$. In what follows, we take for simplicity $r = 0.7 R_\odot$, which is well within the tachocline for all cases considered. Furthermore, GM98 assume that the tachopause is much thinner than the tachocline, so they did not need to differentiate between $\Delta$ (the tachocline thickness only) and $\Delta + \delta$ (the sum of the tachocline and tachopause thicknesses). Their original derivation leads to Equation (\ref{tacho_ur3}). Here, $\delta$ can be of the same order of magnitude as $\Delta$, so we must decide whether Equation (\ref{tacho_ur3}) or
\begin{equation}
\label{tacho_ur4}
u_r \approx  K u_{\rm theor} =  K \frac{\Omega_\odot^2}{\bar{N}^2}  \frac{\kappa_{\rm rz}}{r_{\rm cz}}  \frac{r_{\rm cz}^3}{D^3}  \mbox{  where } D = \Delta + \delta
%Checked ok
\end{equation}
is more appropriate. We actually believe that the latter is, since the tachopause is {\it also} in thermal equilibrium and in thermal-wind balance, and since both the shear and the temperature perturbations should vanish at the base of the tachopause rather than the base of the tachocline (so $\partial \tilde{\Omega}/\partial z \approx \alpha \Omega_\odot/D$ and $\partial \tilde{T}/\partial r \approx \beta \tilde{T}/D$ instead).  
%Checked ok

%%%%%%%%%%%%%%%%%% FIGURE X
  \begin{figure}%[h]   
                                % includes only ONE panel 
%              \centerline{\includegraphics[width=8cm]{fig12.eps}}
    \centerline{\includegraphics[width=8cm]{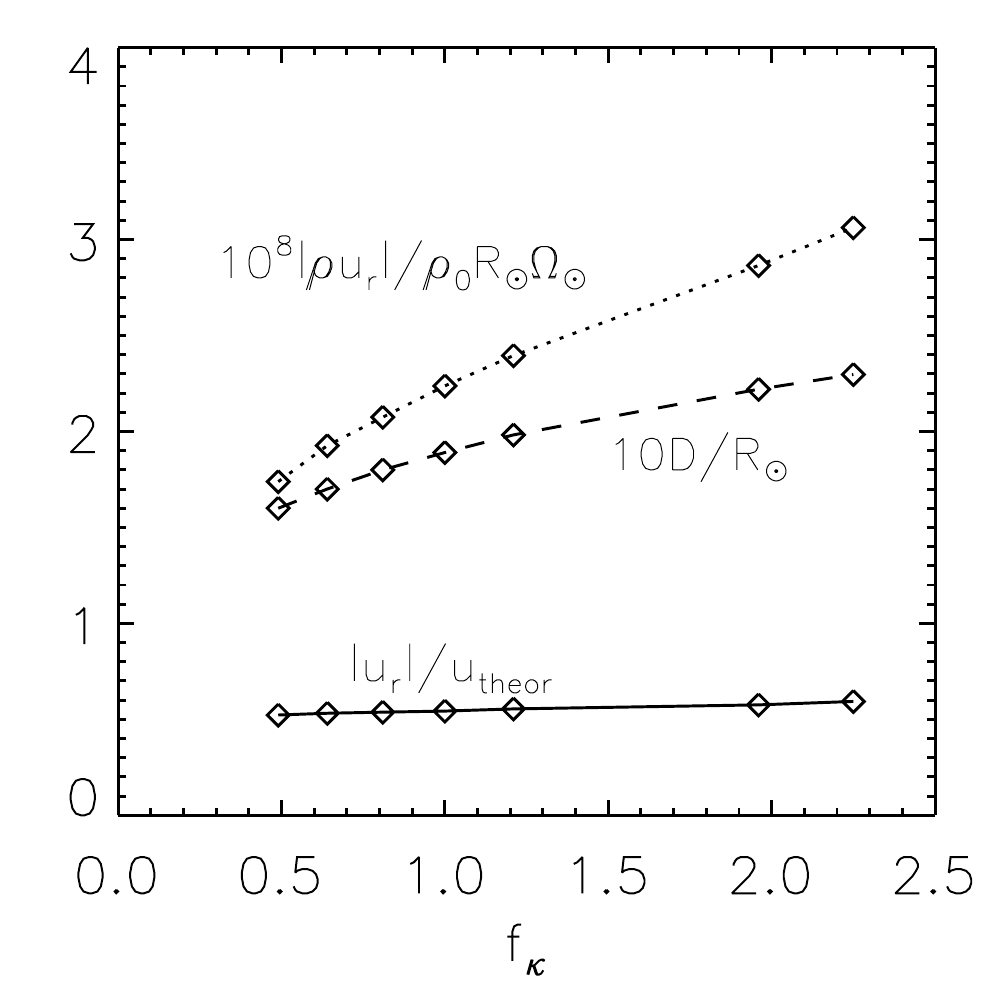}}
\caption{Variation with respect to $f_{\kappa}$ of the mass flux $|\bar{\rho} u_r|$ in the tachocline (dotted line), of the combined thicknesses of the tachocline and tachopause $D$ (dashed line), and of $K = |u_r|/u_{\rm theor}$ (solid line), all evaluated at $r=0.7R_\odot$. Our results show that Equation (\ref{tacho_ur4}) provides a good estimate for $|u_r|$, and that $K \simeq 0.5$. }
   \label{Ek_ratio}
   \end{figure}
%Checked ok.
%%%%%%%%%%%%%%%%%% FIGURE X 

The relationship between $f_{\kappa}$, the amplitude of the mass flux in the tachocline $|\bar{\rho} u_r|$, and the thickness of the tachocline-tachopause region $D= \delta + \Delta$ is shown in Figure \ref{Ek_ratio}.  As expected, increasing $f_\kappa$ effectively weakens the stable thermal stratification, allowing for a larger meridional mass flux in the tachocline.  We find that $\bar \rho u_r$ changes by a factor of about two when $f_{\kappa}$ changes by a factor of five. The stronger flows push the field down further so $D$ increases too, although by a smaller amount (about 25\% only). Using this information, we can check the estimate for $u_r$ given in Equation (\ref{tacho_ur4}). This is also shown in Figure \ref{Ek_ratio}. We see that the ratio $K = |u_r|/u_{\rm theor}$ varies very little with $f_\kappa$, taking a value of about 0.5.  This is a few times larger than the estimate provided by GM98, but is overall remarkably consistent with their predictions. Finally, note that if we use $\Delta$ instead of $D$, that is, Equation (\ref{tacho_ur3}) instead of (\ref{tacho_ur4}), the ratio $|u_r|/u_{\rm theor}$ varies more with $f_\kappa$. This suggests that Equation (\ref{tacho_ur4}) is the better estimate for $u_r$.
%Checked ok.

In summary, we find that our simulations confirm the estimates of GM98 and \citet{Woodal11} concerning the relationship between the downwelling flow velocities and the thickness of the tachocline, albeit with a minor modification. However, we note that the range of $f_\kappa$ tested is very limited, owing to computational limitations. With the other parameter values and resolution fixed, we were unable to find any solution for larger $f_\kappa$ than 2.25, and for smaller $f_\kappa$ than 0.5. For larger $f_\kappa$, the tachocline is increasingly deep and the meridional flows interact more strongly with the magnetic field. Our steady state solver has difficulties following the solutions in that limit. For smaller $f_\kappa$, $\sigma$ increases above unity and the tachopause begins to overlap with  the convection zone, both of which lead to a qualitative change in the nature of the solutions.
%Checked ok.

\subsection{The effect of the magnetic diffusivity}
\label{eta_section}

We now study the effect of varying the magnetic diffusivity of the radiation zone on the dynamics of the system. Starting from the reference model described in Section \ref{results}, we vary $\eta_{\rm rz}$ (or equivalently, $E_\eta$), by multiplying the reference value (see Table 1) by a factor $f_\eta$. All other parameters remain the same as in the reference model. This time, the magnetic diffusivity profile within the convection zone does not change when varying $f_\eta$, which implies that the magnetic Reynolds number within the convection zone remains approximately the same across all simulations. This ensures that the pre-confinement process is the same for all cases.
%Checked ok.

\begin{figure*}    
                                % includes the two top panels 
   \centerline{\hspace*{0.015\textwidth}
     \includegraphics[width=8cm]{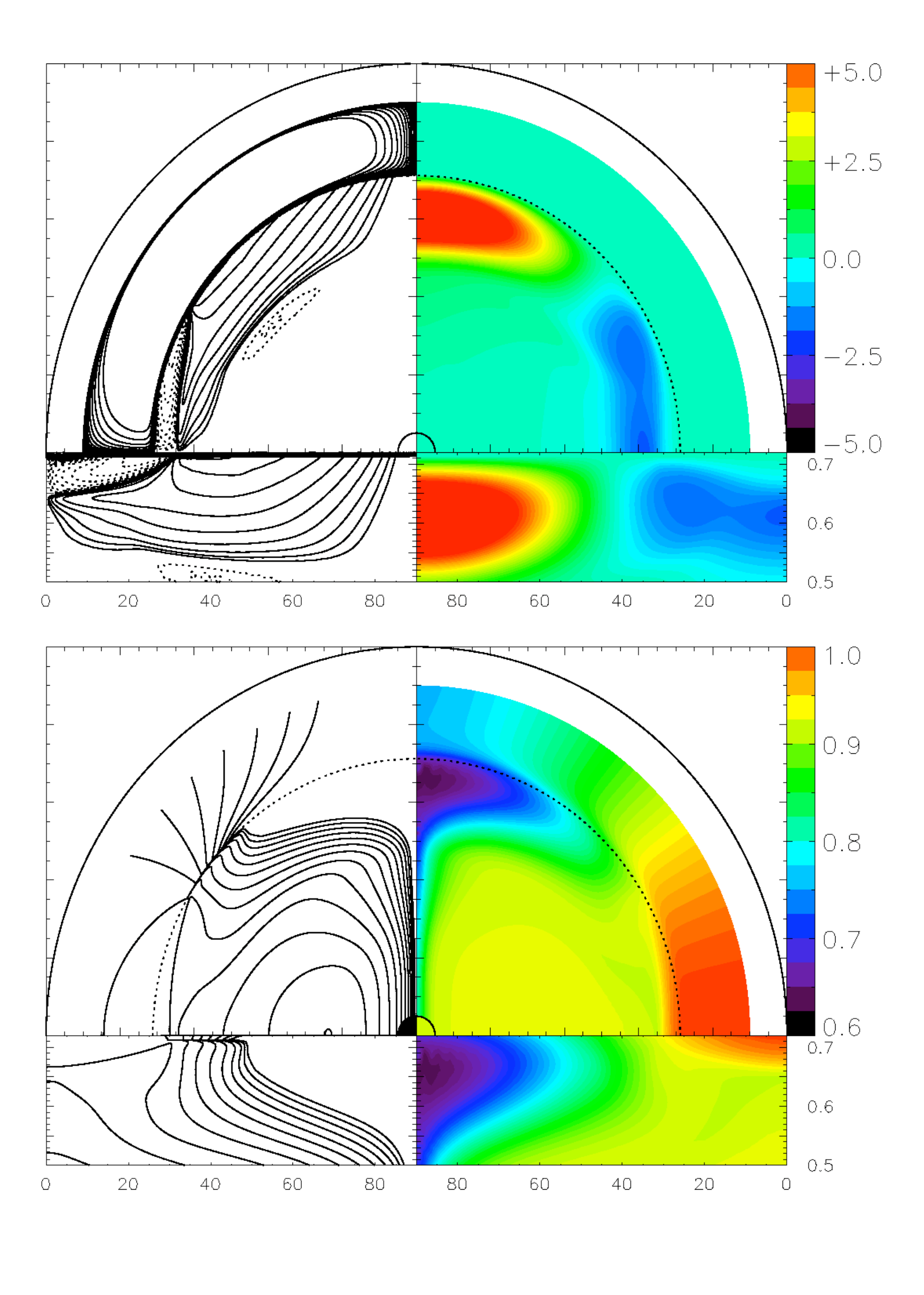}
     \hspace*{-0.03\textwidth}
     \includegraphics[width=8cm]{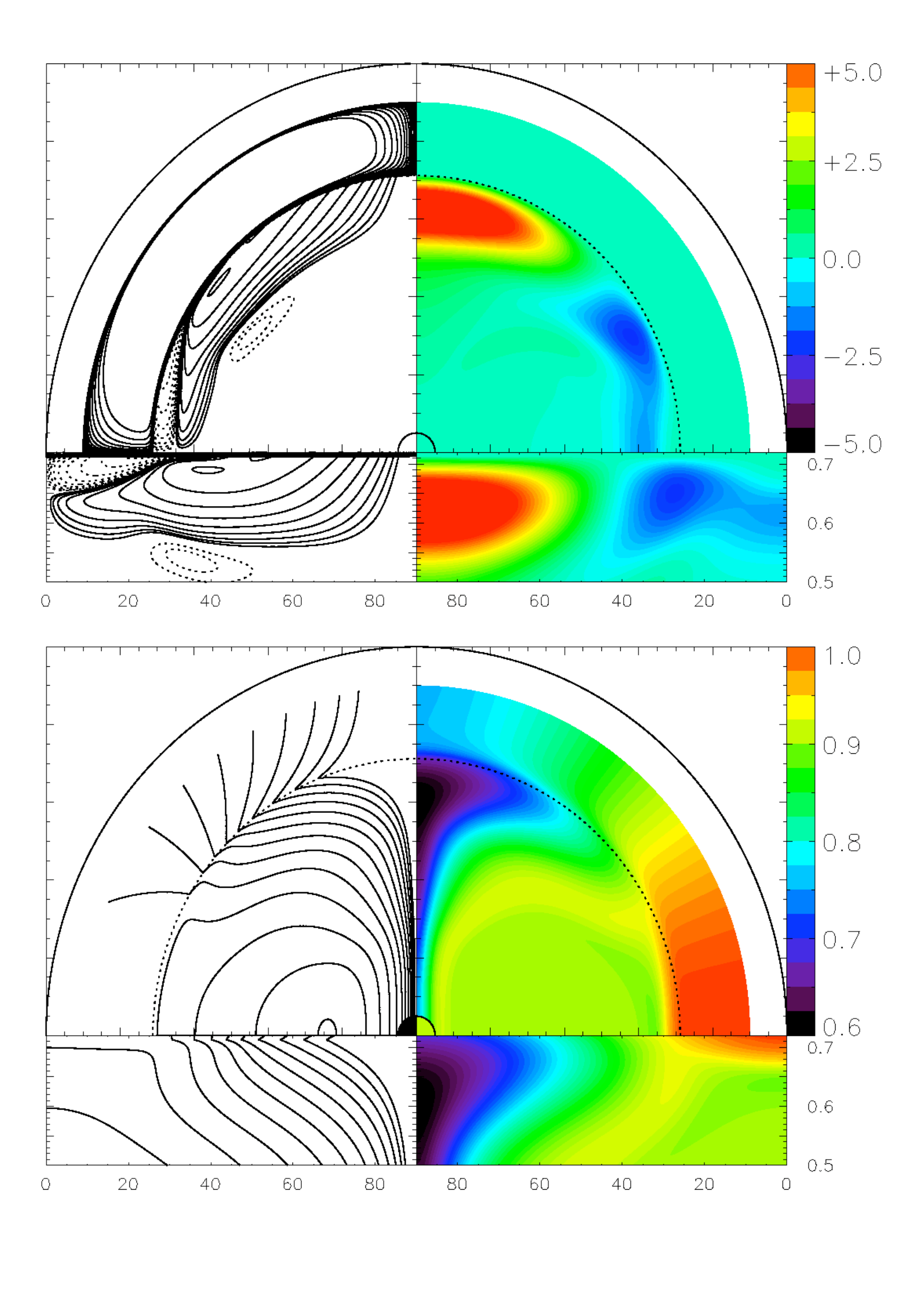}
              }
     \vspace{-0.35\textwidth}   % Shift close to the panel top 
     \centerline{\Large \bf     % Includes the labels (here needs the color 
                                %   package, see beginning of this file)
      \hspace{0.0 \textwidth}  % \color{blue}{(a)}
      \hspace{0.415\textwidth} % \color{blue}{(b)}
         \hfill}
     \vspace{0.31\textwidth}    % Shift back to the panel bottom 

\caption{Simulations with a magnetic diffusivity $\eta_{\rm rz}$ that is half that of the reference model (left, $f_\eta = 0.5$), and 3.33 times that of the reference model ($f_\eta = 3.33$, right). All other parameter values are otherwise unchanged from those of Table 1. The lower-diffusivity simulation has a thicker tachocline, and the magnetic field lines are more distorted.}
   \label{sigma_bestsim2}
   \end{figure*}

\begin{figure*}%[h]    
                                % includes only ONE panel 
%              \centerline{\includegraphics[width=14cm]{fig14.eps}}
    \centerline{\includegraphics[width=14cm]{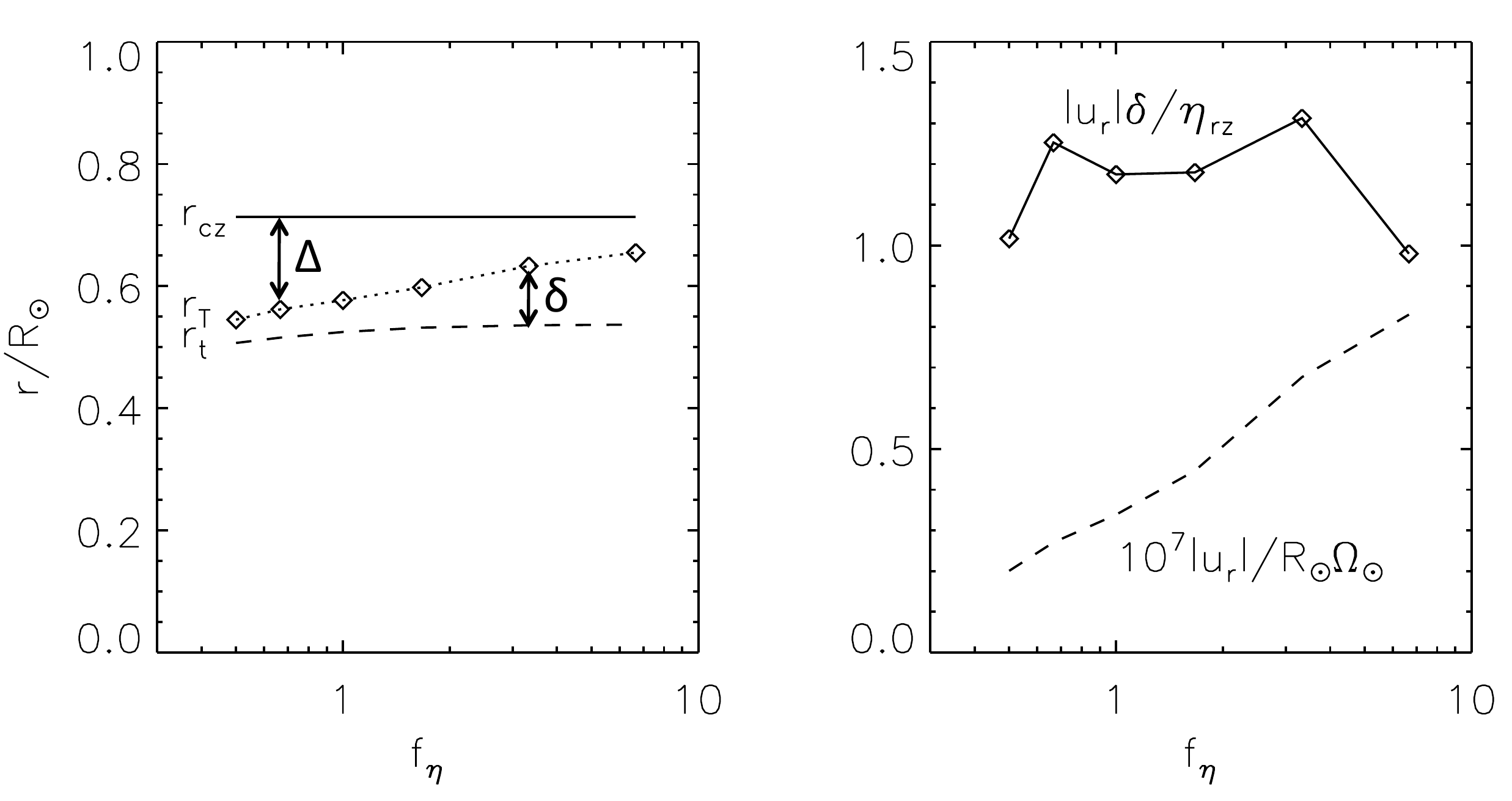}}
\caption{Left: Variation of the position of the base of the tachocline (dotted line with symbols) and tachopause (dashed line) with magnetic diffusivity, as measured by $f_\eta$. Right: While the  downwelling flow velocity steadily increases with $f_\eta$ (dashed line), the magnetic Reynolds number of the tachopause, defined as ${\rm Rm}_T = |u_r(r_T)| \delta/\eta_{\rm rz}$, remains close to unity (solid line with symbols).}
   \label{Eeta_ratio}
   \end{figure*}

Figures \ref{sigma_bestsim2}a and \ref{sigma_bestsim2}b show our results for $f_{\eta} = 0.5 $ and $f_{\eta}= 3.33$ respectively. The low-diffusivity case clearly exhibits a much stronger degree of field confinement within the radiation zone than the high-diffusivity case (see lower left panels). From the streamline plots (top left panels), we also see that the tachocline is slightly thicker when $f_\eta$ is lower. Figure \ref{Eeta_ratio}a shows that the geometry of the tachocline and the tachopause, estimated using the method described in Section \ref{tachocline_tachopause}, changes significantly with $f_\eta$. As $f_\eta$ decreases, we find that the base of the tachocline moves deeper into the radiation zone, but the base of the tachopause remains roughly at the same place. This implies that the tachopause becomes thinner whereas the tachocline thickens, and the ratio $\delta/\Delta$ decreases. At low enough magnetic diffusivity (e.g. for $f_\eta = 1$ and lower), the numerical solution satisfies the condition $\delta \ll \Delta$ assumed by GM98. %Checked ok.

%%%%%%%%%%%%%%%%%% FIGURE XX
%Checked, ok. 
%%%%%%%%%%%%%%%%%% FIGURE XX

A defining property of the tachopause is balance between advection and diffusion of the magnetic field. GM98 argue that, in a steady state, 
\begin{equation}
\label{ur_mag1}
|u_r| \approx \frac{\eta_{\rm rz}}{\delta} \mbox{ , }
%Checked ok 
\end{equation}
which relates the downwelling meridional flow velocity to the thickness of the tachopause and the magnetic diffusivity. In other words, the magnetic Reynolds number within the tachopause, ${\rm Rm}_T = \delta |u_r(r_T)|/ \eta_{\rm rz} $, where $u_r(r_T)$ is the radial velocity at the base of the tachocline, should be of order unity. To study whether this scaling holds here, we compute ${\rm Rm}_T$ as a function of $f_{\eta}$. The results are plotted in Figure \ref{Eeta_ratio}b.  We see that, although $u_r$ itself changes by a factor of about three, ${\rm Rm}_T$ is essentially constant and close to unity. This, once again, is in good agreement with the GM98 model. Finally, we see that, as a natural consequence of the decrease in $u_r$ when $f_{\eta}$ decreases, the strength and extent of the polar sub-rotating region also decreases, as discussed in Section \ref{results}. 
%Checked ok.

%%%%%%%%%%%%%%%%%% FIGURE X
%Checked ok.
%%%%%%%%%%%%%%%%%% FIGURE X

\subsection{The influence of the Elsasser number} 

The results discussed so far prove that the assumptions explicitly made by GM98 are correct, and strongly suggests that their model is a good representation of a laminar solar tachocline. The scalings we have been able to test appear to hold, at least within factors of order unity. 
Within the scope of that model, GM98 deduce that the thickness of the tachocline should vary with the primordial magnetic field strength as:
\begin{equation}
\frac{\Delta}{r_{\rm cz}} \propto \left( \frac{|B_0|}{4\pi \bar \rho } \frac{r_{\rm cz}}{\sqrt{\kappa_{\rm rz} \eta_{\rm rz}}}  \right)^{-1/9} \left(\frac{\kappa_{\rm rz}}{\eta_{\rm rz}} \right)^{1/3} \left(\frac{\Omega_\odot}{\bar N} \right)^{7/9}  
\label{GMeq}
%Checked ok.
\end{equation}
where the constant of proportionality between the two sides of this expression also depends on $\alpha$, $\beta$ and $l$.
%Checked ok

Unfortunately, we have not been able to test this last relationship directly. The weak dependence on the internal field strength implies that one would need to vary $B_0$ (or equivalently, $\Lambda$) by {\it many} orders of magnitude to detect any significant change in $\Delta$. Here, with all other parameters fixed at their reference model value, we have only 
been able to increase $B_0$ by about one order of magnitude, and, within the range of $B_0$ tested, have not detected any significant variation in $\Delta$. Any attempt to increase $B_0$ further results in the failure of our relaxation algorithm to find steady-state solutions. We believe the source of the problem is similar to the one encountered by GG08; as $B_0$ increases,
the system probably becomes unstable to various MHD instabilities associated with unrealistically large field strengths near the artificial inner boundary of the domain, and a steady-state solution can no longer be found. Numerically speaking, the matrix inversion problem involved in the steady-state search becomes increasingly ill-conditioned, and the solver no longer converges.
By selecting other values of the diffusivities, it is possible to find solutions for larger field amplitudes. However, in no cases have we been able to span a range for $B_0$ large enough to test the validity of Equation (\ref{GMeq}), for the same reasons as the ones described above. Nevertheless, since Equation (\ref{GMeq}) is possibly the least robust result of the GM98 model anyway (see Section \ref{conclusion} for details), we do not view our failure to validate it directly in our simulations as particularly troublesome.
%Checked ok

Beyond interfering with the relaxation algorithm in our simulations, one should certainly question whether hydrodynamic or magnetohydrodynamic instabilities could actually prevent the GM98 model from applying to the Sun altogether. \citet{BrunZahn06} investigated this issue, with their 3D time-dependent model of the radiation zone. They found two emerging modes of instability. MHD instabilities with high azimuthal wavenumber $m$ seem to develop deep in the radiation zone, and similar features are reported by \citet{Strugarekal11}. As discussed by both studies, these are in fact  expected since the simulations are initiated with a field in a purely poloidal configuration, which is known to be unstable to such modes \citep{Wright73,MarkeyTayler73}. \citet{BrunZahn06} also report that instabilities with a characteristic $m=1$ azimuthal wavenumber \citep{Tayler73} are found in regions with large toroidal fields, in particular close to the pole. 
%Checked ok

Interestingly, however, both \citet{BrunZahn06} and \citet{Strugarekal11} find that the deeply-seated, high-$m$ instabilities do not disrupt significantly the structure of the internal poloidal field. The growing instabilities generate a non-axisymmetric toroidal field whose addition rapidly stabilizes the system \citep{BraithwaiteSpruit04}, and there is no indication that other MHD instabilities play any further significant role in the dynamics. From this we conclude that even though they could be the reason behind the failure of our steady-state solver to find solutions for larger field strengths and lower diffusivities, MHD instabilities do not necessarily imply a failure of the GM98 model itself. The potential effect of turbulence and other instabilities on the GM98 model is discussed in more detail in Section \ref{conclusion}. 
%Checked ok

\subsection{The effect of the viscosity and the failure of previous models.}
\label{enu_section}

As described in Section \ref{intro}, attempts at finding numerical representations of the GM98 model using large-scale 2D or 3D time-dependent simulations have been made by \citet{BrunZahn06}, \citet{Strugarekal11} and \citet{Rogers11}. In all three cases, the attempts failed to exhibit any evidence for confinement of the magnetic field below the base of the convection zone. \citet{Strugarekal11} concluded that the 2D laminar nature of the GM98 model is too simplistic, and cannot capture the dynamics of the solar tachocline adequately. 
%Checked ok

We believe this conclusion is premature. While the GM98 model neglects any effects of turbulence and instabilities (see Section \ref{conclusion} for detail), its 2D laminar nature is {\it not} the reason for the failure of previous models. A strong clue to this effect comes from the analysis performed by \citet{Strugarekal11} and \citet{Rogers11} of their own simulations, which demonstrates that momentum transport therein is essentially viscously dominated, whereas viscosity is assumed to be negligible in the GM98 model. 
These numerical simulations clearly operate in a very different physical regime from that envisaged by GM98.
%Checked ok

We argued earlier that the source of the problem lies in the parameter $\sigma$, which is $\gg 1$ in the work of \citet{Strugarekal11} and \citet{Rogers11} but $<1$ in the solar tachocline. To demonstrate this effect more directly, we now present an axisymmetric steady-state solution using very similar parameters to those of \citet{Strugarekal11}, and compare our results with theirs.
%Checked ok

\begin{figure*}%[h]
                                % includes the two top panels 
   \centerline{\hspace*{0.015\textwidth}
               \includegraphics[width=7cm]{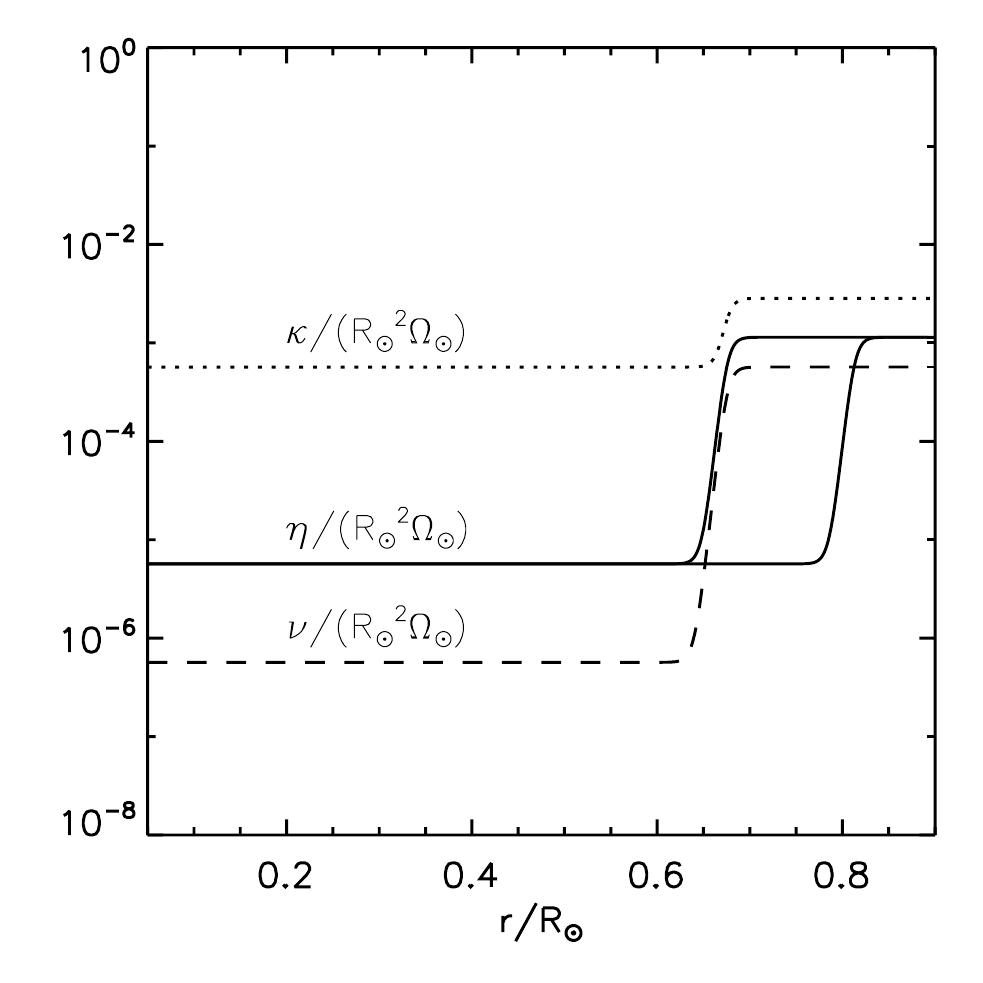}
               \hspace*{0.1\textwidth}
               \includegraphics[width=7cm]{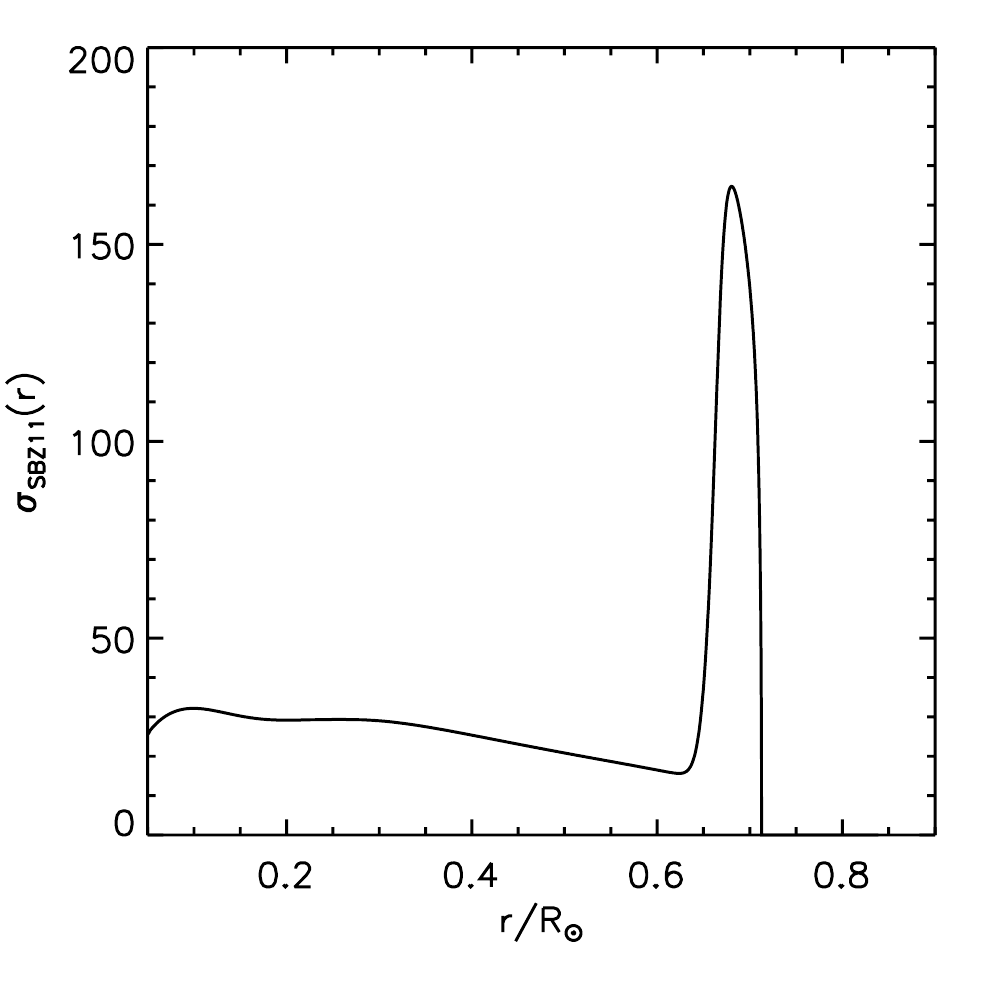}
              }
     \vspace{-0.35\textwidth}   % Shift close to the panel top 
     \centerline{\Large \bf     % Includes the labels (here needs the color
                                %   package, see beginning of this file)
      \hspace{0.0 \textwidth}  % \color{blue}{(a)}
      \hspace{0.415\textwidth} %  \color{blue}{(b)}
         \hfill}
     \vspace{0.31\textwidth}    % Shift back to the panel bottom 

\caption{Left: Diffusivity profiles actually used by \citet{Strugarekal11}, as well as one where $\eta(r)$ is shifted upward to allow pre-confinement (see main text for detail). Right: The $\sigma$ profile used by \citet{Strugarekal11}. Note how it is $\gg 1$ everywhere in the radiation zone. }
   \label{diff_sigma_SBZ11}
   \end{figure*}

Following \citet{Strugarekal11}, we use solar profiles for all background thermodynamical quantities, including the buoyancy frequency $\bar N(r)$. We also use their diffusivity profiles:
\begin{eqnarray}
\nu(r) &=& 8.0 \times 10^{9} + 8.0 \times 10^{12} \left[ 1.0 + \tanh \left( \frac{r - 0.6753 R_{\odot}}{0.01 R_{\odot}} \right) \right] \mbox{ , } \nonumber \\
\eta(r)  &=& 8.0 \times 10^{10} + 1.6 \times 10^{13} \left[ 1.0 + \tanh \left( \frac{r - 0.6753 R_{\odot}}{0.01 R_{\odot}} \right) \right] \mbox{ , } \nonumber \\
\kappa(r) &=& 8.0 \times 10^{12} + 3.2 \times 10^{13} \left[ 1.0 + \tanh \left( \frac{r - 0.6753 R_{\odot}}{0.01 R_{\odot}} \right) \right] \mbox{ , }
\end{eqnarray}
which are given in cgs units here, and shown in non-dimensional form in Figure \ref{diff_sigma_SBZ11}a. Note that as a result of this choice, the corresponding profile for $\sigma_{\rm SBZ11}(r)$, shown in Figure \ref{diff_sigma_SBZ11}b, is quite different from $\sigma_{\odot}(r)$. In particular, $\sigma_{\rm SBZ11} \gg 1$ in the entire radiative region, while $\sigma_{\odot}$ should be smaller than one in the tachocline. 
%Checked text not eqs.

%%%%%%%%%%%%%%%%%% FIGURE XX
%Checked ok. 
%%%%%%%%%%%%%%%%%% FIGURE XX 

To emulate the 3D, time-dependent simulations of \citet{Strugarekal11} with our steady-state solver, we use a similar method to the one presented in Section \ref{ourmodel}.
We impose the differential rotation in the convection zone using the same forcing term as in our reference model (see Section \ref{forcing1}). 
We must also assume the presence of an azimuthal current to maintain the primordial magnetic field, as discussed in Section \ref{forcing2}. We take its functional form to be the same as in our reference model, and choose\footnote{This large value is needed in order to have an Elsasser number that is comparable to that of  \citet{Strugarekal11} {\it in the region of the tachocline}. Indeed, $\Lambda$ is defined according to the magnetic field strength near the core, while the amplitude of the primordial field rapidly drops with radius so that its value in the tachocline region is much smaller than $B_0$ (see Appendix A). } $\Lambda = 6  \times 10^7$. Note that since the magnetic field continuously decays by Ohmic diffusion in a time-dependent calculation, it is difficult to know exactly what value of $B_0$ most appropriately describes the simulation of \citet{Strugarekal11} at any point in time. As such, the following comparison between their results and ours necessarily remains mostly qualitative. %Checked ok

\begin{figure*}    
                               % includes the two top panels 
   \centerline{\hspace*{0.015\textwidth}
               \includegraphics[width=8cm]{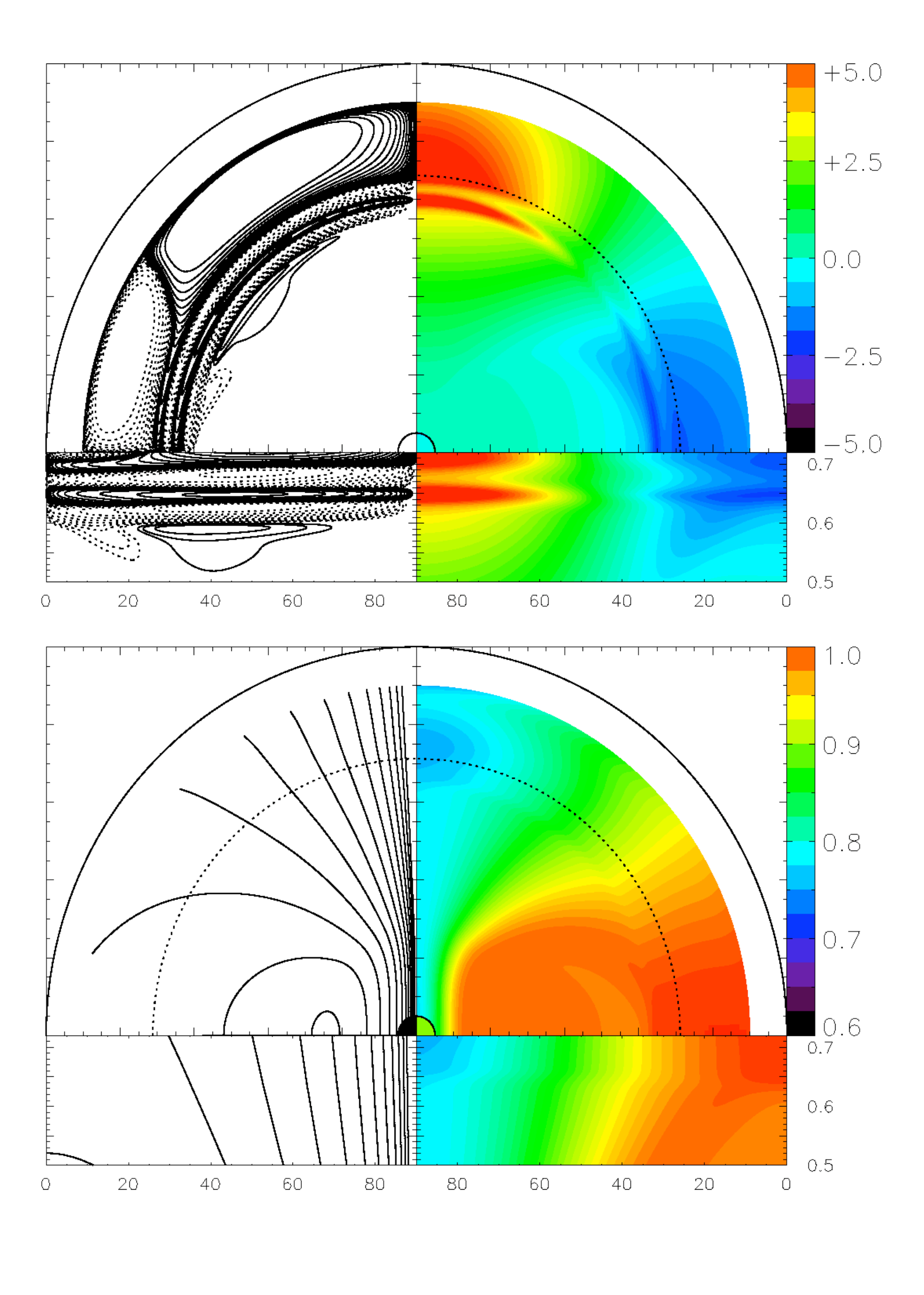}
               \hspace*{-0.03\textwidth}
               \includegraphics[width=8cm]{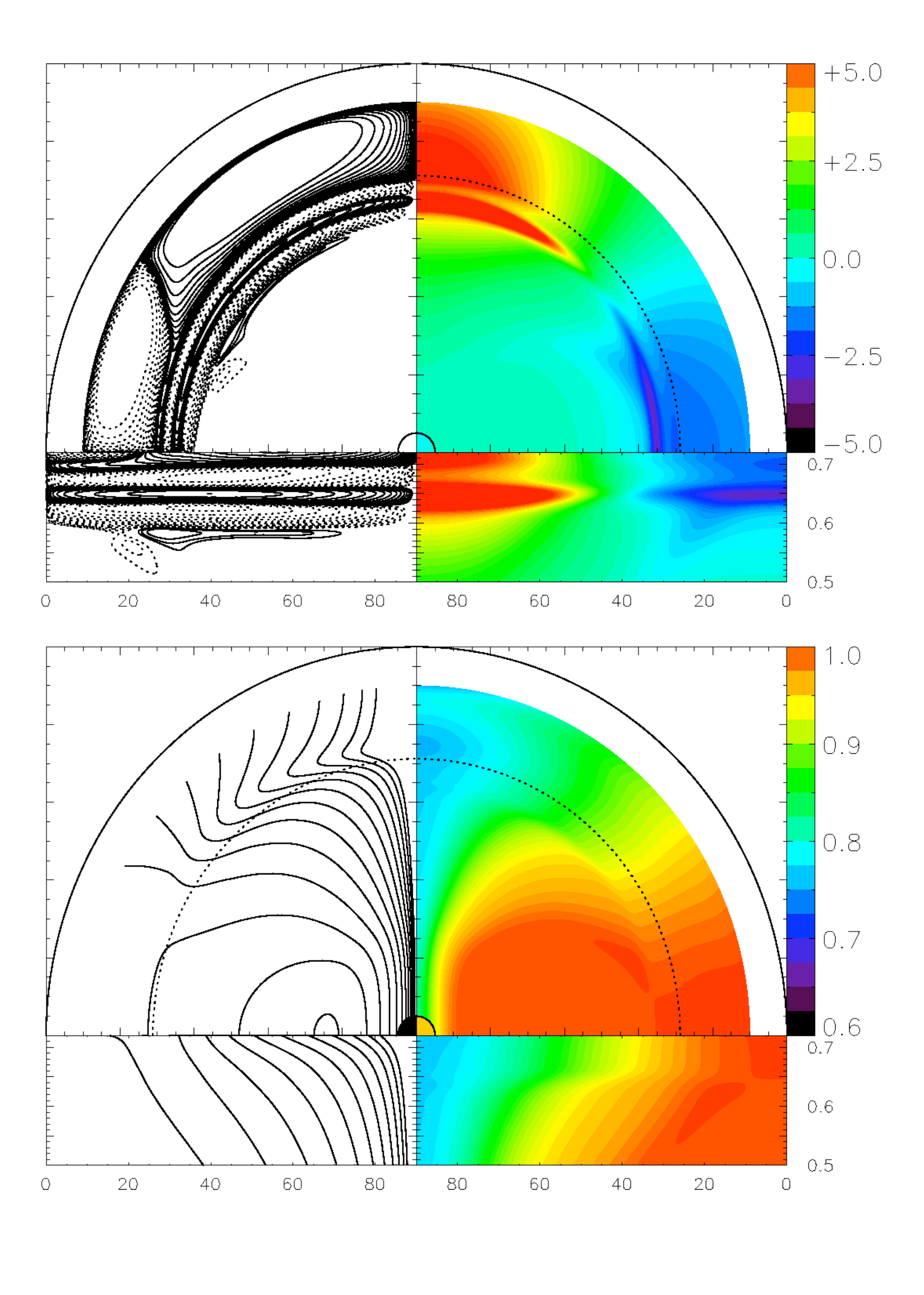}
              }
     \vspace{-0.35\textwidth}   % Shift close to the panel top 
     \centerline{\Large \bf     % Includes the labels (here needs the color 
                                %   package, see beginning of this file)
      \hspace{0.0 \textwidth}  % \color{blue}{(a)}
      \hspace{0.415\textwidth} % \color{blue}{(b)}
         \hfill}
     \vspace{0.31\textwidth}    % Shift back to the panel bottom 
     \caption{Steady-state solutions obtained in a parameter regime close to the one used by \citet{Strugarekal11}, without (left) and with (right) pre-confinement (see main text for detail).}
   \label{sigma_bestsim_SBZ11}
   \end{figure*}

The resulting steady-state solution for the system described above is shown in Figure \ref{sigma_bestsim_SBZ11}a. It is both strikingly different from the solutions presented in Section \ref{results}, and quite similar to the final stages of evolution of the simulation presented by \citet{Strugarekal11}. The downwelling flows entering the radiation zone from above decay rapidly within the interior and are clearly unable to confine the internal magnetic field. The latter diffuses into the convection zone, and, by Ferarro's isorotation law \citep{Ferraro37}, causes the radiation zone to rotate differentially. We also see that the meridional flow pattern just below the base of the convection zone is very different from that of our reference model (see Figure \ref{sigma_bestsim_ref}), but rather similar to that found by \citet{Strugarekal11} and by \citet{Rogers11}, with shallow alternating layers of opposite circulation. Hence, despite being qualitative, the comparison between our 2D steady-state model and the 3D time-dependent model of \citet{Strugarekal11}  appears to be meaningful. 
%Checked ok

%%%%%%%%%%%%%%%%%% FIGURE XX
%Checked ok. 
%%%%%%%%%%%%%%%%%% FIGURE XX

Of course, one may correctly argue that our reference model facilitates the field confinement process through {\it pre}-confinement (see Section \ref{issues}). To test whether pre-confinement can produce a confined magnetic field even for $\sigma \gg 1$, we present a second simulation, set up exactly as above, with the exception of the magnetic diffusivity profile. The latter is now shifted upward, as shown in Figure \ref{diff_sigma_SBZ11}a, thus allowing for pre-confinement to take place. The resulting 2D steady-state numerical solution is shown in \ref{sigma_bestsim_SBZ11}b. Even though the magnetic field is now indeed somewhat pre-confined in the convection zone, the field lines in the radiation zone remain mostly unaffected by the downwelling flows, and significant differential rotation persists. In other words, pre-confinement is not sufficient, on its own, to produce a confined magnetic field at these parameter values.
%Checked ok

\begin{figure}%[h]
                                % includes only one panel 
%              \centerline{\includegraphics[width=12cm]{fig17.eps}}
    \centerline{\includegraphics[width=8cm]{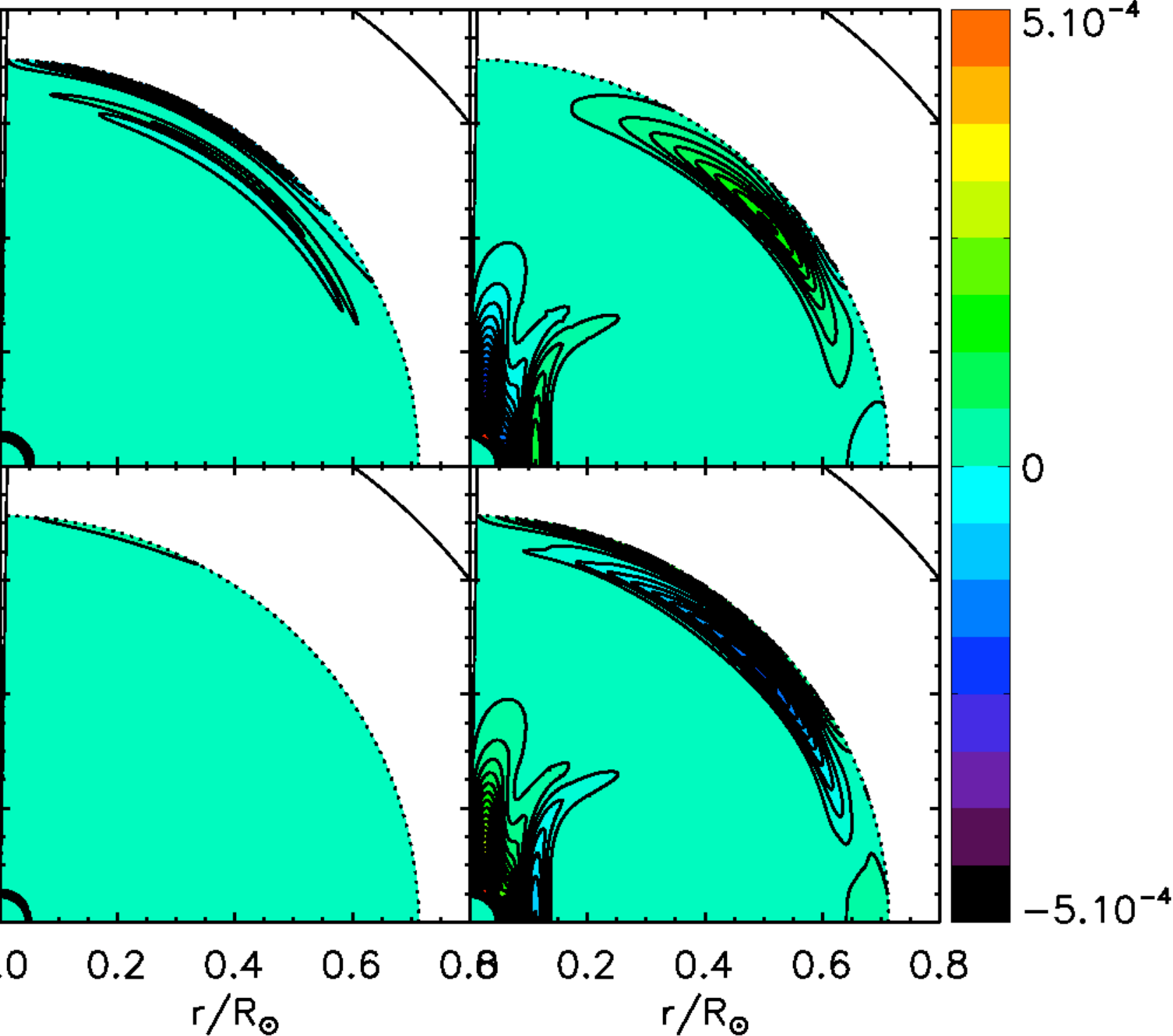}}
\caption{Comparison of the various terms that contribute to the azimuthal force balance in the radiation zone for the simulation presented in Figure \ref{sigma_bestsim_SBZ11}b, in units of $\rho_0 R_\odot \Omega_\odot^2$. Shown are the Coriolis force (top left), Lorentz force (top right), the inertial term (bottom left) and the viscous torque (bottom right). By contrast with our reference model, the dominant balance below the base of the convection zone is between the Lorentz force and the viscous torque.}
   \label{momentumforces_SBZ11}
   \end{figure}

We now analyze the results of this second simulation, which includes pre-confinement, more quantitatively. Figure \ref{momentumforces_SBZ11} shows the relative amplitude of the various forces operating in the azimuthal component of the momentum equation. It is clear that, by contrast with our reference model, the viscous stresses are dominant throughout the radiation zone. They balance the Coriolis force at high latitudes, and the Lorentz force at mid- to low latitudes. This is consistent with the results of  \citet{Strugarekal11}, who observe viscous stresses to be significant in their simulations, even though the Ekman number is $\ll 1$. %As a consequence of the importance of viscosity, meridional flow velocities decay rapidly below the radiative--convective interface,a nd are unable to confine the field.  
%Checked ok

%As suggested by GAA09, a low $\sigma$ value implies that flows crossing the interface must decay on a lengthscale $2R_{\odot}/\sigma$ XXX well this is no longer true it it's in Lorentz - viscous balance XXX. This is confirmed in Figure \ref{massflux_SBZ11}, which shows the variation of the downwelling mass flux $|\bar{\rho} u_r|$ below the interface at $80^\circ$ latitude. This figure is again strikingly different from the typical behavior of our reference model. The downwelling mass flux regularly changes sign owing to the presence of the alternating meridional cells, and its amplitude appears to vary roughly exponentially below $0.65R_\odot$, which is the point at which the diffusivity profiles selected by \citet{Strugarekal11} achieve their respective radiation zone values. To study this more quantitatively, we compare $|\bar{\rho} u_r|$ to the function $\exp(\sigma_t (r-r_{\rm cz})/2R_\odot)$, where $\sigma_t = 150$ is taken to be close to the peak value of $\sigma_{\rm SBZ11}(r)$. The fit is reasonably good, and shows that the meridional flow velocities decay exponentially with depth below the convection zone, as predicted by GAA09. Since $\sigma_{\rm SBZ11}  \gg 1$, the flows are so rapidly damped that they barely have any effect on the internal field below $r=r_{\rm cz}$. 
% To be checked... 

In summary, our axisymmetric laminar and steady-state models of the solar interior, yield solutions that are comparable to the long-term evolution of more realistic 3D time-dependent simulations, when run at the same parameters. Our results demonstrate that the failure of prior numerical attempts to model magnetic confinement and a self-consistent tachocline can be attributed to the fact that $\sigma \gg 1$ in all those simulations. In this ``high $\sigma$" regime, viscosity plays a significant role in the transport of angular momentum, the meridional flow pattern consists of alternating shallow cells with a flow velocity that decays exponentially with depth, and the magnetic field is unconfined. 
%Checked ok

%%%%%%%%%%%%%%%%%% FIGURE XX
%Checked ok
%%%%%%%%%%%%%%%%%% FIGURE XX

%%%%%%%%%%%%%%%%%% FIGURE XX
%  \begin{figure}    
                                % includes only one panel 
%              \centerline{\includegraphics[width=8cm]{fig18.eps}}
%    \centerline{\includegraphics[width=8cm]{fig18-eps-converted-to.pdf}}
%\caption{Downwelling meridional mass flux $|\bar{\rho} u_r|$ at 80$^\circ$ latitude for  the simulation presented in Figure \ref{sigma_bestsim_SBZ11}b. This model fails to develop a tachocline because the large-scale meridional flow decays rapidly to zero at the radiative--convective interface. The dash-dotted line is the exponential function $e^{-\sigma_t (x_{\rm cz}-x)/2}$, where $\sigma_t = 150$, which illustrates the rapid exponential decay of downwelling flows when $\sigma \gg 1$. XXX Note: what do we do with this? XXX}
%   \label{massflux_SBZ11}
 %  \end{figure}
%%%%%%%%%%%%%%%%%% FIGURE XX

\section{Discussion and Prospects} %%%%%%%%%%%%%%%%%%%%%%%%%%%%%%%%%%%%%%%%
      \label{conclusion} 

\subsection{Summary}

The GM98 model provides a plausible explanation for the uniform rotation of the bulk of the solar radiation zone, and various other observations related to the tachocline, including its angular velocity profile, sound speed anomaly, and the rate of lithium depletion. It also provides scaling laws that relate the thickness of the tachocline and its mixing timescale to the strength of the assumed internal primordial field. However, the GM98 model makes a number of simplifying assumptions, and describes only the laminar, axisymmetric, and time-independent dynamics of the tachocline. As such, its predictions can only be fully verified by self-consistent, nonlinear numerical modeling. Unfortunately, recent attempts at simulating the solar interior have failed to reproduce the dynamical regime predicted by the GM98 model, which have led some to conclude that it is unworkable \citep{Strugarekal11} . 
%Checked ok

In this work, we also study the GM98 model numerically, still assuming an axisymmetric steady state, but solving the full set of governing nonlinear MHD equations in global spherical geometry. 
Although our model cannot self-consistently describe all the effects of instabilities and turbulence in the tachocline, it can certainly be used to validate (or invalidate) the GM98 model under these more general conditions. The limitations of our approach are discussed in depth below. Within this framework, however, we not only present the first numerical simulations of the solar interior to exhibit a tachocline and a tachopause with properties that are qualitatively and quantitatively consistent with most aspects of the GM98 model, but also provide a simple explanation for the failure of previous models to do so.  
%Checked ok

For low-enough magnetic diffusivity (see Figures \ref{sigma_bestsim_ref} and  \ref{sigma_bestsim2}a for instance), we obtain numerical solutions of the governing equations that have a radiation zone held close to uniform rotation by a confined primordial magnetic field. As in the GM98 model, we find that the field is confined by meridional flows downwelling from the convection zone, within a thin tachopause in magnetic advection-diffusion balance (i.e. with Rm $\sim 1$, see Figure \ref{Eeta_ratio}). The tachopause is also in magnetostrophic balance (see Figure \ref{momentumforces}), and the Lorentz forces generated as the field is wound up by the rotational shear provide just enough angular momentum to deflect to the downwelling flows equatorward on their path back to the convection zone. As assumed by GM98, the bulk of the tachocline is ``magnetic free" (in the sense that the magnetic forces are negligible) and in thermal-wind balance (see Figure \ref{vorticityterms}). The strength of the downwelling flow is determined by this balance together with thermal equilibrium (see Figure \ref{Ek_ratio}). 
%Checked ok

By contrast, when we use the same governing parameters as \citet{Strugarekal11}, we do not obtain a confined magnetic field, or a tachocline. As discussed in Section \ref{intro}, the problem is linked to the parameter $\sigma = \sqrt{\rm Pr} \bar N/\Omega_\odot$, which is $\gg 1$ in the models of \citet{Strugarekal11} (as well as \citet{Rogers11} and \citet{BrunZahn06}), but is $<1$ in the solar tachocline. When $\sigma \gg 1$, meridional flows are strongly suppressed beneath the convection zone, and therefore unable to confine the field, and the angular-momentum balance is dominated by viscosity. When $\sigma < 1$, however, meridional flows are able to extend (or "burrow") much more deeply into the radiation zone, and with an amplitude sufficient to confine the magnetic field.
%Checked ok

Within the scope of our axisymmetric, steady-state model, our results confirm the predictions of  GM98, and guide parameter selection for future 3D numerical simulations.  We now discuss the caveats and implications of this model and of our findings in more detail, and lay out future theoretical and observational prospects for solar and stellar astrophysics. 
%Checked ok

\subsection{Model caveats and their implications}

The axisymmetric, steady-state approach to studying the solar tachocline proposed by GM98 and adopted throughout this paper has clear numerical advantages, but neglects by construction any time-dependent, non-axisymmetric dynamics. In what follows, we loosely refer to these neglected dynamics as ``turbulence", although they may also include any effects of Alfv\'en waves or internal waves. We now consider whether turbulence in and around the tachocline can qualitatively affect %the confinement of the magnetic field and, if not, whether it can otherwise affect 
the scaling laws derived by GM98.
%Checked ok

\subsubsection{The validity of thermal-wind balance and thermal equilibrium}

The main tachocline scaling law given in Equation (\ref{tacho_ur4}) results from the assumptions of thermal-wind balance and thermal equilibrium. As we now show, this scaling probably continues to hold regardless of the presence or absence of turbulence, given reasonable upper bounds for the turbulent intensity. 
%Checked ok

Indeed, one can estimate under which conditions a turbulent flow with energy-bearing eddies of vertical scale $l_r$ and horizontal scale $l_h$ can be present in the tachocline without upsetting either balance. First note that in strongly stratified flows $l_r$ is typically much smaller than $l_h$. If $u_{\rm rms}$ is the mean eddy velocity, then their typical horizontal velocity will be of the order of  $u_{\rm rms}$ whereas the vertical velocity $w_{\rm rms}$ is of the order of $w_{\rm rms} \sim (l_r / l_h) u_{\rm rms}$. %For such flows, the thermal diffusion timescale $l_r^2/\kappa$ is much shorter than the characteristic turnover time $l_r/w_{\rm rms}$, so the turbulent entropy perturbations $s_{\rm rms}$ can be estimated from local thermal equilibrium using $w_{\rm rms}  (d\bar s/dr) \simeq \kappa s_{\rm rms} /l_r^2$.
%Checked text, not eqs
Second, note that the momentum (or vorticity) equation involves the spatial derivative of the Reynolds stress tensor. This derivative involves vertical scales of the order of the thickness of the tachocline $\Delta$, and horizontal scales of the order of the solar radius $R_\odot$. Using these estimates, it can then be shown that for thermal-wind balance to hold, one must have 
\begin{equation}
\left|  \nabla \times (\nabla \cdot( {\bf u}{\bf u} ) ) \right|_\phi  \sim \frac{u^2_{\rm rms}}{\Delta^2} \max\left( \frac{l_r}{l_h}, \frac{\Delta}{R_\odot} \right) \ll 2 \Omega_\odot R_\odot \frac{\partial \tilde{\Omega}}{\partial z} \sim 10^{-11} s^{-2} \mbox{   . }
\label{limit1}
\end{equation}
%Checked text, not eqs

Finally, note that the development of any instability in the strongly stratified tachocline is conditional on the fact that their vertical scale be small enough {\it not} to be constrained by the buoyancy restoring force. In other words, vertical turbulent fluid motions are by nature thermally diffusive, and do not contribute much to the vertical heat flux\footnote{Helioseismic observations also confirm this statement.}. The thermal energy Equation (\ref{eq:Teq}) is thus expected to hold in the presence of self-consistently generated turbulence. 
%for thermal equilibrium to hold in the sense described by Equation (\ref{eq:Teq}), the divergence of the turbulent heat flux must be negligible compared with the flux advected by the large-scale flows, i.e. $w_{\rm rms} s_{\rm rms} / \Delta \ll u_r (d\bar s/dr) $ where $u_r$ is the downwelling velocity of the mean meridional flows, and $ s_{\rm rms}$ is derived above. This yields the second constraint:
%\begin{equation}
%\frac{w_{\rm rms}^2 l_r^2}{\kappa \Delta} \ll u_r \Rightarrow  u_{\rm rms}^2  \ll \frac{u_r \kappa }{\Delta} \frac{ l_h^2\Delta^2}{l_r^4} \mbox{   . }
%\label{limit2}
 %\end{equation}
%Checked text, not eqs
Using $\Delta \simeq 0.02 R_\odot$ in (\ref{limit1}), we then find that Equation (\ref{tacho_ur4}) holds as long as 
%Equations (\ref{limit1}) and (\ref{limit2}) provide two independent upper limits on the amplitude of turbulent motions above which the GM98 tachocline scalings may be invalidated. Using $\kappa_{\rm rz} = 1.4 \times 10^7$cm$^2$/s in the tachocline, $\Delta \simeq 0.02 R_\odot$ and $u_r \simeq 10^{-5} $cm/s (GM98), the constraints become
%\begin{eqnarray}
%&& u^2_{\rm rms} \ll  \frac{2 \times 10^7}{ \max\left( \frac{l_r}{l_h}, \frac{\Delta}{R_\odot} \right) } {\rm cm}^2/s^2 \nonumber \\
%&& u_{\rm rms}^2  \ll 10^{-7} \frac{ l_h^2\Delta^2}{l_r^4} {\rm cm}^2/s^2
%\end{eqnarray}
\begin{equation}
u^2_{\rm rms} \ll  \frac{2 \times 10^7}{ \max\left( \frac{l_r}{l_h}, \frac{\Delta}{R_\odot} \right) } {\rm cm}^2/s^2 \mbox{   .}
\label{limit3}
\end{equation}

%The latter is the strongest of the two constraints. Given that the tachocline is strongly stratified (for fast-timescale dynamics) and that the turbulence is rotationally constrained, the horizontal scale of the eddies is likely to be large, possibly as large as the scale of the overlying convective eddies. Hence, $l_h$ is probably of the order of $0.1 R_\odot$. This then implies $u_{\rm rms} \ll 6 \times 10^{-7} (R_\odot/l_r)^2$cm/s. Whether the constraint holds thus depends sensitively on $l_r$. 
 %Checked text, not eqs

Meanwhile, a strict upper limit on the turbulence intensity $u_{\rm rms}$ comes from assuming that it cannot be larger than a fraction of the convective velocities in the overshoot layer, so that $u_{\rm rms} \lesssim N_{\rm cz} (0.1 H_p) \sim 7 \times 10^{2} $cm/s, where $N_{\rm cz} \sim 10^{-6}$s$^{-1}$ is the imaginary part of the buoyancy frequency just above the base of the convection zone, and where we have assumed that the overshoot layer height is no larger than $0.1H_p \sim 0.01 R_\odot$, where $H_p$ is the pressure scaleheight at the same location. In other words, we expect $u_{\rm rms}$ to be at most a few meters per second, and to decrease rapidly with depth below the base of the overshoot layer. 
%We conclude that for thermal equilibrium to hold as in Equation (\ref{eq:Teq}), the vertical size of the turbulent eddies should be no larger than about $10^{-2} \Delta$, which seems reasonable given the strong stratification of the tachocline. 
 %Checked text, not eqs
This implies that (\ref{limit3}) is readily satisfied, and that Equation (\ref{tacho_ur4}) can be used with confidence to model the dynamics of the tachocline. 

\subsubsection{The effect of turbulence on the angular momentum balance in the tachocline and tachopause.} 

By contrast with thermal-wind balance, angular-momentum balance is much more easily disrupted  by turbulent motions since the only source of angular-momentum transport considered by GM98 and in this paper is advection by slow, large-scale meridional flows. 
In many tachocline models, such as the one proposed by \citet{SpiegelZahn92}, turbulence is assumed to dominate the transport of angular momentum throughout the tachocline.  If the turbulent transport is very efficient, and also frictional (i.e. down-gradient in angular velocity), then it could by itself maintain a thin tachocline even in the absence of a primordial magnetic field\footnote{We note, however, that turbulent transport in the tachocline cannot explain the uniform rotation of the rest of the radiation zone while the Sun is undergoing spin-down. This by itself provides strong evidence for the existence of a primordial magnetic field \citep[][GM98]{MestelWeiss87}.}.
In that case the scaling law (\ref{tacho_ur4}) would still hold, but any scalings constructed from angular-momentum balance, in particular Equation (\ref{GMeq}), would be invalid.
%Checked ok

However, for reasons discussed by GM98 and \citet{McIntyre07} \citep[see also][]{Tobiasal07}, turbulence in the tachocline is unlikely to be frictional. As a result, even if it contributes significantly to the transport of angular momentum in the tachocline and in the tachopause, the thickness of the tachocline must ultimately still be constrained by the strength of the magnetic field.  In that case, Equation (\ref{GMeq}) should be replaced by a law that incorporates the contribution from turbulence to the angular-momentum balance, but the qualitative tachocline picture would be unchanged.

\subsection{Theoretical and observational prospects}

\subsubsection{Computational prospects}

%In this work, we clearly identified the reason for the failure of previous numerical models, such as the ones by \citet{Strugarekal11} and \citet{Rogers11}, to find solutions in which the primordial magnetic field is confined by the meridional flows. These simulations are run at values of the stratification parameter $\sigma = \sqrt{\rm Pr} N/\Omega$ which are much larger than one, while $\sigma$ should, as in the real solar tachocline, be markedly below one. When $\sigma$ is large, meridional flows are damped exponentially with depth below the base of the convection zone (which explains the lack of confinement), and viscous stresses play an unphysically large role in the global angular momentum balance. In other words, simulations with large $\sigma$ have little to do with solar interior dynamics.  Of course, it has always been the goal of the authors of such simulations to decrease Pr progressively towards solar parameter values, as the availability of computing power increases. Figure \ref{theygotitwrong}, however, clearly shows that this approach will not yield any interesting result until Pr can be decreased to $10^{-5}$ or less -- a value which is so low that there is little prospect of achieving it within the next 20 years or so. 

As discussed above, a complete validation of the GM98 model can only come from 3D simulations.  Clearly, 3D numerical models are still far from being able to achieve solar parameter values. The question thus arises of whether it is possible to model solar-like dynamics with non-solar parameters. In this work, we have shown that it is possible, as long as model parameters are chosen such that $\sigma = \sqrt{\rm Pr} \bar N/\Omega < 1$ everywhere in the tachocline. 
%Checked ok.

  \begin{figure}    
                                % includes only one panel 
    \centerline{\includegraphics[width=8cm]{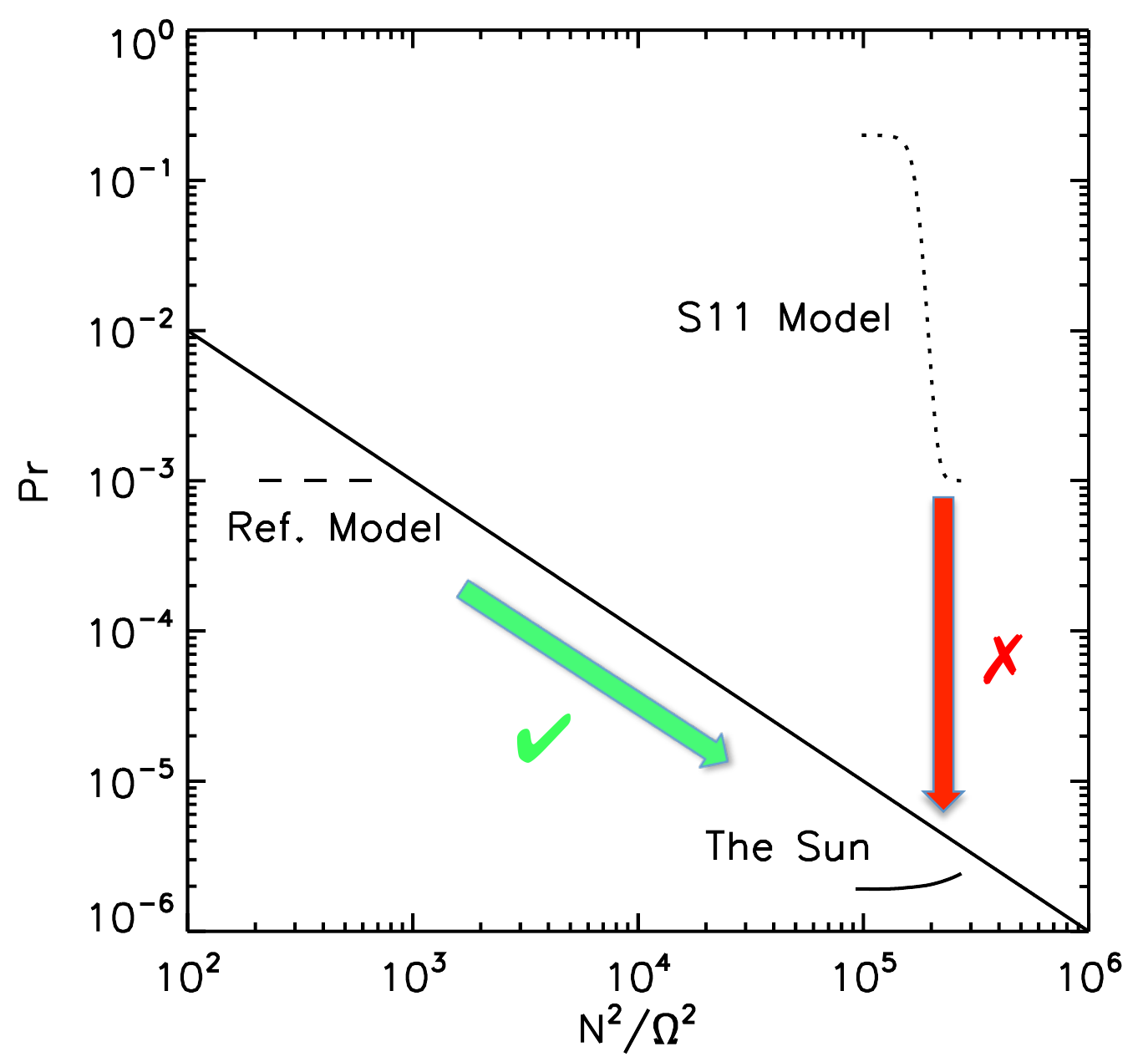}}
\caption{A 2D slice of parameter space showing the variation of the Prandtl number and $N^2/\Omega^2$ in the region of the tachocline (i.e. for $r \in [0.6,0.7]R_\odot$), in our reference model simulation, in the model of \citet{Strugarekal11} and in the Sun. The oblique line represents the limit Pr$N^2/\Omega^2 = 1$. Lines of constant $\sigma$ are parallel to this line. Simulations above this line are viscously dominated, while simulations below this line are not, and thus reside in the same region of parameter space as the Sun. }
   \label{theygotitwrong}
   \end{figure}

%Figure \ref{theygotitwrong} compares the range of values of ${\rm Pr} $ and $\bar N/\Omega$ in our own reference model simulation, in the \citet{Strugarekal11} simulation and in the Sun. 
%The simulations of \citet{Strugarekal11} are clearly far into the viscously dominated region of parameter space. Moving them into the correct regime while maintaining a solar value of $\bar N$  (see red arrow) would require a decrease in Pr of three orders of magnitude at least, and none of the intermediate simulations (i.e. simulations with $10^{-6}<{\rm Pr}<10^{-3}$) are expected to yield meaningful results. By contrast, Figure \ref{theygotitwrong} also reveals a much better approach to the problem. 
Figure \ref{theygotitwrong} illustrates our proposed route to ensure that this condition is always satisfied. While our reference model is not yet at solar parameters, it is at least in the correct ``solar" region of parameter space. As computational power continues to increase, it will become possible to decrease the Prandtl number further. If one gradually increases $\bar N$ back towards its original solar value at the same time, thus ensuring that $\sigma$ remains close to its solar value, then every simulation along the path described by the green arrow remains in the correct region of parameter space, and numerically-motivated asymptotic scalings (to confirm or replace Equation (\ref{GMeq}) for instance) could in principle be derived. We shall present the results of 3D simulations of the full Sun, following the path described above, in the future. 
%Checked ok.

%Of course, now that a tachocline solution has been found, one may be interested in determining whether its dynamics can influence those of the convection zone in return, as discussed for instance by Miesch (XXX can't remember the year) and \citet{Rempel05}. Indeed, the presence of a significant latitudinal entropy gradient in the tachocline is known to be necessary to break away from cylindrical rotation in the convection zone, as required by observations. We find that an entropy gradient naturally arises from thermal equilibrium in our model, to compensate for the advection of the background stratification by the downwelling flows. In order to study this effect, one needs to model more carefully the coupled transport of heat and angular momentum in the convection zone. This can be done for instance using a Reynolds-stress average model \citep{Rempel05,Garaudal10}, in lieu of our Darcy friction law, and could be the subject of a subsequent publication.

\subsubsection{Observational prospects}

Meanwhile, the partial validation of the GM98 model presented here opens interesting prospects for applications of this theory to other solar-type stars. In particular, it can be used to deduce a number of conditions under which a tachocline could be expected to exist, and scaling laws for mixing of chemical species and angular-momentum transport within. 
%Checked ok.

First and foremost, the star must host a primordial magnetic field of sufficient amplitude to impose uniform rotation. For this to be the case, the field must be strong enough everywhere in the radiation zone to satisfy $\Lambda(r) > 1$, that is,
\begin{equation}
B^2(r,\theta) > 4 \pi \eta(r) \rho(r) \Omega_\star
\label{EqBsuff}
\end{equation}
where $\Omega_\star$ is the star's mean rotation rate, and $ \eta(r)$ and $\rho(r)$ are its magnetic diffusivity and density profiles. The magnetic field plays no significant role in the long-term dynamics of stars that do not satisfy this constraint, as found by \citet{Rogers11} (see Section \ref{intro}). Whether (\ref{EqBsuff}) holds or not at any point throughout stellar evolution depends primarily on two competing processes. On the one hand, magnetic braking spins the star down, and weakens the constraint over time. On the other hand, the internal field also decays with time through Ohmic dissipation. 
%Checked ok.

While the rotation rate of a star can be measured, its internal field strength cannot, and one must rely on rough arguments to estimate its amplitude. Since the Ohmic dissipation timescale for a confined dipolar field is of the order\footnote{It can be shown that for a dipolar field confined in a sphere of radius $r_{\rm cz}$ which has a uniform magnetic diffusivity $\eta$, then $ t_{\rm Ohm} = r_{\rm cz}^2/ \xi^2  \eta$ where $\xi \simeq 4.49$ is the first zero of the $J_{3/2}$ Bessel function. For variable diffusivity, a good approximation for $\bar \eta$ is  $\bar \eta = \left(\int_0^{r_{\rm cz}} \eta^{1/2}(r) dr  \right)^{2} $ \citep[Gough, personal communication, see also][]{Garaud99}.} of $t_{\rm Ohm} = r_{\rm cz}^2/ \xi^2 \bar \eta$, where $\bar \eta$ is a mean value of the magnetic diffusivity across the radiation zone, and $\xi$ some geometrical factor, a necessary condition for the field to be non-negligible is $t_{\rm Ohm} > t_\star$ where $t_\star$ is the age of the star. %Within a given cluster (i.e. for stars of the same age), this much more easily satisfied for higher-mass stars, which have a larger radiative core, than lower-mass stars. For field stars, the condition is also more likely to have been broken in lower-mass stars than in higher-mass ones, since the latter are statistically significantly younger. 
%Checked ok.

Next, one must study whether the condition $\sigma < 1$ holds or not. If it does, then the field is likely to be ``deeply" confined by the large-scale meridional flows, as discussed by GM98. If it does not, the field may still be confined, but this time only by turbulent magnetic pumping. The implications for the depth of the mixed layer and its ventilation timescale in both cases are quite different. If $\sigma > 1$, the mixed layer is likely as deep as the convective overshoot layer -- but no deeper -- and its mixing timescale is very short. This implies that angular-momentum transport between the convection zone and the radiative interior is very efficient (with implications for the stellar spin-down problem), and that chemical species are efficiently mixed down to the base of the overshoot layer, but probably no deeper. 
%Checked ok.

If $\sigma < 1$ on the other hand, the star is likely to have a tachocline that has a similar structure to the one described by GM98 and in this paper, albeit probably with a different tachopause (see the previous Section). Since the (unknown) details of the tachopause structure relate the thickness of the tachocline to the (unknown) internal field strength, it is difficult to place any constraints purely from theory on the plausible depth of a stellar tachocline. However, if the latter can be measured via asteroseismology, then the velocity of the large-scale meridional flows mixing it can be estimated using Equation (\ref{tacho_ur4}). This estimate should then be used, together with the measured mixed-layer depth, to infer the expected surface abundance of various elements. Comparing these predictions with observations can then serve as an independent validation (or invalidation) of the GM98 model. Finally, the associated estimate for the angular-momentum transport timescale across the tachocline can be used to study stellar spin-down (see Oglethorpe \& Garaud, in prep.), and statistical comparison with observations of the angular-velocity distributions of various stellar clusters can serve as yet another independent test of the theory, as in the work of \citet{MacGregorBrenner91} and \citet{Denissenkov10} for instance.
%Checked ok.

To conclude, our results lay the foundations for future quantitative comparisons between models and observations, of the differential rotation profile of course, but also of the light-element abundances and of the multi-dimensional sound-speed structure of the solar interior. It will also provide better dynamical constraints, and perhaps new paradigms, for solar dynamo theory. Finally, it lays a clear path forward in the study of the internal dynamics of other solar-type stars, with natural implications for asteroseismic data from CoRoT and KEPLER. 
%Checked ok.

\section*{Acknowledgements}

L. Acevedo-Arreguin, P. Garaud and T. Wood gratefully acknowledge funding from the NSF (CAREER-0847477). Financial support for T. Wood was also provided by the Center for Momentum Transport and Flow Organization sponsored by the US Department of Energy - Office of Fusion Energy Sciences and the American Recovery and Reinvestment Act 2009. We thank Nic Brummell, Douglas Gough, C\'eline Guervilly, and Michael McIntyre for very fruitful discussions. 

%%%%%%%%%%%%%%%%%%%%%%%%%%%%%%%%%%%%%%%%%%%%%%%%%%%%%%%%%%%%%%%%%%%%%%%%%%%

\appendix   

\section{Derivation of the background magnetic field properties}
\label{appA}

The expression for the primordial magnetic field ${\bf B}_0$ generated by the imposed azimuthal current ${\bf j}_0$ is most easily derived by 
introducing the magnetic potential $A_0(r,\theta)$ \citep[see][]{GaraudGuervilly09} defined as 
\begin{equation}
{\bf B}_0 = \nabla \times \left( \frac{A_0}{r \sin \theta}\hat{{\bf e}}_{\phi}\right) \mbox{  .} \label{eq:A0def}
%Checked ok. 
\end{equation}
We then have, using (\ref{eq:A0def}) in conjunction with (\ref{eq:J0def}),  
\begin{equation}
 \nabla^2 \left( \frac{A_0}{4 \pi r \sin \theta} \right) - \frac{A_0}{4 \pi r^3 \sin^3 \theta}  = - 4 \pi J_0 \frac{(r-r_a)(r-r_b)}{R_\odot^2}  H(r-r_a)H(r_b-r)\mbox{ , }
%Checked ok. 
\end{equation}
Seeking dipolar solutions of the form
\begin{equation}
A_0(r,\theta) = \hat A(r) \sin^2 \theta 
%Checked ok. 
\end{equation}
implies
\begin{equation}
\frac{d^2 \hat A}{dr^2} - 2 \frac{\hat A}{r^2} = - 4\pi J_0 r \frac{(r-r_a)(r-r_b)}{R_\odot^2} H(r-r_a)H(r_b-r) \mbox{  ,} 
%Checked ok. 
\end{equation}
which can be solved to yield
\begin{equation}
A_0(r,\theta) = \left\{ \begin{array}{lc}
\frac{B_0}{2}r^2 \sin^2 \theta & \mbox{if } 0 \leq r \leq r_a \mbox{  ,} \\
- 4 \pi J_0 \frac{r^2}{R_\odot^2}  \left[\frac{1}{18}r^3-\frac{1}{10}(r_a+r_b)r^2 +\frac{1}{4}r_a r_b r+ \frac{c_1}{3} + \frac{c_2}{r^3}\right] \sin^2 \theta & \mbox{if } r_a \leq r \leq r_b \mbox{  ,} \\
\frac{c_3}{r}\sin^2 \theta  & \mbox{if } r_b \leq r \mbox{  .}  \end{array} \right. 
%Checked ok. 
\end{equation}
Note that some of the integration constants were set to zero to guarantee that the field in the core and at infinity be regular. This then implies that ${\bf B}_0 = B_0 \hat{{\bf e}}_z$ for $r < r_a$, where $B_0$ is related to $J_0$ (see below). 

The constants $J_0$, $c_1$, $c_2$, and $c_3$ are obtained by requiring continuity of $A_0$ and $\partial A_0/\partial r$ at $r = r_a$ and $r = r_b$, which results in:
\begin{eqnarray}
c_1 &=& \frac{r_b^2}{6}\left( r_b - 3r_a \right) \mbox{  ,}  \\ %- \left [ \frac{1}{3} r_b^3 - \frac{1}{2}(r_a + r_b) r_b^2 + r_a r_b^2 \right ]\mbox{  ,}  \\
c_2 &=& \frac{r_a^5}{30} \left( \frac{r_b}{2} - \frac{r_a}{3} \right) \mbox{  ,} \\ %\frac{1}{3} \left(\frac{1}{6}r_a^6-\frac{1}{5}(r_a+r_b)r_a^5 + \frac{1}{4} r_a r_b r_a^4\right)  \mbox{  ,}  \\
c_3 &=& \frac{4 \pi J_0}{R_\odot^2}  \left( \frac{r_a^6 - r_b^6}{90} + \frac{r_ar_b(r_b^4-r_a^4)}{60} \right)  %\frac{4 \pi J_0}{3R_\odot^2} \left[\frac{1}{6} (r_b^6-r_a^6) -\frac{1}{5} (r_a +r_b) (r_b^5-r_a^5) + \frac{1}{4} r_a r_b (r_b^4-r_a^4) \right] \mbox{ , }
%Checked ok. 
\end{eqnarray}
and 
\begin{equation}
J_0 = \frac{-9B_0R_\odot^2}{ 4 \pi (r_b-r_a)^3} \mbox{ . }
%Checked ok. 
\end{equation}
%Having selected $r_a = 0.1R_\odot$ and $r_b = 0.3R_\odot$, we have $J_0 = -1125B_0/4 \pi R_\odot$, $c_1 = 0$, $c_2 = 3.8 \times 10^{-8}R_\odot^6$, and $c_3 = 8.25556 \times 10^{-6}R_\odot^4$. 
Note that this last expression can also be derived directly and more easily through the use of Biot-Savart's law. 

We can then derive the components of the background magnetic field ${\bf B}_0$ using (\ref{eq:A0def}). We find, for instance, that its radial component is
\begin{equation}
B_{0r}(r,\theta) = \left\{ \begin{array}{lc}
  B_0 \cos \theta & \mbox{if } 0 \leq r \leq r_a \mbox{  ,}  \\
- \frac{8 \pi J_0}{R_\odot^2}  \left[\frac{1}{18}r^3-\frac{1}{10}(r_a+r_b)r^2 +\frac{1}{4}r_a r_b r+ \frac{c_1}{3} + \frac{c_2}{r^3}\right] \cos \theta & \mbox{if } r_a \leq r \leq r_b \mbox{  ,}  \\
\frac{2c_3}{r^3}\cos \theta  & \mbox{if } r_b \leq r \mbox{  .}  \end{array} \right.
%Checked ok. 
\end{equation}
Note that the solution guarantees that the amplitude of the magnetic field at the center of the sphere to be finite, an advantage from a computational perspective over the expression for a point-dipole used by GG08.

% Note: tami's eta = 5 times 10^12
% B0 = 4000G
% Btypical = B_0 * 0.5^4 = 0.1B_0 = 400G. 
% Elsasser = B0^2 / (4pi rho eta Omega) = 400^2/ (10*1* 10^12 * 3 10^-6) =  160000/3x10^7 

%%%%%%%%%%%%%%%%%%%%%%%%%%%%%%%%%%%%%%%%%%%%%%%%%%%%%%%%%%%%%%%%%%%%%%%%%%%
\bibliography{tachocline}

\end{document}